\documentclass{article}

\usepackage{ifthen}
\usepackage[dvips]{graphicx}
\def\DMS#1,#2,#3{\includegraphics[width=#1pt,height=#2pt]}

\newcommand{\Fig}[3]
{\bigskip
\begin{figure}\begin{center}
\ifthenelse{\equal{#2}{}}
{\includegraphics[width=200pt,height=200pt,draft]{#1.eps}}
{\DMS#2,0{#1.eps}}
\caption{{\footnotesize{#3}}}\label{#1}\end{center}
\end{figure}
\bigskip
}

\newcommand{\Figeps}[3]
{\bigskip
\begin{figure}\begin{center}
\ifthenelse{\equal{#2}{}}{\includegraphics{#1.eps}}
{\DMS#2,0{#1.eps}}
\caption{{\footnotesize{#3}}}\label{#1}\end{center}
\end{figure}
\bigskip
}

\def\be{\begin{eqnarray}}
\def\ee{\end{eqnarray}}
\def\nn{\nonumber}

\hoffset= - 1.2in         

\begin{document}

\hfill ITEP/TH-01/07

\bigskip

\centerline{\Large{On the shapes of elementary domains
or}}
\smallskip
\centerline{\Large{why Mandelbrot Set is made from
almost ideal circles?
}}

\bigskip

\centerline{\it V.Dolotin and A.Morozov}

\bigskip

\centerline{ITEP, Moscow, Russia}

\bigskip

\centerline{ABSTRACT}

\bigskip

Direct look at the celebrated "chaotic"
Mandelbrot Set in Fig..\ref{Mand2} immediately reveals that
it is a collection of almost ideal circles and
cardioids, unified in a specific {\it forest} structure.
In /hep-th/9501235 a systematic algebro-geometric
approach was developed to the study of generic
Mandelbrot sets,
but emergency of nearly ideal circles in the special case
of the family \ $x^2+c\ $\ was not fully explained.
In the present paper the shape
of the elementary constituents  of Mandelbrot
Set is explicitly {\it calculated}, and difference
between the  shapes of {\it root} and {\it descendant}
domains (cardioids and circles respectively)
is explained. Such qualitative difference persists
for all other Mandelbrot sets: descendant domains
always have one less cusp than the root ones.
Details of the phase transition between different
Mandelbrot sets are explicitly demonstrated, including
overlaps between elementary domains and dynamics of
attraction/repulsion regions.
Explicit examples of $3$-dimensional sections of
Universal Mandelbrot Set are given.
Also a systematic small-size approximation is developed
for evaluation of various Feigenbaum indices.

\bigskip

\bigskip

\tableofcontents

\section{Introduction}

The question of how dynamics of a physical system depends
on the choice of its Hamiltonian is one of the most important
in theoretical and mathematical physics.
Its significance is only enhanced by the fact that
in modern theory dynamics is considered not only in physical
time, but in many other variables, including the coupling constants
of a theory and the shape of functional-integration domain
(the so-called renormalization-group dynamics \cite{RG}).
Normally dynamics
is described in terms of a phase portrait
or of eigenstates configuration for classical and
quantum systems respectively, and the question is how
these portraits and configurations change under
variation of the Hamiltonian.
It is well known that this change is not everywhere smooth:
at particular "critical" or "bifurcation" points
in the space of Hamiltonians the phase portraits get
reshuffled and change {\it qualitatively}, not only
{\it quantitatively}: this phenomenon is also known as
"phase transition".
Normally these bifurcations are described in terms of
the change of stability properties of various periodic
orbits (including fixed points, cycles and "strange
attractors").
More delicate information is provided by the study of
intersections of {\it unstable} orbits, but it is
a little more difficult to extract.

Dynamical systems are much better studied in the case of
discrete dynamics: this reveals many properties,
which get hidden in transition to continuous evolution.
In other words, this resolves ambiguities of continuous
dynamics: there are many different discrete dynamics
behind a single continuous one, and to reveal the properties
of the latter it is often needed to look at the whole
variety of the former.
In classical case discrete dynamics (with a time-independent
"Hamiltonian") is the theory of iterated maps:
\be
x \rightarrow f(x) \rightarrow f^{\circ 2}(x) = f\big(f(x)\big)
\rightarrow \ldots \rightarrow f^{\circ p}(x) =
f\Big(f^{\circ(p-1)}(x)\Big) \rightarrow \ldots
\ee
and is actually a branch of algebraic geometry \cite{DM}
(for generalization of \cite{DM} to discrete
dynamics of many variables see  sections 7 and 8 of \cite{nolal}).
According to \cite{DM},
the structure of the phase portrait is controlled by the
{\it Julia set}: collection of all periodic orbits
of the map $f$ in the $x$ space, i.e. of all roots of
all functions
\be
F_p(x) = f^{\circ p}(x)-x
\label{Fpdef}
\ee
Therefore the {\it Universal Mandelbrot set} (UMS),
consisting of all points
$f(x)$ in the space of functions (Hamiltonians) where some
two periodic orbits coincide, can be alternatively
characterized as the {\it Universal Discriminantal variety}
formed by the roots of various resultants
$R_x\Big(F_{pm}(x)/F_p(x),F_p(x)\Big)$.
This is almost a tautological identification,
since by definition the resultant of two functions vanishes
whenever they have a common zero,
still it establishes relation between {\it a priori} different
sciences: the theory of phase transitions and
algebraic geometry.

Usually considerations are restricted to
particular {\it sections} of the Universal
Mandelbrot set, by choosing specific one-dimensional families of
functions: see Figs.\ref{Mand2}-\ref{0Man4} for the three famous
examples, $f(x;c) = x^d + c$ with $d=2$, $d=3$ and $d=4$.
In Fig.\ref{symmebre} we show also the result of a deviation from
this simple form.
We keep the name "Mandelbrot set" for any of such
one-complex-dimensional sections of infinite-dimensional UMS,
while "Mandelbrot Set" (with two capital letters) refers to the
original example in Fig.\ref{Mand2}. Today all kinds of {\it
experimental data} about these sets can be obtained with the help
of available computer programs, like {\it Fractal Explorer}
\cite{FE}, which is used to make Figures \ref{Mand2}-\ref{0Man4}
in the present paper\footnote{
However, one should be careful in using this program for
non-canonical families, like in Fig.\ref{symmebre},
see \cite{DM} and s.\ref{interpol} below for explanations.}.

\bigskip

\Figeps{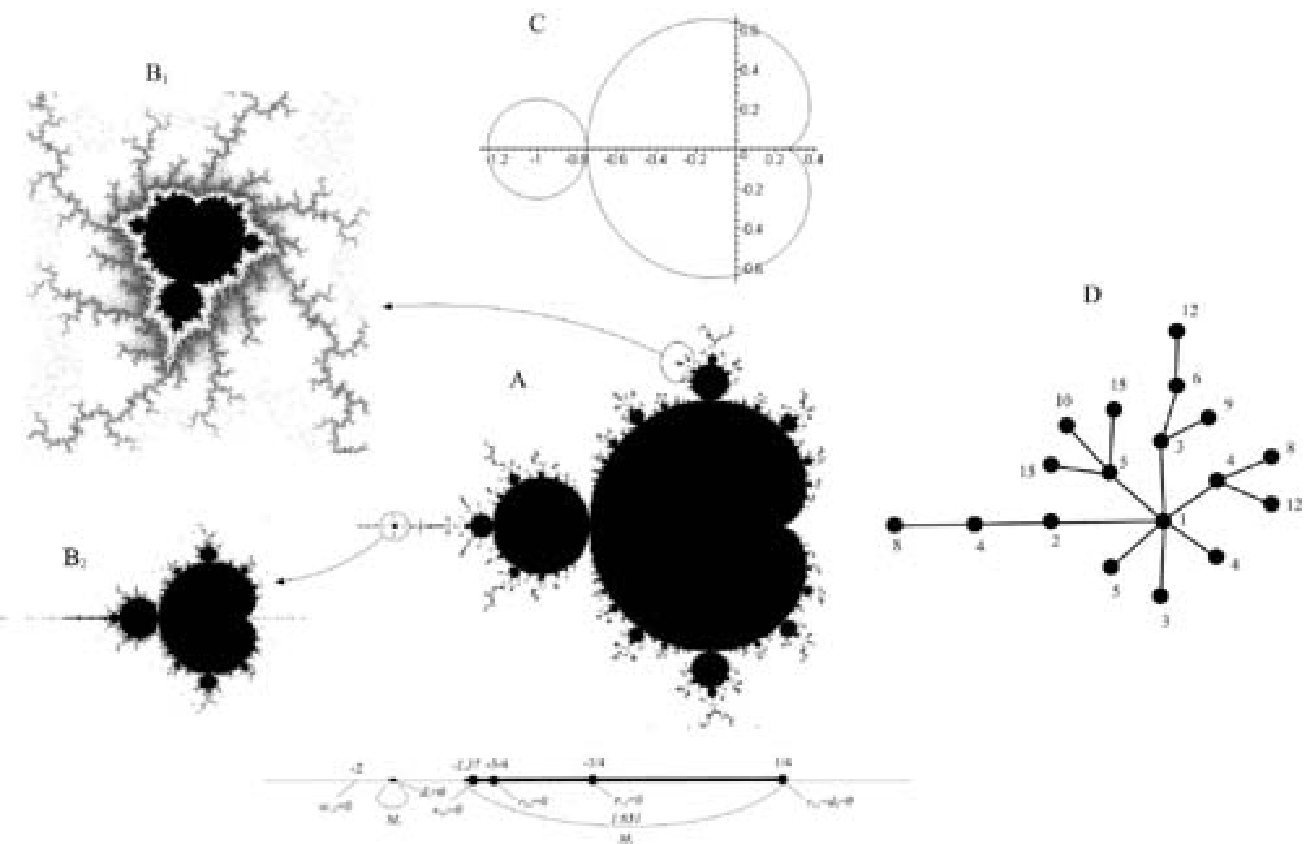}
{500,328}
{\footnotesize {\bf A.} Mandelbrot set
${\cal M}_2$ for the family $f(x;c)=x^2+c$ \cite{Mand}. The
boundary of the black domain in the complex plane of $c$-variable
consists of all values of $c$ where {\it Julia set} is reshuffled:
as explained in \cite{DM} this happens when a stable orbit crosses
an unstable one. A given orbit ${\cal O}$ is stable within and
{\it elementary domain}, which is -- with incredibly good accuracy
-- either a cardioid $c-c_{{\cal O}} = r_{{\cal O}}e^{i\phi}\left(
1 - \frac{1}{2}e^{i\phi}\right)$ at the center (root) of a cluster
or a circle $c-c_{{\cal O}} = r_{{\cal O}}e^{i\phi}$ at the
non-root nodes of the tree. Explanation of this fact (that only
these two shapes occur and exactly in these roles: at roots and
higher nodes respectively) is the task of the present paper.
Projection to the line of real $c$ is also shown. \ \ \ \ {\bf B.}
Two pieces of the Mandelbrot set ${\cal M}_2$ under microscope:
exactly the same structures are seen as the central cluster in
{\bf A}, with the same central cardioids and attached circles.
There are infinitely many such structures of different sizes
$r_{{\cal O}}$ in ${\cal M}_2$, since $r_{{\cal O}}$ are very
small, they are not actually seen in {\bf A}, but can be easily
studied with the help of the {\it Fractal Explorer} \cite{FE}.
More numerical characteristics of the lowest elementary domains
are collected in a Table in s.\ref{accu}. \ \ \ \ {\bf C.} Domains
$(1)$ and $(2,1)$, obtained as numerical solutions of exact
equations (\ref{shapeeq}), see s.\ref{homsol}. They coincide with
domains, seen in {\bf A}. This is non-trivial, because pictures
{\bf A} and {\bf B} are obtained by absolutely different procedure
(actually, black region consists of points $c$, with limited
sequences $f^{\otimes n}(c)$), and there is no {\it a priori}
reason for any parts of them to satisfy any kind of algebraic
equations. Separation of Mandelbrot sets into domains, possessing
an algebraic description, in particular the relation between
pictures ${\bf A}$ and ${\bf C}$, is important property of
iterated maps. \ \ \ \ {\bf D.} Tree structure of the central
cluster (only a few lowest branches are shown), each branching
occurs at the center of a new elementary domain, and the number at
the vertex is the order of the periodic orbit, which is stable
inside this domain. Thus elementary domains are naturally labeled
by sequences of divisors, leading to the root of the tree. All
other clusters are represented by exactly the same trees, only
numbers are multiplied by the order of the root orbit. Thus entire
${\cal M}_2$ has a natural {\it forest} structure. }

\bigskip

\Figeps{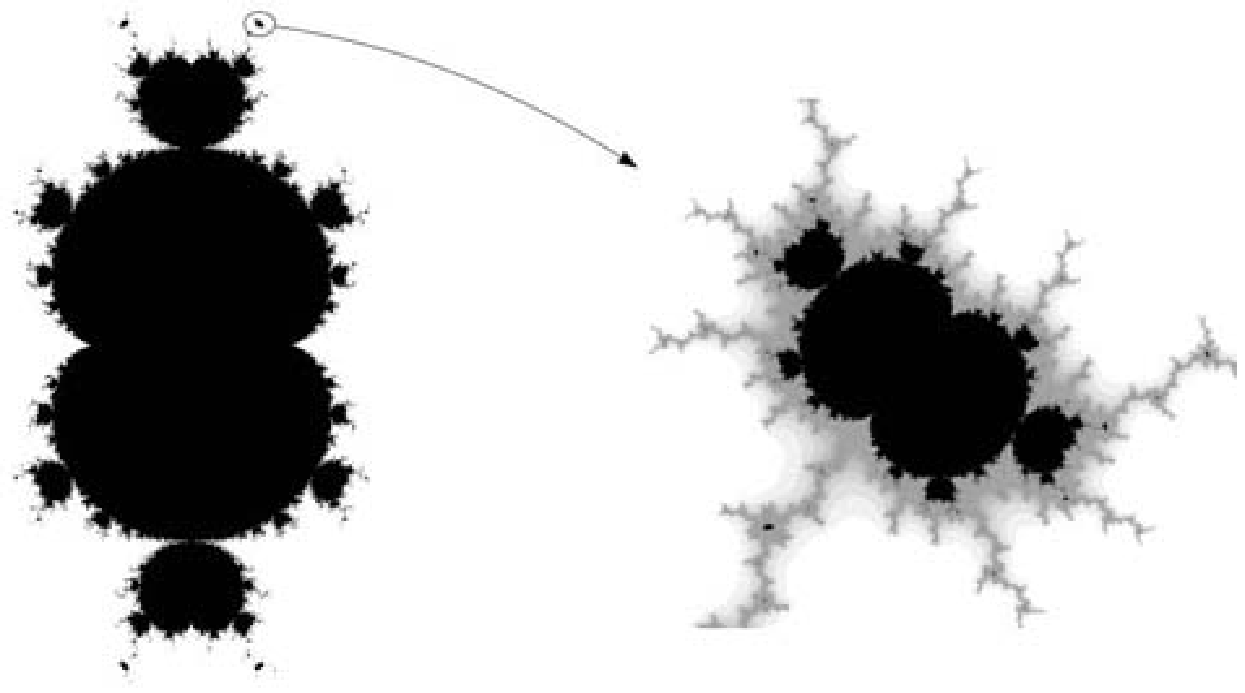}
{450,255}
{\footnotesize
Mandelbrot set ${\cal M}_3$ for the family $f(x;c)=x^3+c$.
Everything said about ${\cal M}_2$ is true in this case,
only the place of simple cardioids at  roots of clusters
is taken by the two-cusp ones
$c-c_{{\cal O}} \approx r_{{\cal O}}e^{i\phi}\left(
1 - \frac{1}{3}e^{2i\phi}\right)$.
Descendant domains are nicely approximated by single-cusp
cardioids
$c-c_{{\cal O}} \approx r_{{\cal O}}e^{i\phi}\left(
1 - \frac{1}{2}e^{i\phi}\right)$, see eq.(\ref{cuca}).
No circles are present.}

\bigskip

\Fig{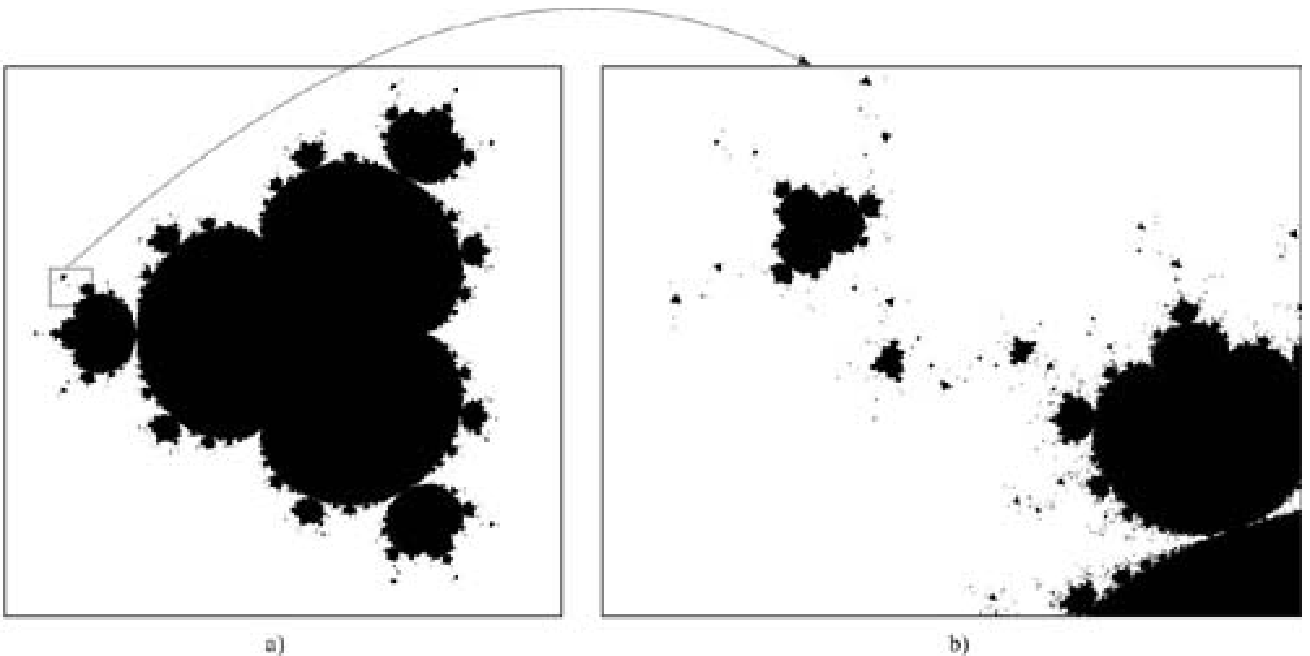}
{450,224}
{\footnotesize
Mandelbrot set ${\cal M}_4$ for the family $f(x;c)=x^4+c$.
Everything said about ${\cal M}_2$ is true in this case,
only the place of simple cardioids at  roots of clusters
is taken by the three-cusp ones
$c-c_{{\cal O}} \approx r_{{\cal O}}e^{i\phi}\left(
1 - \frac{1}{4}e^{3i\phi}\right)$.
Descendant domains have the shape of deformed 2-cusp
cardioid,
$c-c_{{\cal O}} = r_{{\cal O}}e^{i\phi}\left(
1 - ae^{i\phi} + \frac{b}{3}e^{2i\phi}\right)$
with $a = 2^{-4/3} \approx 0.40\ldots$ and
$b \approx 11/(9\cdot 2^{2/3}) = 0.77\ldots$,
see eq.(\ref{quca}).
No circles or single-cusp cardioids are present.
}

\bigskip

\Figeps{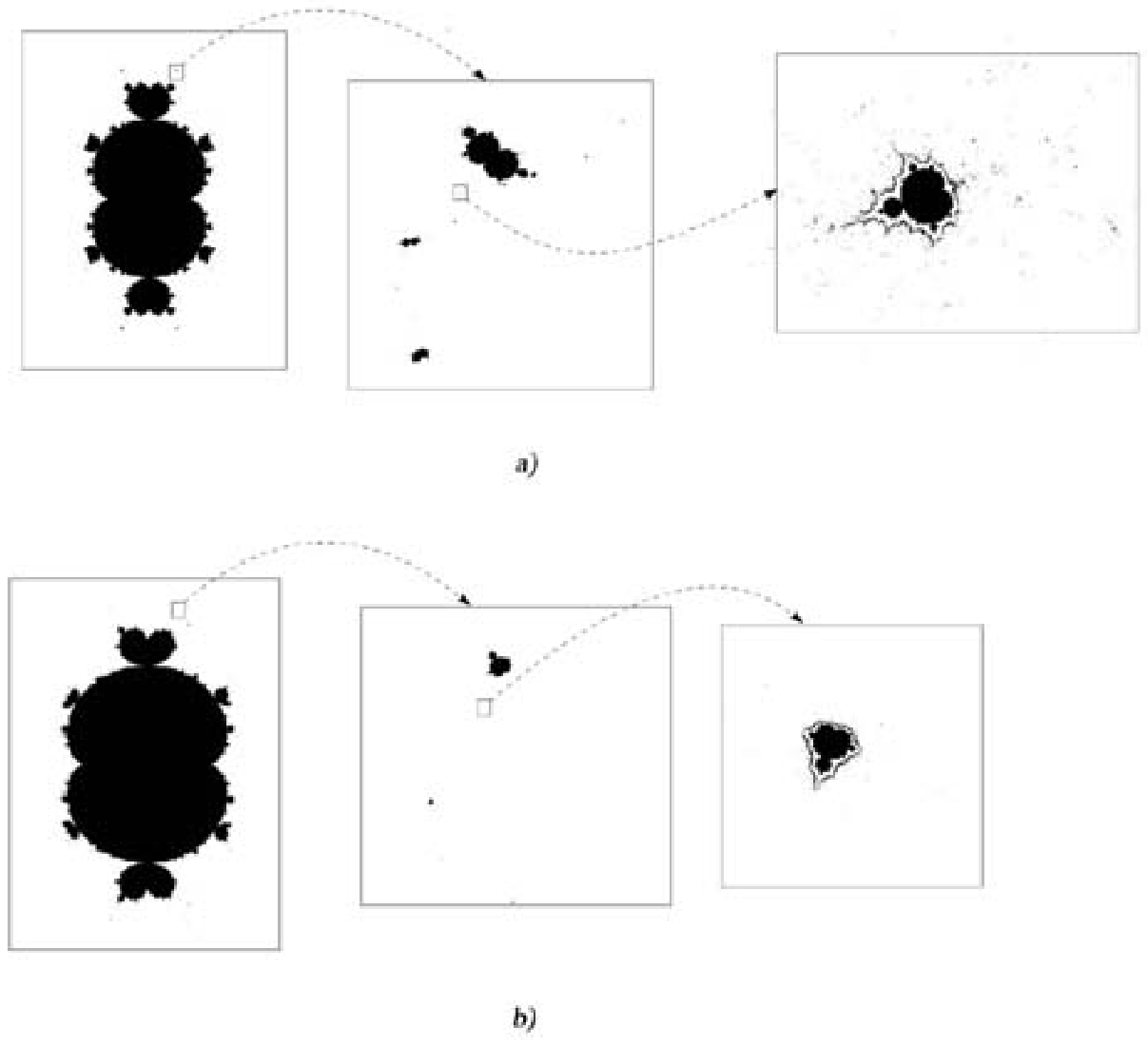}
{450,385}
{\footnotesize
Mandelbrot set ${\cal M}_{3-1}$ for the family
$f(x;c)=ax^3+(1-a)x^2+c$
with two different values of additional
parameter $a=4/5$ {\bf (A)} and $a=2/3$ {\bf (B)}.
Everything said about ${\cal M}_2$ is true in this case,
only in addition to the simple single-cusp cardioids
$c-c_{{\cal O}} = r_{{\cal O}}e^{i\phi}\left(
1 - \frac{1}{2}e^{i\phi}\right)$
in the role of central (root) domains there are also
two-cusp ones,
$c-c_{{\cal O}} = r_{{\cal O}}e^{i\phi}\left(
1 - \frac{1}{3}e^{2i\phi}\right)$.
Moreover, for distinguished value $a=1$, see Fig.\ref{0Man3},
simple cardioids do not appear as central (root)
domains of individual clusters at all --
their place at roots is taken by
2-cusp curves. Instead for $a=1$ the single-cusp cardioids
fully replace circles in the role of descendant domains.
For smaller values of $a$ the simple cardioids start to emerge
as roots (and circles -- as descendants), but in remote clusters,
at large distances from the central domain.
\ \ \ The central cluster in these pictures look a little
asymmetric: this is wrong, and is an artefact of the
erroneous algorithm, used to construct Mandelbrot sets
by {\it Fractal Explorer}, see introductory remarks to
s.\ref{interpol} below.
}

Mandelbrot sets are often considered as typical examples of
"fractal structures", serving mostly for admiration, philosophical
speculations and, perhaps, numerical exercises.
However, as explained in \cite{DM}, they can actually be
subjected to systematic scientific investigation, in the style
of {\it experimental mathematics}, with questions coming from
direct observations and numerical experiments, and rigorous answers
provided by knowledge of underlying algebro-geometric structures.
Our presentation below can be considered as an example
of this increasingly important approach to modern mathematical
physics problems.

As explained in \cite{DM} -- and clearly seen in
Figs.\ref{Mand2}-\ref{symmebre}, -- the Mandelbrot set consists
of infinitely many separated {\it clusters},
of which only the central one is well seen in the main picture,
while examples of smaller clusters
are shown in auxiliary pictures with enhanced resolution.
Though separated, clusters form a well organized
structure: they are connected by "trails", populated with
other clusters.
Further, each cluster has its own {\it tree} structure,
Fig.\ref{Mand2}.D, with two types of {\it elementary domains}:
one type at the root of the tree and another type at all higher
nodes (we call them {\it descendants}).
Fig.\ref{symmebre} demonstrates that a given Mandelbrot set
can contain different types of clusters, while
for special families\ $f(x;c) = x^d+c$, where maps possess
additional $Z_{d-1}$ symmetry $x \rightarrow e^{2\pi i n/(d-1)}x$,
all clusters are of the same type.
In Mandelbrot Set (i.e. for $d=2$) the root
{\it elementary domains} are nearly ideal cardioids,
Fig.\ref{cardi}.A,
\be
c-c_{{\cal O}} = r_{{\cal O}}e^{i\phi}\left(
1 - \frac{1}{2}\,e^{i\phi}\right)
\label{card2}
\ee
while descendants are nearly ideal circles
\be
c-c_{{\cal O}} = r_{{\cal O}}e^{i\phi}
\label{circdef}
\ee
For $d>2$ we have nearly ideal $(d-1)$-cusp cardioids,
Fig.\ref{cardi},
\be
c-c_{{\cal O}} = r_{{\cal O}}e^{i\phi}\left(
1 - \frac{1}{d}\,e^{i(d-1)\phi}\right)
\label{carddef}
\ee
at roots, while descendant domains are
{\it deformed} cardioids with $d-2$ cusps
and some non-vanishing coefficients $a_k$ in
\be
c-c_{{\cal O}} = r_{{\cal O}}e^{i\phi}\left(
1  + \sum_{k=1}^{d-3}a_ke^{ik\phi} -
\frac{1}{d-1}\,e^{i(d-2)\phi}\right)
\label{defcarddef}
\ee
Actual shapes of the domains slightly deviate from these
ideal cardioids and circles and depend on particular cluster
and node, but deviations are at the level of a few percents
at most.

\bigskip

\Fig{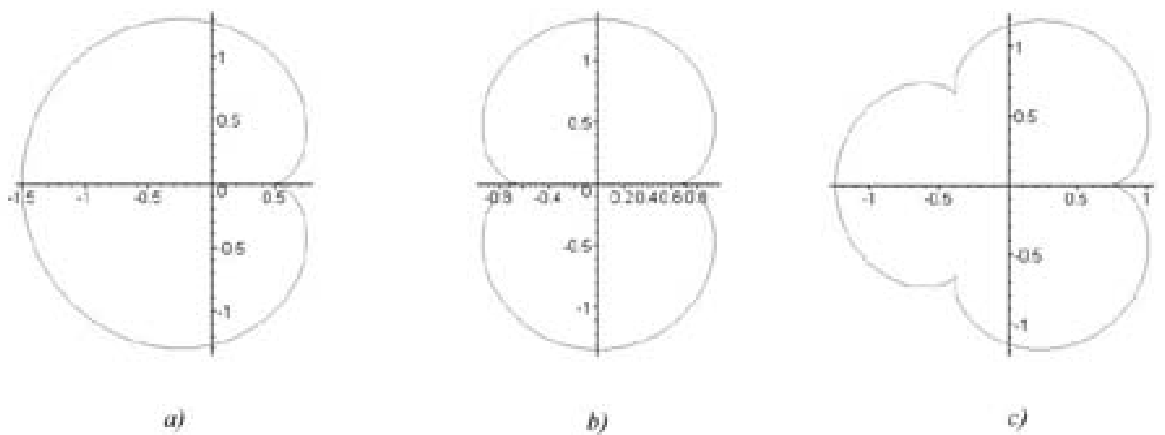}
{400,130}
{\footnotesize Cardioid curves described by
the equations (\ref{card2}) and (\ref{cardi}). These pictures
reproduce the shapes of the central domains in Mandelbrot sets
${\cal M}_d$. \ \ {\bf A.} $d=2$. This is the central domain of
${\cal M}_2$ in Fig.\ref{Mand2}. \ \ {\bf B.} $d=3$. This is the
central domain of ${\cal M}_3$ in Fig.\ref{0Man3}. \ \ {\bf C.}
$d=4$. This is the central domain of ${\cal M}_4$ in
Fig.\ref{0Man4}. }

\bigskip

Each elementary domain is associated with some periodic orbit
${\cal O}$ of the map $f$,
which is stable exactly within the domain.
Thus the domain is labeled by the order $p$ of this orbit.
This fact allows one to give an explicit {\it analytical}
description of the domain's shape, see eq.(\ref{shapeeq}) below.
However, there are many different periodic orbits with the same
$p$ and thus many elementary domains with the same $p$.
They differ by the choice of the root $c_p$ of the equation
$f_p(c) \equiv f^{\circ p}(x=x_{cr};c) = x_{cr}$, which lies in the
"center" of the domain
($x_{cr}$ is the root of $f'(x)$, $f'(x_{cr},c)=0$, in most examples
below $x_{cr}=0$).
Some of these order-$p$ domains are at roots of some clusters,
some at higher-level nodes, but in that case the root
of the cluster should be associated with a divisor of $p$.
Actually the tree underlying the cluster is the {\it divisor tree},
and the entire forest structure
(i.e. collection of trees, associated with all clusters)
is that of {\it divisor forest} of natural numbers.
Accordingly, elementary domain is labeled by a sequence of
integers $\Big((p_r\cdot m_1\cdot \ldots\cdot m_k)_{\alpha_{r,k}},
\ \ldots,\ (p_r\cdot m_1)_{\alpha_{r,1}},\ (p_r)_{\alpha_r}\Big)$,
to be read from rights to left.
Sequence of multiples
\ $[m_k,\ldots, m_1]\ $ characterizes domain's position
in the tree, $k$ is the distance from the root,
$p=(p_r\cdot m_1\cdot \ldots\cdot m_k)$
is the period of the corresponding orbit,
and $p_r$ is the order
of the orbit, associated with the root domain.
Since there can be many different root domains with the same $p_r$
in particular Mandelbrot set
and many different descendants with the same $p$ in a given tree,
there are additional labels $\alpha$, distinguishing between
different trees in the forest with the same order $p_r$ at roots
and between different branches with the same $p$ in each tree.
While divisor trees are the same for all particular Mandelbrot sets,
collections of ${\alpha}$'s are different: they are defined by
the way the given section crosses the $p$-variety in the
Universal Mandelbrot set, since it can be crossed many times,
there are many traces of the same variety in the given section
(in the role of either root or descendant domains)
-- and this is the origin of the {\it forest} and of the $\alpha$
parameters, which at the level of particular Mandelbrot set look
somewhat arbitrary.
At the level of UMS there is a single divisor tree and a section
intersects it many times and
cuts it into many similar trees: any cut-off branch looks like a
separate tree and gives rise to a separate cluster.
The memory of their common origin at UMS level is preserved
in the {\it trail} structure, connecting the clusters, but its
detailed description is not yet available.

Two adjacent elementary domains {\it touch} at exactly one point
(i.e. along a complex-codimension-one variety in UMS),
where their corresponding orbits cross and exchange stability.
This important statement, however, needs to be treated carefully:
as we shall see, in general (beyond the $x^d+c$ families)
the elementary domains can {\it overlap}: there can be several
stable orbits at the same value of $c$.
This means that arbitrarily chosen $c$ is not a good
coordinate on a Mandelbrot set, which is actually a fibration
over a region in the complex-$c$ plane than a region {\it per se}.
Fibration structure is inherited from {\it Julia sheaf} over the
Mandelbrot set \cite{DM}.
In this general situation the word {\it touch} is not fully
informative:
when different domains seem to overlap,
they rather lie in different
fibers over the same region on $c$-plane, and these fibers are
sewed exactly at a single point, where the orbis cross.
What we can show in simple $2d$ pictures are {\it projections}
of the domains, these projections can overlap and {\it touch}
at the orbit-crossing point.
{\it Touch} means that the tangents to two domains are collinear,
in practice this can be either a smooth touching (typical for
crossing of orbits of different orders) or a cusp (when the orders
are the same).

Crossing of orbits is possible only when the order of the smaller
one (with domain lying closer to the root in the cluster) divides
the order of the bigger one. Analytically, crossing takes place
at the root of associated {\it resultant}.
If the two orders differ by a factor of $2$, this is the celebrated
period-doubling bifurcation \cite{pdb,LL} -- and the chain of exactly
such bifurcations occurs along the real line in Fig.\ref{Mand2},--
but in fact {\it doubling} is in no way distinguished: bifurcation
can multiply period by arbitrary integer $m$.
Crossing of {\it unstable} orbits is not seen at the level of
Mandelbrot sets shown in Figs.\ref{Mand2}-\ref{0Man4},--
to study these phenomena
(also essential for bifurcations of Julia sets) the full
(or {\it Grand}) Mandelbrot set should be considered.
Actually, behind UMS there are even more fundamental entities:
the Universal Julia Sheaf (UJS), consisting of all periodic
orbits of all orders "hanged" over the UMS, and
the Grand UJS, including also all {\it pre-orbits} of periodic orbits.
UMS is a projection of UJS, obtained by neglect of the phase-space
dimension $x$, where the orbits live, and, as any projection, it
can and does suffer from overlap ambiguities, namely, when
two different stable orbits coexist at the same point of UMS.

\bigskip

We refer to \cite{DM} for further details and explanations.
The task of this paper is to provide close-to-Earth
illustrations of somewhat abstract formulations from the
previous paragraphs and to fill some of the gaps
left in \cite{DM}, which concern three closely related
subjects:

(i) the shape of elementary domains;

(ii) Feigenbaum indices,
defining the ratio of sizes of adjacent elementary domains
(immediate descendant as compared to its parent)
from the ratio of the corresponding orders;

(iii) reshuffling of Mandelbrot set and its elementary sets
under the change of selected family of functions, i.e.
new properties of 
$2_C$-dimensional sections of
Universal Mandelbrot set as compared to the 
$1_C$-dimensional sections.

In fact these subjects capture the main aspects of the general
theory and at the same time they can be considered by simple
methods of theoretical physics with minimal involvement of
abstract algebro-geometric constructions.
Even resultant and discriminant analysis, which was the
main machinery in \cite{DM}, will be at periphery of our
simplified presentation in this paper.

As a key puzzle and a starting point for all considerations
we choose the question, posed in the title of this paper:
why exactly the cardioids (\ref{card2})
and circles (\ref{circdef}) and exactly in the right places
in the {\it divisor forest} emerge as the shapes of elementary
domains of the Mandelbrot Set in Fig.\ref{Mand2}, and
how this picture is continuously deformed
into Figs.\ref{0Man3} and \ref{0Man4}.

As already shown in \cite{DM}, cardioids (\ref{carddef})
{\it exactly} describe the {\it central} domains $(1)$
for the families $x^d+c$, while in description of all other
domains they  straightforwardly appear in the
"small-size approximation" (SSA) to exact shape-defining
eqs.(\ref{shapeeq}).
In what follows we

-- explain, why for the\ $x^d+c\ $ families the cardioids
(\ref{carddef}) do not exhaust all possible shapes:
deformed cardioids (\ref{defcarddef}) with one less cusp
are also allowed;

-- explain, why (\ref{carddef}) appear exactly at {\it roots}
of clusters, while all descendants have one cusp less:
instead of this lacking cusp a descendant domain has a
merging-point to the parent domain;

-- provide a detailed description of interpolation between
Mandelbrot sets and Julia sheafs
for the families $x^2+c$ and $x^3+c$
(actually only the orbits of two lowest orders $p=1$ and $p=2$
will be analyzed, but this is already enough to reveal many
interesting details of the process);

-- demonstrate inaccuracy of {\it Fractal Explorer} \cite{FE}
(and thus the underlying text-book interpretations of
Mandelbrot sets) in application to UMS studies and appeal
for the writing of corrected and fully adequate computer programs
on the base of improved knowledge provided by \cite{DM};

-- demonstrate high accuracy of the small-size approximation (SSA)
in evaluation of various characteristics of the Mandelbrot Set
by comparing its predictions for the (complex-valued) sizes
$r_{{\cal O}}$ and Feigenbaum indices with exact answers
(when they are already available) and
experimental data provided by the {\it Fractal Explorer} \cite{FE}.

Concerning the last story -- about the SSA -- it deserves saying
that no {\it theoretical} explanation for its spectacular accuracy
is known:
particular corrections are not small, but various
corrections always combine into a small quantity, whenever
characteristics of Mandelbrot Set are evaluated.
At the same time, as explained in \cite{DM},
SSA fails completely in description of Julia sets; still it
describes reasonably well the Mandelbrot sets for the families
$x^d+c$ with $d>2$ (though some qualitative properties are
spoiled, e.g. cusps are somewhat smoothed), but works much worse
for interpolations between different $d$.
In any case, today SSA remains the only available tool
for theoretical investigation of high branches in divisor tree,
in particular for approximate evaluation
(rather than {\it measuring})
of various Feigenbaum indices --
and for this purpose it works spectacularly
well, even for $d > 2$ and even for interpolations.
Still rigorous algebro-geometric theory of Feigenbaum indices
remains to be found.

\bigskip

We begin in s.\ref{qual} from qualitative description of elementary
domains supported by the limited amount of {\it exactly-solvable}
examples in s.\ref{exact}: these include some non-trivial cases
and, as usual, provide the solid ground for future approximate
considerations.
Section \ref{interpol} is devoted to interpolation $\{ax^3+bx^2+c\}$
between the two best-known Mandelbrot sets: $\{x^2+c\}$ and
$\{x^3+c\}$.
Other examples of $3_R$-dimensional sections of UMS
(actually of its central domain) are also given in this section.
Then in s.\ref{ssasec} we introduce the small-size approximation
and present some calculations for the Mandelbrot Set.
Their results are compared with experimental data in s.\ref{accu}.
After some more borrowing SSA calculations in s.\ref{shapes2},
we discuss Feigenbaum indices in s.\ref{Feig}.
Section \ref{dfam} is devoted to SSA consideration of some other
Mandelbrot sets.
Brief conclusions are collected in s.\ref{conc}.

\section{The shape of elementary domains. Qualitative description
\label{qual}}

\setcounter{equation}{0}

\subsection{Defining equations}

According to eqs.(10) and (38) of \cite{DM},
the boundary of an elementary domain satisfies
a pair of equations:
\be
f \in {\cal M} \ \Leftrightarrow \
\left\{\begin{array}{c}
F_p^\prime(x) + 1 = e^{i\theta} \\
F_p(x) = 0
\end{array} \right.
\label{shapeeq}
\ee
Here $F_p(x) = f^{\circ p}(x) - x$,
prime denotes $x$-derivative
and $\theta$ is a {\it real-valued} angle-parameter used to
coordinatize the boundary of the domain (it can actually
vary between $0$ and  a multiple of $2\pi$, see below).
After $x$ is excluded from the pair of equations (\ref{shapeeq}),
we obtain a real-codimension-one hypersurface
in the space of functions $f$, i.e. a collection of curves
$c(\theta)$ in $1_C$-dimensional Mandelbrot set.
Particular curves -- branches of $c(\theta)$ -- are boundaries
of particular elementary domains of the order $p$,
root and descendant.

\subsection{Cusps}

Even if function $c(\theta)$ is smooth, the corresponding
curve in the complex-$c$ plane can be singular.
Generical singularity is self-intersection, which takes
place when
$c(\theta_1) = c(\theta_2)$ for $\theta_1\neq\theta_2$.
Of interest for us are {\it cusps}:
degenerated self-intersections, appearing in the limit when
$\theta_2 \rightarrow \theta_1$, i.e. when
$\frac{dc}{d\theta}(\theta_0) =0$ at some $\theta_0$.
In the vicinity of such point
$\sigma(\vartheta) = c(\theta) - c(\theta_0) =
a\vartheta^2 + b \vartheta^3 + 
\ldots$, where $\vartheta = \theta - \theta_0$.
This means that
\be
{\rm Re}\left(\frac{\sigma}{a}\right) =
\vartheta^2 + \ldots,\ \ \ \ {\rm while} \ \ \ \
{\rm Im}\left(\frac{\sigma}{a}\right) =
{\rm Im}\left(\frac{b}{a}\right)\vartheta^3 +
\ldots
\ee
i.e.
\be
{\rm Im}\left(\frac{\sigma}{a}\right)\sim
{\rm Im}\left(\frac{b}{a}\right)
\left\{{\rm Re}\left(\frac{\sigma}{a}\right)\right\}^{3/2}
\ee
Thus we see that a cusp emerges at points where $dc/d\theta = 0$
and its orientation in the complex-$c$ plane is defined by the
phase of the complex-valued parameter $a$.

If $\ {\rm Im}(b/a)=0$, i.e.
\be
{\rm Im}\left(\frac{d^3c/d\theta^3}{dc/d\theta}\right)=0
\label{cuspint}
\ee
along with $dc/d\theta = 0$, then a self-intersection point
collides with the cusp and disappears.

\subsection{Cardioids}

Cardioids are represented by polynomials of the unimodular variable,
they form the simplest natural class of curves with cusps.

For {\bf quadratic  cardioid},
\be
c = r\left(e^{i\phi} + ae^{2i\phi}\right) \ = \
\frac{r}{4a}\Big((1+2ae^{i\phi})^2 - 1\Big),
\ee
derivative vanishes, $dc/d\phi = 0$, when $2ae^{i\phi} = -1$.
This never happens if $|a| \neq \frac{1}{2}$.
Thus a cusp (and exactly one) occurs only when $|a|=\frac{1}{2}$,
the curve is everywhere smooth for $|a|<\frac{1}{2}$ and
possesses one self-intersection for $|a|>\frac{1}{2}$.

\bigskip

For {\bf cubic  cardioid},
\be
c = r\left(e^{i\phi} + ae^{2i\phi} + be^{3i\phi}\right),
\label{3card}
\ee
derivative $dc/d\phi = 0$ vanishes when
\be
1+2ae^{i\phi} + 3be^{2i\phi} = 0,
\label{3carder}
\ee
i.e. when the r.h.s. $e^{-i\phi} = -a \pm \sqrt{a^2-3b}\ $
has unit modulus.
If $a$ and $b$ are real, then cusp can occur when either
$1+2a+3b=0$ (then there is one cusp at $\phi = 0$, unless
$a=0$ and $b=-\frac{1}{3}$, when another cusp appears at
$\phi = \pi$, see Fig.\ref{cardicusp})
or $-1<a<1,\ b=+1/3$ (then two cusps arise at $\phi=\pm\phi_0
\neq 0,\pi$).
In general, (\ref{3carder}) defines
a hypersurface of real codimension one in the space
of complex parameters $a$ and $b$ (parameterized by $\phi$),
where cubic cardioids (\ref{3card}) have cusps (one or two).

Transition point (\ref{cuspint}) between a phase with
and without self-intersection is defined by a system of two
equations,
$$
\left\{\begin{array}{c} {dc}/{d\phi} \sim 1+2a+3b = 0 \\
{d^3c}/{d\phi^3} \sim 1+8a+27b = 0
\end{array}\right.
$$
i.e. $a=-4/5$ and $b=1/5$, see Fig.\ref{cardicusp}.

MAPLE program for cardioid studies, which was used to generate
Figs.\ref{cardi} and \ref{cardicusp}, can be found in Appendix to
this paper, see s.\ref{MAPcard}.

\bigskip

\Fig{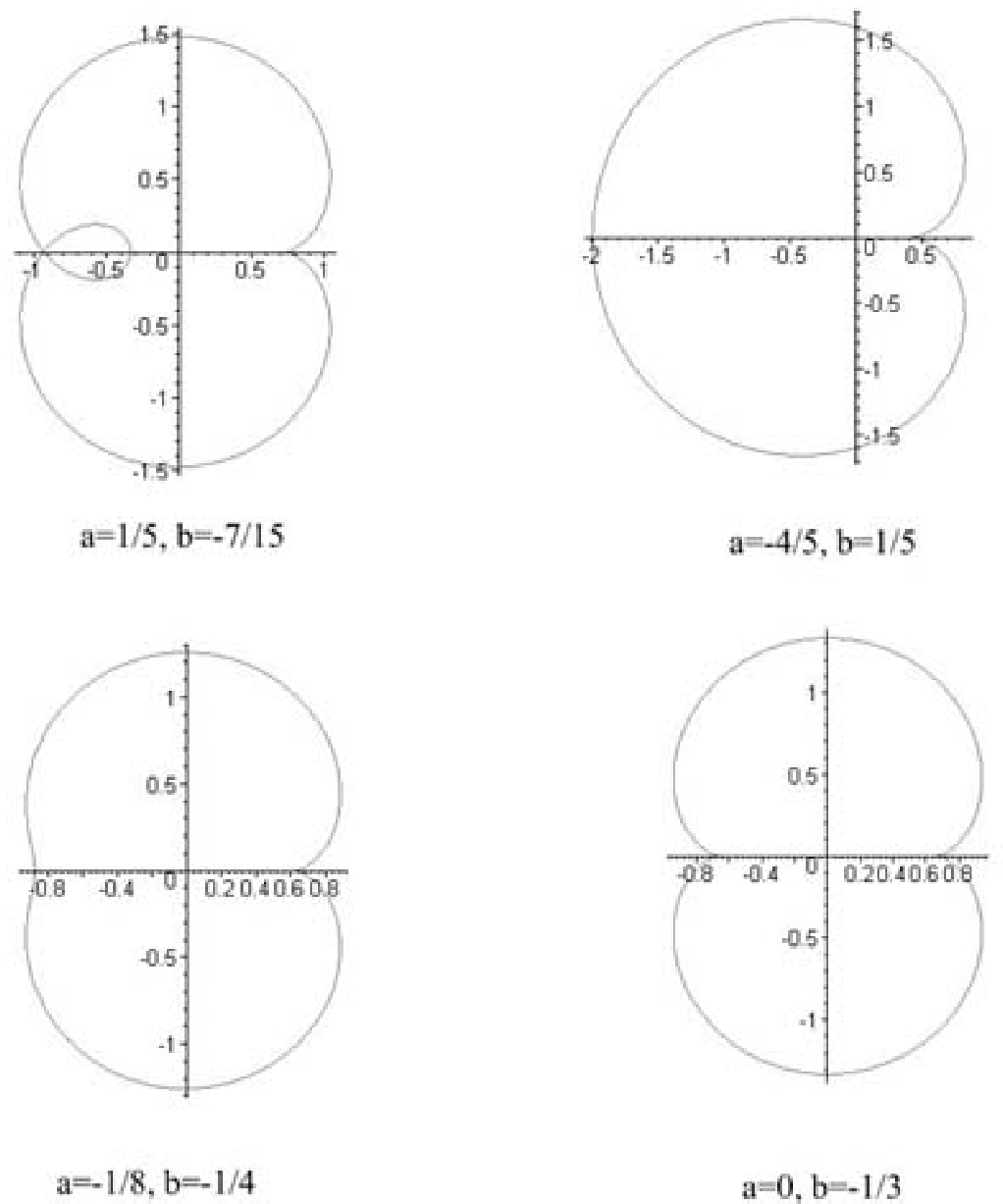}{450,422}
{\footnotesize
Cardioids with cusps and self-intersections.
The cusp-possessing subset in the family (\ref{3card})
with $1+2a+3b=0$ is shown.\ \
{\bf A.} $a=\frac{1}{5}$, $b=-\frac{7}{15}$.
Both cusp and self-intersection are present.\ \
{\bf B.} $a=-\frac{4}{5}$, $b=\frac{1}{5}$. This is the point where
self-intersection point hits the cusp and disappears.\ \
{\bf C.} $a=-\frac{1}{8}$, $b=-\frac{1}{4}$.
A single cusp is present at $\phi = 0$.
\ \
{\bf D.} $a=0$, $b=-\frac{1}{3}$. The second cusp appears at
$\phi = \pi$.
}

\subsection{Cusps in the boundaries of elementary domains
\label{cueldo}}

The second component of eq.(\ref{shapeeq}) implies that
\be
\dot F_p\frac{dc}{d\theta} = -F'_p\frac{dx}{d\theta}
\ee
(dot and prime denote $c$- and $x$-derivatives respectively),
so that $\frac{dc}{d\theta} = 0$ when $F'_p=0$, provided
$\dot F_p\neq 0$ at the same point.
Together with the first eq.(\ref{shapeeq}) this means
that cusp can occur only when $\theta = 0$.
Thus the number of cusps depends essentially on the
range of variation of $\theta$-variable.
If $\theta$ runs from $0$ to $2\pi (d-1)$, we can
expect up to $d-1$ cusps to occur.

\subsection{Why descendant domains have one cusp less
than the root ones? \label{minuscusp}}

Descendant domain differs from the root one, because it
always has one special point at the boundary where $F' =0$
and $\dot F =0$ together.
This means that there is no cusp at this point, and
if the total number of zeroes of $F'$ at the boundary was
$d-1$, but the domain was a descendant,
then the total number of cusps will be $d-2$.

Characteristic feature of any descendant is reducibility of
the corresponding function $F_{mp}$: it is divisible by $F_p$
of a parent domain,
\be
F_{mp}(x) = \tilde G_{mp}(x)F_p(x),
\ee
(in variance with $G_{mp}$ from ref.\cite{DM} such $\tilde G_{mp}$
can still be reducible, but this does not matter for our purposes
in this paper).
Then $F'_{mp} = \tilde G_{mp}' F_p + \tilde G_{mp} F_p'$
and $\dot F_{mp} = \dot{\tilde G}_{mp} F_p + \tilde G_{mp} \dot F_p$
vanish {\it simultaneously} whenever both
$F_p=0$ and $\tilde G_{mp}=0$, i.e. when $x$ belongs simultaneously
to orbits of orders $p$ and $mp$.
According to \cite{DM} the last two equations possess
exactly one common zero at the boundary of {\it descendant} domain:
it is exactly the merging point, where descendant domain is attached
to the parent one, and in the $c$-space it is a zero of the
resultant $R(\tilde G_{mp},F_p)$.

Discriminant $D(\tilde G_{mp})$ also vanishes when
$R(\tilde G_{mp},F_p) = 0$, because different points of the
order-$mp$ orbit (roots of $G_{mp}$)
should merge $m$-wise to merge with the points
of the order $p$-orbit (roots of $F_p$).
Of more interest are {\it other} zeroes of $D(G_{mp})$,
representing crossings of different orbits of order $mp$.

\section{Exact results about elementary domains \label{exact}}

\setcounter{equation}{0}

This section is devoted to exactly-solvable examples.
Here exact solvability  means not obligatory explicit analytical
solutions -- though they will also be considered.
Whenever the problem can be effectively studied by user-friendly
computer tools like MAPLE or Mathematica, it is considered
equally (and may be even better) solvable as if explicit formulas
were derived.
We shall see that sometime the best way to analyze such explicit
formula is to generate its plot with the help of the same MAPLE.
One should keep in mind, however, that the number of problems
solvable in this way is also very limited: in most cases even
clearly formulated algebraic problems can be handled only by
specially designed programs, which usually {\it could} but
{\it never were} written. This makes such problems {\it potentially}
solvable (as many other hot problems in theoretical physics),
but they are clearly different from {\it practically} solvable.
We also distinguish these {\it solvable} problems from those
which are effectively solved, but only {\it approximately}:
under certain additional assumptions or when improving of accuracy
is increasingly difficult
(like it happens, for example, in perturbation theory).
We turn to approximate methods in ss.\ref{ssasec}-\ref{dfam}.
Before we are going to describe what is known today at {\it exact}
level.

Our primary goal is to understand what is the domain of variation
of the $\theta$-variable --
because we already know from s.\ref{cueldo} that it is its
size that defines the number of cusps, both for root and descendant
domains.
Moreover, we want to see how this variation domain is changed in
transition from one Mandelbrot set to another, i.e. to study the
bifurcations of Mandelbrot sets themselves, which happen in
complex-codimension-two in the Universal Mandelbrot space.
Examples will be also used in other sections, where we
derive (approximately) the {\it analytical} shape of the domains.

\subsection{Elementary domains of order $p=1$
for special Mandelbrot sets}

Let us consider a Mandelbrot set for a one-parametric family
\be
f(x,c) = P_d(x) + c
\label{addfam}
\ee
with polynomial $P_d(x)$ of degree $d$ (we do not require it to
be homogeneous $x^d$ at this moment).
Additive dependence on $c$-parameter considerably
simplifies consideration of such families.

For $p=1$ equations (\ref{shapeeq}) state simply that
\be
\left\{ \begin{array}{c} P'_d(x) = e^{i\theta} \\
c = x - P_d(x)
\end{array}\right.
\label{shapeeqp1}
\ee
and for every particular choice of $P_d(x)$ the function
$c(\theta)$ can be easily plotted with the help of MAPLE
or Mathematica.
Moreover, there are two important examples when even
{\it analytical} solution is immediately available.

The first case is homogeneous $P_d(x) = x^d$, associated with
the standard $Z_{d-1}$-symmetric Mandelbrot sets ${\cal M}_d$.
The second case is generic {\it cubic} polynomial
$P_3(x) = x^3+ax^2+bx$: associated family of Mandelbrot sets
${\cal M}_3(a,b)$ interpolates between
${\cal M}_2={\cal M}_3(\infty,0)$ and ${\cal M}_3={\cal M}_3(0,0)$.
For such interpolation one can also use a one-dimensional
and "better" parameterized family ${\cal M}_{2,3}(a)$ with
$P_3(x) = a x^3+(1-a)x^2$
(then ${\cal M}_2 = {\cal M}_{2,3}(0)$
and ${\cal M}_3 = {\cal M}_{2,3}(1)$).\footnote{
It deserves saying that these "families" of Mandelbrot sets
are somewhat artificial entities.
${\cal M}\{a,b,c\}$ is actually
a $3_C$-dimensional section of the Universal Mandelbrot set,
and $1_C$-dimensional Mandelbrot sets with coordinate $c$ are
obtained if $a$ and $b$ are artificially considered as
"external" parameters.
Of course, one can instead take $a$ for coordinate and $b,c$
for parameters: no distinguished choice exists and
all such sets should be studied on equal footing.
It is nothing but a historical casus that particular sets
${\cal M}_d$ are more popularized than the others (and even
for these particular sets the period-{\it doubling} is
better known than tripling etc -- despite it is in no way
distinguished).
Worse than that: the {\it standard} presentations like
\cite{Mand} and even the software like our favorite
{\it Fractal Explorer} \cite{FE}
implicitly exploit specific properties of these maps and
produce errors in application to generic families, say,
when $c$-dependence is not additive like in (\ref{addfam})
and even if $P_d$ in (\ref{addfam}) is non-homogeneous,
see also introductory remarks to s.\ref{interpol} below.
}

\subsection{Analytically solvable examples for $p=1$}

\subsubsection{Homogeneous $P_d(x)$}

For homogeneous $P_d(x)=x^d$ eq.(\ref{shapeeqp1}) converts directly
into (\ref{carddef}):
\be
\left\{\begin{array}{c}
dx^{d-1} = e^{i(d-1)\phi}, \ \ \ \ \ \ {\rm i.e.}
\ \ \ \ \ \ x = r(d)e^{i\phi} \\ \\
c = x\Big(1-x^{d-1}\Big) = r(d)e^{i\phi}
\Big(1 - \frac{1}{d}\,e^{i(d-1)\phi}\Big)
\end{array}\right.
\label{c1phi}
\ee
where $r(d) = d^{-\frac{1}{d-1}}$ and $\phi = \frac{\theta}{d-1}$.
It is obvious that in this case $\theta$ changes from $0$ to
$2\pi(d-1)$ and $\phi$ is the right angle-parameter.

Now we can use another solvable example with $P_3(x)$ in order
to deform these ideal cardioids and see how their order (number of
cusps) can actually change in interpolation between $x^2$ and $x^3$,
see s.\ref{interpol}.

\subsubsection{Cubic polynomial}

For $P_3(x) = ax^3+b x^2$ the first equation in
(\ref{shapeeqp1}),
$\ 3a x^2+2b x = e^{i\theta}$,
is quadratic in $x$ and has explicit analytic solution:
\be
x = \frac{-b \pm \sqrt{b^2 + 3a e^{i\theta}}}{3a}
\ee
(only the "+" branch has a finite limit at $a\  \rightarrow 0$).
Substituting this into the second equation (\ref{shapeeqp1}),
$\ c_1 = x-P_3(x) = -a x^3 - b x^2 + x$,
we obtain the analytical expression for
the boundary of the root domain of the central cluster
for arbitrary values of complex parameters $a\ $ and $b$:
\be
c_1 = \frac{\left(b \mp \sqrt{b^2 + 3a e^{i\theta}}\right)
\left(5b^3-9a + 3ae^{i\theta} \mp 5b
\sqrt{b^2 + 3ae^{i\theta}}\right)}
{27a^2}
\ee
Clearly, the phase transition line, separating the two regimes --
$\theta = 2\phi$ (near $b = 0$) and
$\theta = \phi$ (near $a\  = 0$), --
is $|b|^2 = 3|a\ |$.
If $b = 1-a\ $, see s.\ref{interpol},
it crosses the real-$a\ $ line at
$a_{cr}^\pm = \frac{5\pm\sqrt{21}}{2}$, i.e.
$a^-_{cr} = 0.208712\ldots$ and
$a^+_{cr} = 5-a^-_{cr} = 4.791288\ldots$

\subsection{Solvable examples for $p=2$}

\subsubsection{Equations in case of separated $c$-dependence}

For $p=2$ equations (\ref{shapeeq}) can be rewritten as follows:
\be
\left\{ \begin{array}{c}
f(x) = z; \\
f(z) = x; \\
f'(z)f'(x) = e^{i\theta}
\end{array}\right.
\ee
and when
\be
f(x,c) = P(x) + c
\label{addf}
\ee
with $c$-independent $P(x)$, as
\be
\left\{ \begin{array}{c}
P(z)+z = P(x)+x; \\
P'(z)P'(x) = e^{i\theta}
\end{array}\right.
\label{xzeqs}
\ee
Then $c(\theta)$ can be defined from
\be
c = z - P(x) = x- P(z).
\label{cvsx2}
\ee
Since we did not factor out $F_1$ from $F_2 = G_2F_1$, these
equations describe not only the $(2)$ and $(2,1)$ domains, but
also the $(1)$ ones.
The $(1)$ domains satisfy the system (\ref{xzeqs}) with first
equation substituted by $x=z$, while for the $(2)$ and $(2,1)$
domains it should be substituted by $\frac{P(z)-P(x)}{z-x} = -1$.

\subsubsection{MAPLE-generated solution for homogeneous $P_d(x)$
\label{homsol}}

For $P_d(x) = x^d$ the second equation in (\ref{xzeqs})
can be solved explicitly:
\be
xz = d^{-\frac{2}{d-1}} e^{i\varphi} \equiv \xi
\ee
where we substituted $\theta = (d-1)\phi$.
Then the first equation (\ref{xzeqs}) turns into
\be
x^d - \frac{\xi^d}{x^d} = -x + \frac{\xi}{x}
\label{hompolseqs}
\ee
One solution,  $x=\xi/x$, i.e. simply
$x=z=d^{-\frac{1}{d-1}} e^{i\phi}$ with $\phi = \frac{\varphi}{2}$
changing from $0$ to $2\pi$, provides
\be
c_1(\phi) = z-x^d = x-z^d = d^{-\frac{1}{d-1}}
e^{i\phi}\left(1-\frac{1}{d}\,e^{i\phi}\right)
\ee
which is our familiar eq.(\ref{c1phi}) for the central root
domain $(1)$, with examples shown in Fig.\ref{cardi}.

Remaining solutions, describing the root $(2)$ and
descendant $(2,1)$ domains,
can be solved by MAPLE or Mathematica, see Fig.\ref{hompols}.
In these solutions $\varphi = \phi$.
No root $(2)$ domains occur for homogeneous $P_d(x)=x^d$,
but this is a peculiarity of {\it both} homogeneity and $p=2$:
root domains $(p)$ exist for all $p\neq 2$ even if $P_d(x)=x^d$,
and $(2)$ domains are normally present for generic
non-homogeneous $P_d(x)$, see s.\ref{interpol} for examples.

\bigskip

\Fig{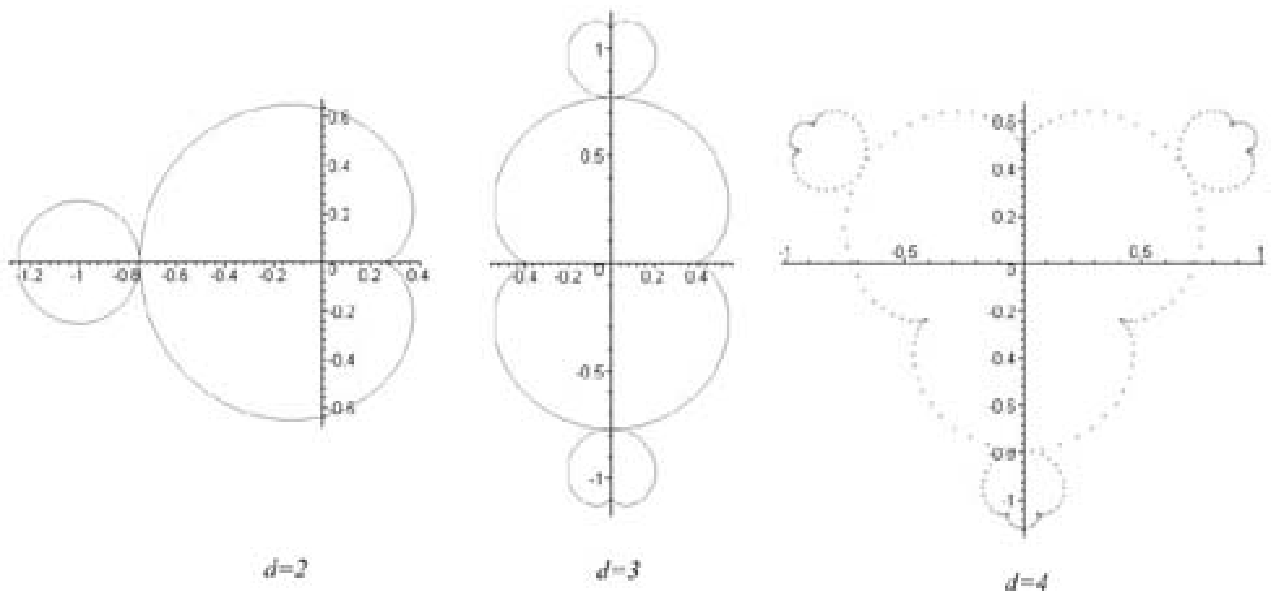}
{450,203}
{\footnotesize
The first two domains $(1)$ and $(2,1)$ of the central cluster,
obtained from solving eq.(\ref{hompolseqs}) with the help of MAPLE.
In the case of $d=2$ and $3$ analytical solutions are also available:
see eqs.(\ref{d2sols}) and (\ref{d3sols}) respectively (and they
are explicitly used by MAPLE).
{\bf A:} $d=2$, $f(x,c) = x^2+c$, {\bf B:} $d=3$, $f(x,c) = x^3+c$,
{\bf C:} $d=3$, $f(x,c) = x^4+c$, {\bf D:}
$d=4$, $f(x,c) = x^{4}+c$.
Mandelbrot set for $f(x,c) = x^d + c$ has $Z^{d-1}$ symmetry under
rotations around the point $c=0$.
}

\bigskip

\subsubsection{Analytical solutions for homogeneous $P_2(x)$ and
$P_3(x)$}

For $d=2$ and $d=3$ {\it analytical} solutions are also available.
Indeed, then eqs.(\ref{hompolseqs}),
after exclusion of solutions $x=z$, turn into
\be
{\bf d=2:}\ \  \ \ \ x+z+1=x + \frac{\xi}{x} + 1 = 0,\ \ \ \
\xi = \frac{1}{4}e^{i\phi}
\ee
and
\be
{\bf d=3:} \ \ \ \  \ x^2+xz+z^2 = x^2+\xi + \frac{\xi^2}{x^2}= -1,
\ \ \ \
\xi = \frac{1}{3}e^{i\phi}
\ee
respectively, which are explicitly solvable quadratic
and biquadratic equations.

Then it follows that for $d=2$
\be
x= -\frac{1}{2} \pm \frac{1}{2} \sqrt{1-e^{i\phi}}, \nn \\
z= -\frac{1}{2} \mp \frac{1}{2} \sqrt{1-e^{i\phi}}
\ee
and
\be
c_{2,1}(\phi) = z - x^2 = x-z^2 = - 1 + \frac{1}{4}e^{i\phi}
\label{d2sols}
\ee
is an ideal circle of radius $r_{2,1} = \frac{1}{4}$
centered at $c_{2,1}=-1$, see Fig.\ref{hompols}.A.

Similarly for $d=3$
\be
x = \pm \frac{1}{\sqrt{2}}\sqrt{-1-\frac{1}{3}e^{i\phi} \pm
\sqrt{1 + \frac{2}{3}e^{i\phi} - \frac{1}{3}e^{2i\phi}}}, \nn \\
z = \pm \frac{1}{\sqrt{2}}\sqrt{-1-\frac{1}{3}e^{i\phi} \pm
\sqrt{1 + \frac{2}{3}e^{i\phi} - \frac{1}{3}e^{2i\phi}}}
\ee
and, see Fig.\ref{hompols}.B,
\be
c_{2,1}(\phi) = z - x^3 = x - z^3
\label{d3sols}
\ee

In both examples $\phi = \varphi$ changes in between $0$ and $2\pi$.

\subsubsection{MAPLE-generated solution for arbitrary cubic
polynomial}

For arbitrary cubic $P_3(x)$ the second equation in (\ref{xzeqs})
is quadratic in $z$ and can be solved explicitly.
After substitution of this $z(x)$ the first equation can be
solved with the help of MAPLE or Mathematica and (\ref{cvsx2})
produces the final answer.
This is how Figs.\ref{a1_10}--\ref{tubesD4} are obtained.

\section{The first two elementary domains in interpolation
between ${\cal M}_2$ and ${\cal M}_3$ \label{interpol}}

\setcounter{equation}{0}

After equations are solved, we can turn to description of solutions.

For particular homogeneous polynomials $P_d(x) = x^d$
we obtained the well known shapes of central
domains in ${\cal M}_d$ Mandelbrot sets, see Fig.\ref{hompols},
-- only this time this is
not a result of computer {\it simulations} by {\it Fractal Explorer},
the shapes are now obtained as solutions
(sometime even analytical) to algebraic equations (\ref{shapeeq}).

Even more interesting is the possibility to study quantitatively
interpolation between the ${\cal M}_2$ and ${\cal M}_3$ sets.
So far only qualitative description was known \cite{DM}, and
the usual computer {\it simulations} fail.
Such simulations \cite{Mand}
are often based on the study of the sequence
$f^{\circ n}(c)$, i.e. the orbit of $x_{cr}=0$.
Interior of the Mandelbrot space, i.e.
the black regions in Figs.\ref{Mand2}--\ref{symmebre},
is {\it assumed} to consist of all functions $f$
where this sequence is bounded  and does not go away to infinity.
However, this assumption is not always true and then
this simple algorithm fails --
and Fig.\ref{symmebre} is the first example.
The reasons for failure can be different:
from $x_{cr}\neq 0$ to
attraction of unstable orbits to finite, rather than
to infinitely remote ones.
There is a strong need to cure this problem and make
a modification of {\it Fractal Explorer} which would treat
properly any kind of Mandelbrot set.\footnote{In the
absence of such modification we had to make use of various pictures,
which are at best qualitatively, but not fully correct:
this is the case with Fig.\ref{symmebre} in this paper and with
numerous Figures in \cite{DM}, including even the picture at
the cover of that book. Below in this section we provide much
better views of the $2$-parametric section of the Universal
Mandelbrot Set, these pictures will be fully correct, but
instead only order-$1$ and $2$ domains will be shown.
}

\subsection{Particular Mandelbrot sets ${\cal M}_{2,3}(a)$
for the families $ax^3+(1-a)x^2 + c$ at different values of $a$
\label{23interp}}

At $a=0$ the Mandelbrot Set acquires its standard form,
Fig.\ref{Mand2}, and its first two domains, $(1)$ and
attached $(2,1)$,
are shown in Figs.\ref{Mand2}.C and \ref{hompols}.A.

\subsubsection{Vicinity of the Mandelbrot Set: small $|a|$}

However, as soon as $a$ infinitesimally deviates from $a=0$,
it gets clear that Fig.{\ref{Mand2} has a twin: an exact copy
of the same shape and size, but with opposite orientation
-- a mirror twin, -- located at infinity of the complex-$c$ plane.
As $|a|$ grows, the twin moves closer, and Figs.\ref{a1_10}
and \ref{a-1_10} show its location at $a=\pm 1/10$
(the sizes of the domains are practically the same as in
Fig.\ref{Mand2} -- just the scale of the picture is different,
because the twin of Fig.\ref{Mand2} is still far away).
Moreover, it appears that additional mirror pair of $(2)_\pm$
domains -- roots of two more clusters --  were hidden at
infinity of $c$ plane and are now located in between the
two root domains $(1)_\pm$ for positive $a>0$ and on the
opposite sides of those for negative $a<0$.
Since for small $a$ these domains are tiny as compared
to the $(1)$ and $(2,1)$, they can be easily
overlooked, therefore one of them is marked by a circle and shown
in a bigger scale in a separate picture at the right low corner.
Clearly this root $(2)_-$ domain has cardioid shape and is
exact copy of the root domain $(1)_-$, only smaller.
In fact it has a $(4,2)_-$ domain attached to it in exactly
the same manner as $(2,1)_-$ is attached to $(1)_-$ -- it is not
shown, because we explicitly construct only domains of
orders $p=1$ and $2$.
For interested reader we add also slices of the Julia sheaf:
show behavior of the orbits in the $x$ space\footnote{
Since $f(x)$ is cubic, there are three order-$1$ orbits and
up to three branches will be seen in the pictures.
Since $G_2(x) = F_2(x)/F_1(x)$ has degree $6$ in $x$, there are
$6/2=3$ orbits of order $p=2$ and
up to six branches will be seen in the pictures.
}
with the change
of $c$, which becomes more and more interesting as we go far
from the "pure" points $a=0$ and $a=1$.
The problem is that Julia sheaf is embedded into a $4_Rd=2_Cd$
space, with complex $x$ and $c$,
and can not be shown {\it in full}, even if $a$ is fixed.
Therefore different sections and projections
are presented, $2_R$- and $3_R$-dimensional.
$3_R$-dimensional are especially informative, but only when
presented on computer screen, where they can be rotated and
regarded from different angles.
This advantage is lost in the printed version of the text,
but one can either use simple MAPLE programs, collected in
s.\ref{MAProgs} below or directly look at the results in
\cite{maplesamples}.

\bigskip

\Fig{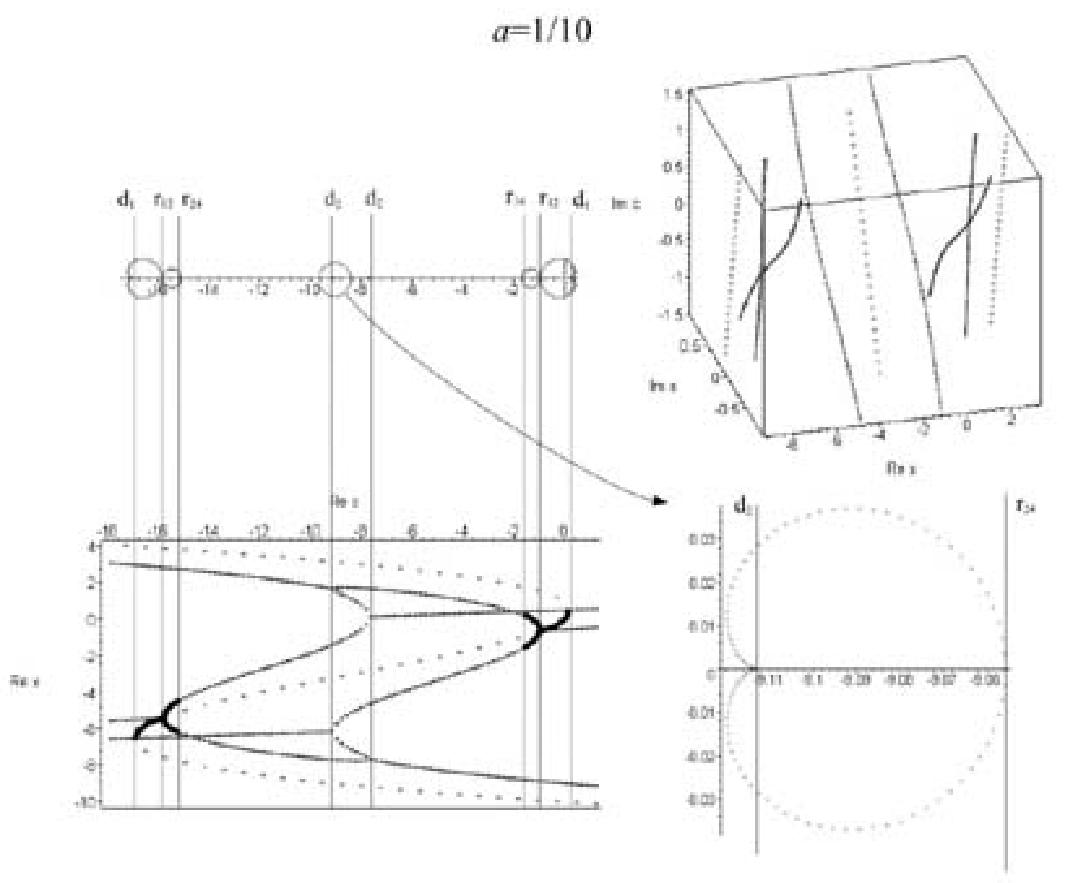}
{500,367}
{\footnotesize
The picture in the left upper corner represents the Mandelbrot
set for the family $ax^3+(1-a)x^2 + c$ with $a=1/10$.
This is the picture in complex $c$ plane and shown are only
domains, associated with the order $p=1$ and $p=2$ orbits.
The orbits themselves lie in the complex $x$ plane and
form the {\it Julia sheaf} over Mandelbrot set.
Julia sheaf itself is a $1_Cd$ complex variety, embedded into
a $2_Cd$ complex space and can not be shown in an ordinary
drawing.
Instead two different {\it views}, one $2_R$-dimensional,
another $3_R$-dimensional (at fixed ${\rm Re}(c)=C(a)$)
are shown in the low left and upper right corners respectively.
As all other three dimensional sections in this paper, it
can be rotated and looked from different angles: this
clarifies the pattern a lot, but can be done only on
computer screen, see ref.\cite{maplesamples}.
(In particular, there are NO intersecting orbits in this
section -- the seeming intersection is an artefact of the
drawing, resolved by rotation of the picture).
Dilute lines represent the three order-$1$ orbits,
while dense lines -- the six order-$2$ orbits.
In $2_Rd$ picture solid segments
show {\it stable} orbits: those of order $1$ are stable
inside the $(1)_\pm$ domains of the Mandelbrot set,
those of order $2$ -- inside the $(2,1)_\pm$ and
$(2)_\pm$ domains (the last two are two small to have
the corresponding solid segments seen in our pictures).
Enlarged picture of the $(2)_-$ domain -- a root of
a new cluster -- is shown in the
low right corner, and it is clear that it is an exact
diminished copy of the root $(1)_-$ domain.
$(2)_+$ is its exact mirror copy, in accordance with
$Z_2$ symmetry of the Mandelbrot set w.r.t. the vertical
line ${\rm Re}(c) = C(0.1) = -8.4$.
Shown are also the roots $c=d_{1,2}$ of discriminants
$D_{1}$ and $D_{2}$ and $c=r_{12}$, $c=r_{24}$ of
the resultants $R_{12}$ and $R_{24}$
(they lye at intersections of vertical lines with
the real-$c$ axis).
According to \cite{DM}, the last two are the crossing
points of the orbits of orders $1$ and $2$ and $2$ and $4$,
define merging points between the domains $(2,1)$ and $(1)$
and between $(4,2)$ and $(2,1)$ respectively, and thus
define the stability segments of orbits of orders $1$ and $2$.
Similarly, discriminant zeroes are intersection points of
the orbits of the same order:
$p=1$ with $p=1$ and $p=2$ with $p=2$.
$2_Rd$ view in the low left corner is in fact a {\it section}
of the $c$ plane with given ${\rm Im}(c)=0$ and
{\it projection} on the ${\rm Re}(x)$ plane.
Accordingly, when two real-valued orbits intersect and
become complex-valued, they remain shown in the picture,
but since they are complex conjugate, two lines are
projected onto one -- this should be taken into account
in analysis of the figure.
}

\Fig{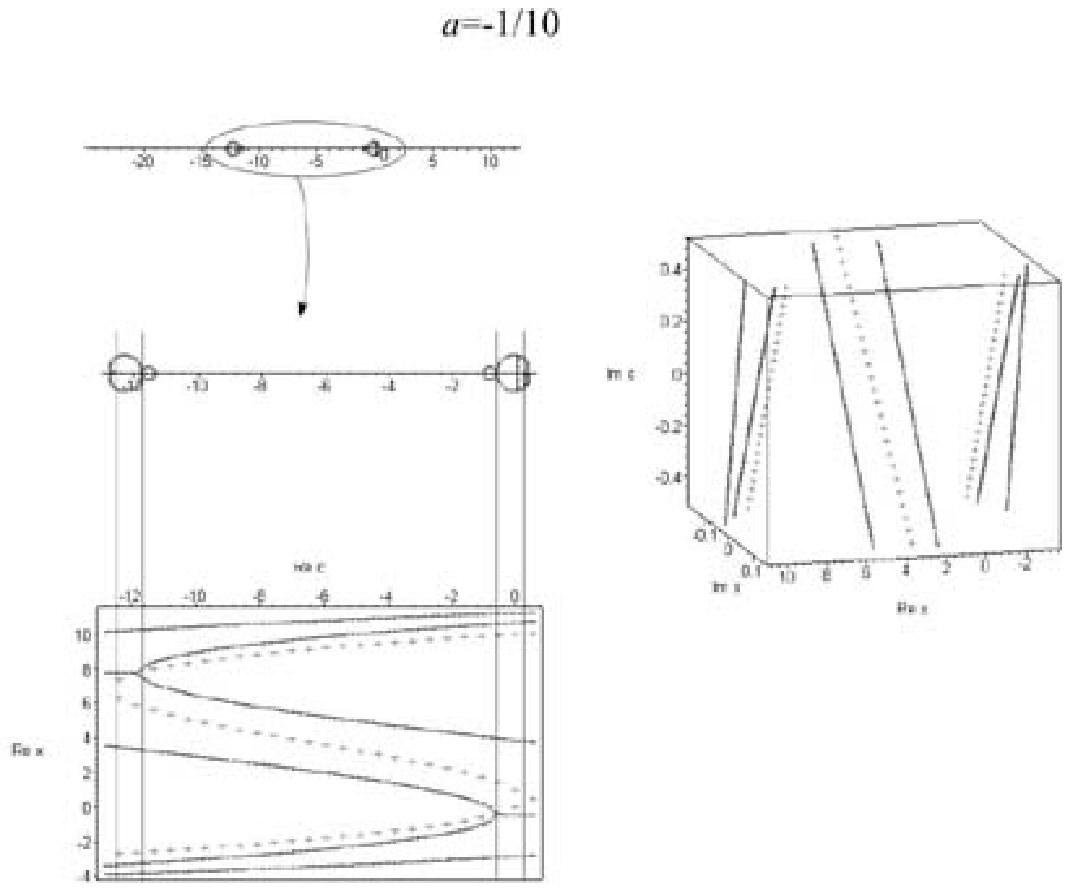}
{500,402}
{\footnotesize
Analogous picture for negative real $a=-1/10$.
The only essential difference from Fig.\ref{a1_10}
is that the two root domains
$(2)_\pm$ are not in between the two $(1)_\pm$, but
on the opposite sides of those.
In other words, at $a=0$ the $(2)$ domains pass through
$(1)_-$, so that $(2)_+$ re-appears from $c=+\infty$,
while $(2)_-$ returns back from $c=-\infty$,
but {\it slower} than $(1)_-$, see also Fig.\ref{speeds}.
Because of this, when $|a|$ increases further in the
direction of negative $a$, the overlap will occur
between the domains $(2,1)_-$ and $(2,1)_+$, unlike
in the positive-$a$ case, where $(2)_-$ and $(2)_+$
will be the first to meet.
}

\Fig{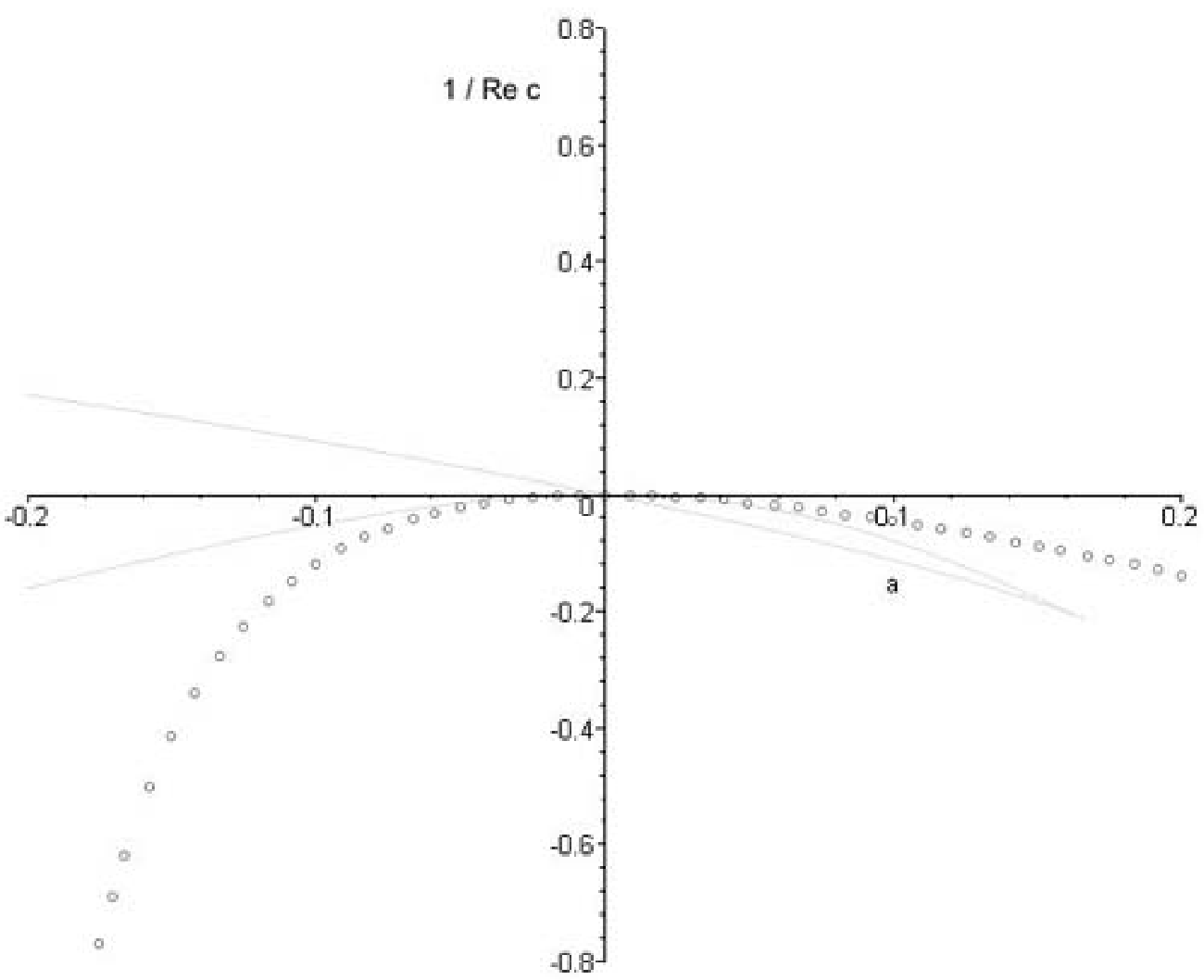}
{450,363}
{\footnotesize
Behavior in the vicinity of $a=0$
of the {\it root} domains $(1)_-$ and $(2)_\pm$,
denoted respectively by circles and solid lines???.
As $a \rightarrow 0$, these three domains, together with
their entire clusters, travel to infinity in the complex-$c$
plane, therefore the picture is drawn in coordinate $1/c$
(only real values of $c$ are plotted).
The Mandelbrot Set from Fig.\ref{Mand2}, including the
central root domain $(1)_+$, stays in the vicinity of $c=0$
and is not shown in this picture.
Behavior of the $(2)_+$ and $(2)_-$ domains is somewhat
different: the former one re-appears from infinity
at the opposite end of the ${\rm Re}(c)$ axis, while the
latter one returns from the same end, only exchanges
positions with the $(1)_-$ domain -- in full accordance
with Figs.\ref{a1_10} and \ref{a-1_10}.
All the domains and in fact the entire clusters shrink to
a single point at $a=0$, the reason for this is the
choice of a very singular map in homogeneous coordinates:
$\big(x,\ y\big) \longrightarrow
\big(ax^3+bx^2y +cy^3,\ y^3\big)$.
Such singular behavior at $a=0$ would be smoothed and
become similar to bifurcations at finite $a$,
shown in forthcoming pictures, if
$y^3$ is substituted by generic cubic polynomial.
See \cite{nolal} for related considerations.
}

All Mandelbrot sets ${\cal M}_{2,3}(a)$ possess a discrete
$Z_2$ symmetry under reflection w.r.t. the vertical line
\be
{\rm Re}(c)=C(a)=-\frac{b}{3a}\left(1+\frac{2b^2}{9a}\right)
\ee
with $b=1-a$,
which is lifted to entire Julia sheaf over ${\cal M}_{2,3}(a)$:
\be
\begin{array}{ccc}
\tilde x = x+\frac{b}{3a}& \ \longrightarrow\ & -\tilde x \\
\tilde c = c-C(a)& \ \longrightarrow\ & -\tilde c
\end{array}
\label{symmman}
\ee
For example, the equation $F_1(x,c)=0$ for the first-order
periodic orbits is obviously symmetric, since
$F_1(x,c) = ax^3 + bx^2 + c - x = a\tilde x^3 -
\left(1+\frac{b^2}{3a}\right)\tilde x + \tilde c$.
In accordance with this symmetry, the $(1)_-$ domain -- the
twin of the $(1)_+$ domain, centered at $c=0$,-- is the
mirror-reflected cardioid with center at $c = 2C(a)$.
Similarly, the two next root domains $(2)_\pm$ are centered
at $c_{2+} = -\frac{1}{a} + 2 + O(a)$ and $c_{2-} = 2C(a)
- c_{2+}$, see s.6.4.5 of ref.\cite{DM}.
Since $c_{2+}$ is {\it odd} function of $a$ for
small $a$, while $c_{1-}$ is {\it even}, it is clear that
they exchange order when $a$ goes through zero --
in accordance with what is shown in Figs.\ref{a1_10},
\ref{a-1_10} and \ref{speeds}.

\subsubsection{Overlapping domains \label{doverlap}}

As $|a|$ increases, the two domains $(1)_\pm$ move closer.
The speed of approaching is somewhat different for positive
and negative $a$.
At some stage of this movement two different clusters
unavoidably {\it meet}.
There are, however, two sorts of meeting:
{\it overlap} and {\it collision}.
For above-explained reasons it is still difficult to
analyze the behavior of entire clusters.
{\it Approximate} results about clusters
(or, better, possible approach to their future derivation)
will be discussed in the last sections \ref{ssasec}-\ref{dfam}
of this paper.
{\it Exact} results to be considered right now
concern meetings of the low ($1$ and $2$) order domains.
{\it Overlap} of these domains takes place soon after it happens
to the upper $(p=\infty)$ leaves of the clusters, while
{\it collision} can start at the $p=\infty$ leaves, but can
also begin at the low-$p$ level.

Overlaps and collisions of particular domains occur when zeroes
of the corresponding resultants collide, i.e. are controlled by
zeroes of double-resultants, like $c$-discriminants of the
$x$-resultants, listed in the following table
(italic lines are quotations from MAPLE program, boldfaced are
{\it real}-valued roots, belonging to the segment $0\leq a \leq 1$).

\bigskip

\begin{tabular}{|l|}
\hline
Two zeroes of $D1$ merge at
$a = -\frac{1\pm i\sqrt{3}}{2}$. \\
$discrim(discrim(F1,x),c) =
96*a^2+112*a^3+96*a^4+48*a+48*a^5+16+16*a^6$\\
\hline
Two zeroes of $R_{12}$ merge at
$a=\frac{5\pm\sqrt{21}}{2}$ and at
$a = -2 \pm \sqrt{3}$. \\
Only one of these points,
$a = \frac{5-\sqrt{21}}{2} = 0.20871215\ldots$ belongs to the interval
$0<a<1$ on a real-$a$ line. \\
$discrim(resultant(G2,F1,x),c) =
16*a^{10}*(a^2+4*a+1)^2*(a^2-5*a+1)$\\
\hline
Two zeroes of $D2$ merge at
$a = 4 \pm \sqrt{15}$. \\
$discrim\Big(\sqrt{discrim(G2,x)/R_{12}}\Big) \sim (a^2-8*a+1)^3$\\
\hline
Two zeroes of $R_{24}$ merge at \\
$a = 0.1487496031\pm 0.03597329725i$,\
$a = 6.351250397\pm 1.535973297i$,\\
$a = -4.485250968, -0.2229529645,\ {\bf 0.1163898556},\
8.591814077 = 1/0.1163898556$,\\
$a = {\bf 0.1497297977},\  6.678697327 = 1/0.1497297977$,\\
$a=0 .5857864376\pm 0.8104654524i$,\\
$a = \frac{9\pm\sqrt{65}}{4}$.\\
The four series correspond to zeroes of the four factors in\\
$discrim\Big(\sqrt{resultant(G2,G4,x)},c\Big) \sim
(2*a^4-26*a^3+93*a^2-26*a+2)\cdot $\\
$\cdot(a^4-4*a^3-39*a^2-4*a+1)^2
(a^4-8*a^3+10*a^2-8*a+1)^2(2*a^2-9*a+2)^4 $\\
\hline
\end{tabular}

\bigskip

Derivative $\partial d_1^-/\partial a$
for the zero $d_1^-$ of discriminant $D_1$, which defines the
position of the cusp in the $(1)_-$ domain, defines the speed
of motion of the cluster $(1)_-$ with the change of $a$.
Since for small $a$ all the clusters are diminished copies of
the central one in Fig.\ref{Mand2}, with all the same proportions
one can actually predict what happens to clusters from the
data about their root domains.

The first event to happen on our way from $a=0$ is
{\it overlap}. If $a>0$ this is the overlap of domains
$(2)_+$ and $(2)_-$, while if $a<0$ it is that of
$(2,1)_+$ and $(2,1)_-$.
The two domains of order $2$
{\it overlap} when the two zeroes of $R_{24}$ coincide.
If we move from $a=0$  along the real-$a$ line
this first happens at $a=0.1163898556\ldots$ if $a>0$
and at $a=-0.2229529645\ldots$ if $a<0$.\footnote{
In fact, as explained in the previous paragraph,
we can approximately find the values of $a_{cluster}$,
when the overlap of the {\it clusters} occurs.
From Fig.\ref{Mand2} we know, that
the total size of the cluster is $\approx 1.65$
times bigger than the size of the root domain,
and the latter size is nothing but the difference
$d_2 - r_{24}$, we get a rough estimate:
$a_{do} - a_{cluster} \approx 0.65(a_{co}-a_{do})$
where $a_{do}$ and $a_{co}$ are moments when
$r_{24}^-(a_{do}) = r_{24}^+(a_{do})$
and $d_2^-(a_{co}) = d_2^+(a_{co})$, i.e.
$a_{do}=0.1163898556\ldots$ and
$a_{co} = 4-\sqrt{15}=0.12701665\ldots$.
Then $a_{cluster} \approx 1.65 a_{do}-0.65 a_{co} \approx
0.109$. It is assumed that the speed of motion
of the cluster with the change of $a$ and the cluster's
size are approximately constant, appropriate corrections
can be easily taken into account.
\label{estcluoverlap}}
Figs.\ref{a1_9s}, and \ref{a-1_4s} show Mandelbrot sets
soon after these points are passed.

It is clear from these pictures, that when overlap occurs,
nothing interesting happens to the orbits -- and this is
what makes {\it overlap} different form {\it collision},
when intersection of orbits takes place, see below.
Overlap means simply that two (or more) different orbits
are simultaneously stable at the same value of $c$:
in this case these are two order-$2$ orbits.
When overlap increases, it involves the $(1)_\pm$ domains
and at the same values of $c$ can coexist two
stable orbits of other orders: $2$ and $1$ (Fig.\ref{a-1_4})
and $1$ and $1$ (Fig.\ref{a-1}).
For $a<-1$ the further increase of $|a|$ leads to
{\it diminishing} of the overlap: the story repeats in the
opposite order, the overlap picture for $a=-4$ ressembles
that for $a=-0.25$ (shown in Fig.\ref{a-1_4}), for
$a = -4.(4)$ -- that for $a=-0.225$ (shown in Fig.\ref{a-1_4s}),
and after that the overlap disappears.

The reason for reversed evolution is that the family
${\cal M}_{2,3}(a)$ and even its Julia sheaf
has a discrete "symmetry" w.r.t.
inversion of parameter $a\rightarrow 1/a$:
\be
x(a^{-1},\theta) = -ax(a,\theta), \nn \\
c(a^{-1},\theta) = -ac(a,\theta)
\label{symmemaf}
\ee
This symmetry complements the
(\ref{symmman}) of particular Mandelbrot sets
at fixed $a$, and it allows to consider only the variation of
$a$ within interval $|a|\leq 1$, all Mandelbrot sets outside
this interval are exact rescaled copies of those inside.

More interesting things are taking place for positive $a$.

\Fig{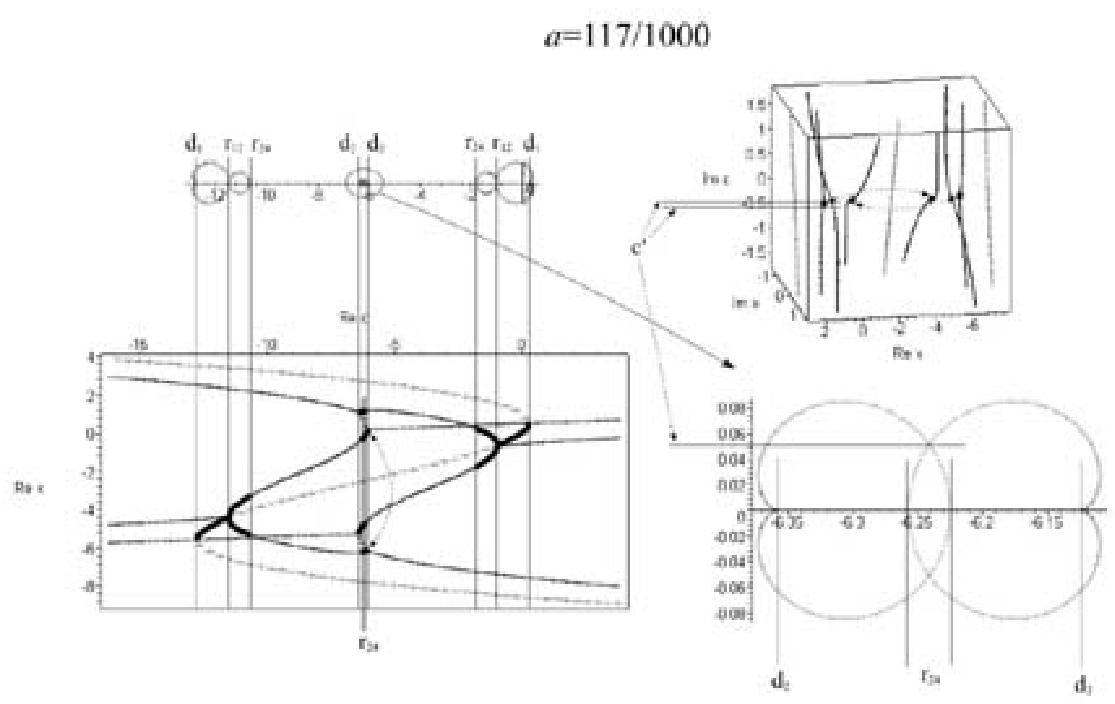}
{500,279}
{
\footnotesize Domains of the orders $p=1,2$ of the
Mandelbrot set ${\cal M}_{2,3}(a)$ (domains of orders $p=1,2$)
at $a=0.117$, immediately after the two root
domains $(2)_\pm$ touched at $a=0.1163898556\ldots$.
Now they {\it overlap}:
for $c$ in close vicinity of $c=-6.24$ there are two stable
order-$2$ orbits at once.
However, as clear from the Julia-sheaf pictures around,
nothing special happens to the orbits themselves.
}

\Fig{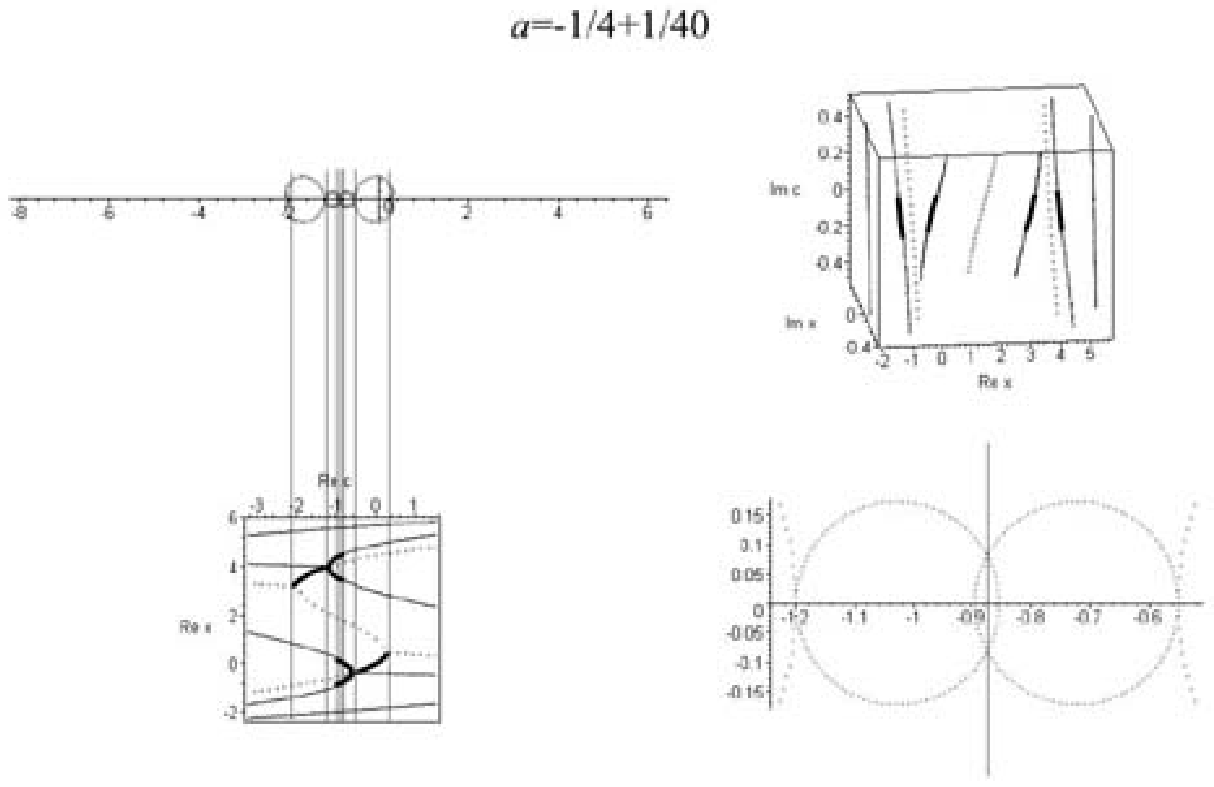}
{500,336}
{\footnotesize
Domains of the orders $p=1,2$ of the
Mandelbrot set ${\cal M}_{2,3}(a)$
at $a=-1/4+1/40 = -0.225$, soon after the two
descendant domains $(2,1)_\pm$ touched at
$a=-0.2229529645\ldots$.
Two order-$2$ orbits are simultaneously stable in the close
vicinity of $c= -0.88$.
Again, nothing special happens to the orbits themselves
when the overlap occurs.
}


\Fig{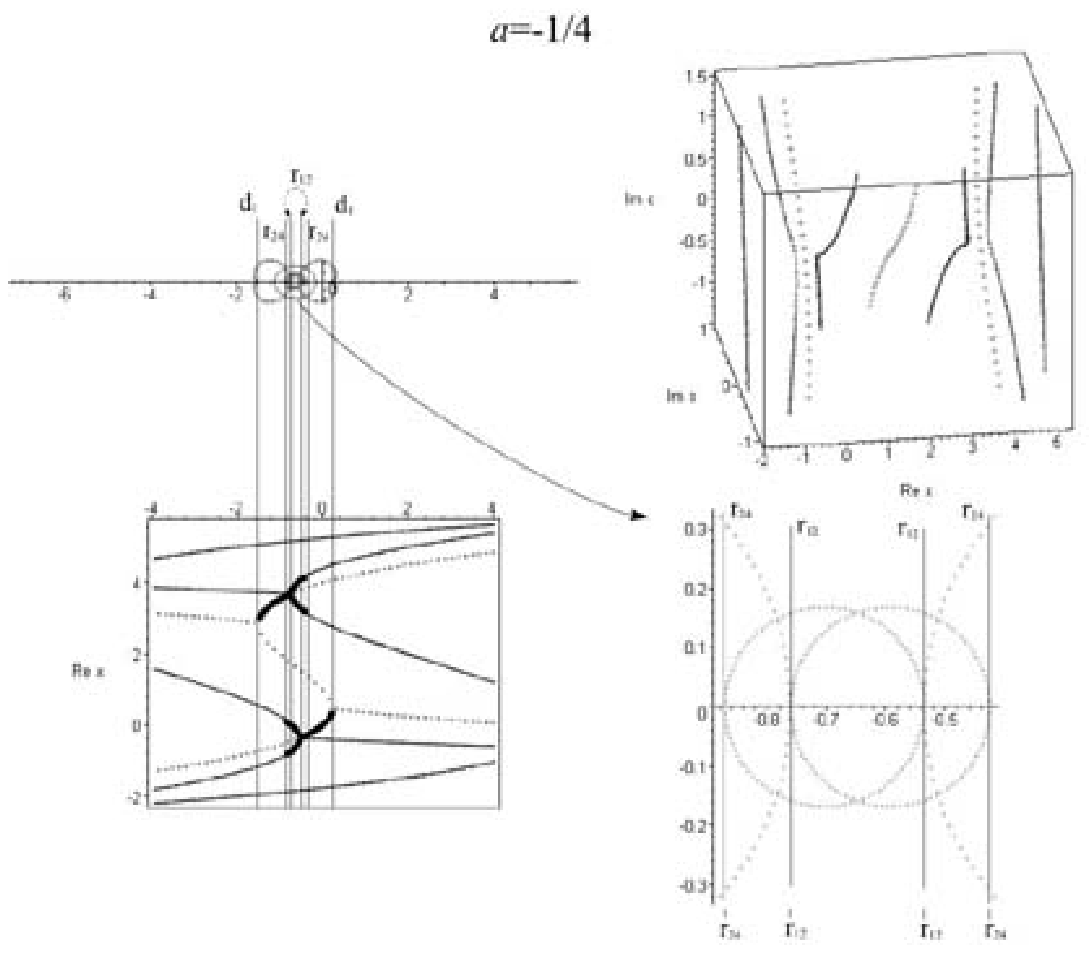}
{500,379}
{\footnotesize
Domains of the orders $p=1,2$ of the
Mandelbrot set ${\cal M}_{2,3}(a)$ at $a=-1/4$.
The overlap increases,
and now the descendant $(2,1)_\pm$ domains overlap not only
each other, but also the twin parent $(1)_\mp$ domains.
This means that in the vicinity of $c=-0.65$ there are
two coexisting stable orbits of order $2$, while in the
vicinities of $c=-0.5$ and $c=-0.8$ coexisting are stable
orbits of orders $2$ and $1$.
Nearly cusp-like behavior of the orbits in Julia-sheaf
view in the upper right corner occurs in the vicinity
of the zeroes $r_{12}$ of the resultant $R_{12}$, where
the orbits of order two cross those of order one and
the $(2,1)_\pm$ domains are attached to their parent
$(1)_\pm$ domains. This singular behavior has nothing
to do with the {\it overlap} of the domains.
}


\Fig{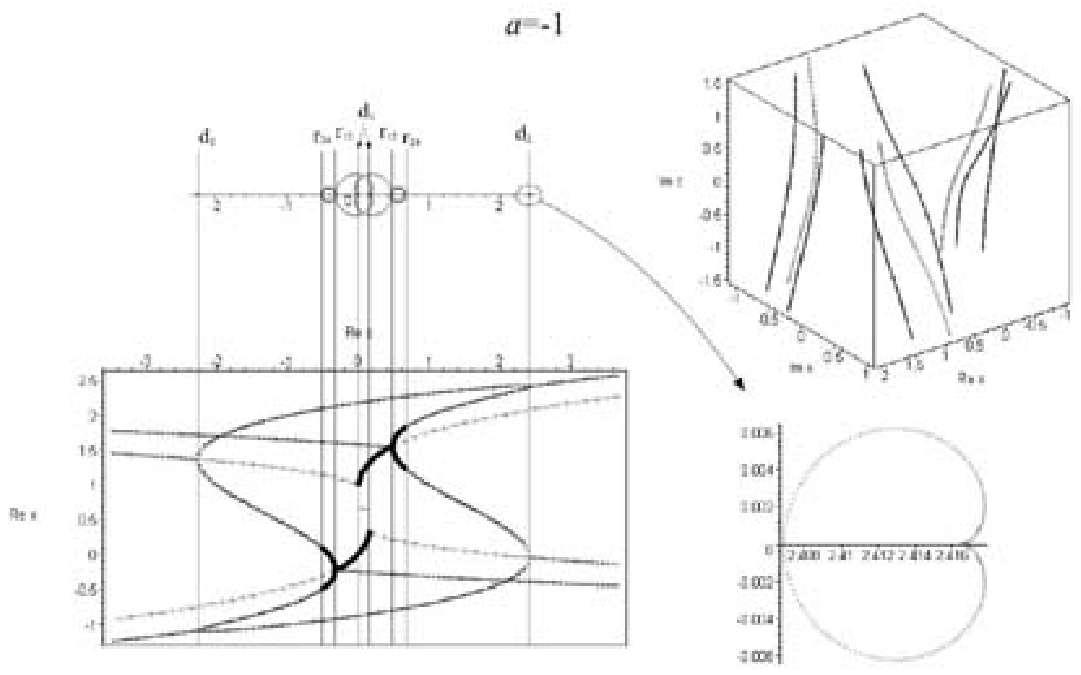}
{500,316}
{\footnotesize
Domains of the orders $p=1,2$ of the
Mandelbrot set ${\cal M}_{2,3}(a)$ at $a=-1$.
The overlap increased even further as compared to
Figs.\ref{a-1_4s} and \ref{a-1_4}.
Descendant domains $(2,1)_\pm$ now passed through
the root ones $(1)_\mp$ and are no longer involved
in the overlap. Instead overlapping are the root
domains $(1)_\pm$ and two stable orbits of order
$1$ coexist in the vicinity of $c=0$.
If $(1)_\pm$ domains continued to move in the same
direction with further increase of $|a|$, the
two order $1$ orbits would cross (when the
two zeroes $d_1^\pm$ of discriminant $D_1$ merge),
and overlap would finally end in a {\it collision}
-- like it happens for positive values of $a$, see
Fig.\ref{a1_8s}.
However, the merging of $D_1$ zeroes occurs at
complex values $a = -\frac{1\pm i\sqrt{3}}{2}$,
and  at $a=-1$ the overlap is the biggest possible
for real negative values of $a$.
As $a$ decreases further (and $|a|$ grows)
along the real-$a$ line, the $(1)_\pm$ domains
start to move in the opposite direction,
see Fig.\ref{a-2}.
We pass again through the overlap
patterns like Fig.\ref{a-1_4} (at $a=-4$)
and Fig.\ref{a-1_4s} (at $a=-4.(4)$)
and finally come back to the no-overlap pattern
like Fig.\ref{a-1_10} (at $a=-10$) -- all this in
accordance with the symmetry (\ref{symmemaf}).
}

\Fig{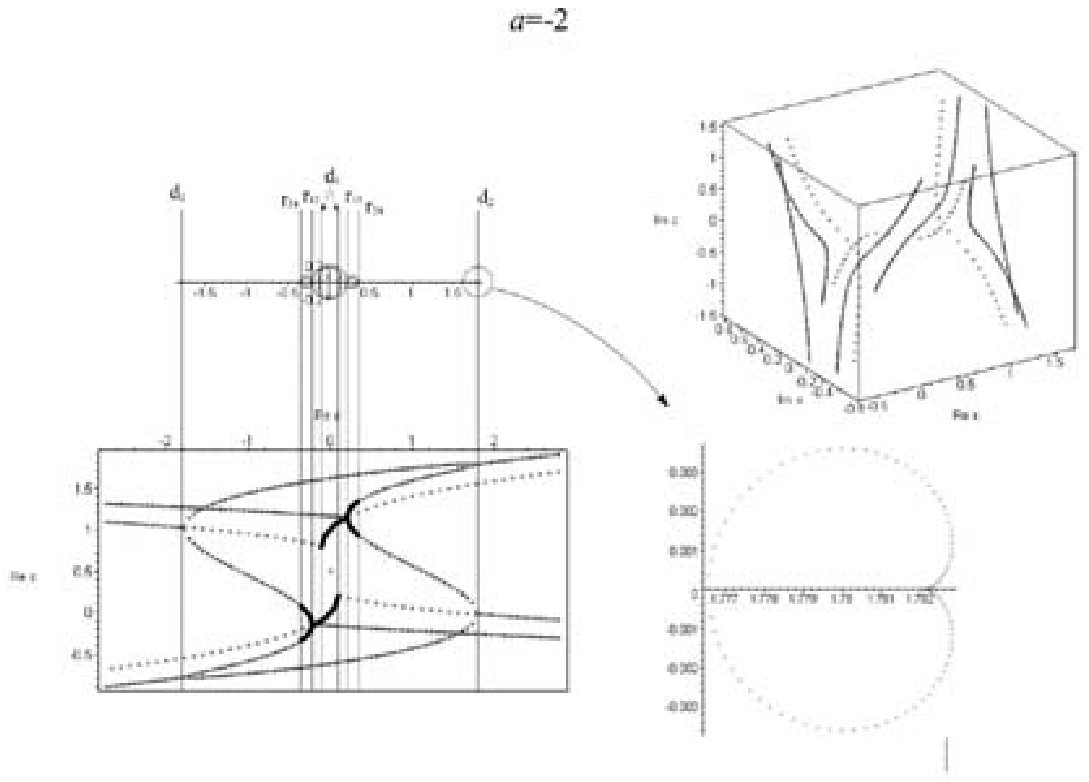}
{500,310}
{\footnotesize
Domains of the orders $p=1,2$ of the
Mandelbrot set ${\cal M}_{2,3}(a)$ at $a=-2$.
The overlapping clusters start to diverge after
maximal approach at $a=-1$, stopping short from
collision of two zeroes of discriminant $D_1$,
which would unify two clusters into a single one
This actually happens at complex values of
$a = \frac{1\pm i\sqrt{3}}{2}$, and we will enconter
this unified cluster in our travel along real-$a$
axis, but at positive values of $a$.
}

\subsubsection{Colliding domains}

We left evolution in the positive-$a$ direction at the stage
of Fig.\ref{a1_9s}, when overlap of the two root domains
$(2)_\pm$ just occured.
In variance with the case of negative $a$, this time
as the overlap increases
the two zeroes $d_2$ of discriminant $D_2$, defining
positions of the cusps of these two domains, coincide
at $a=4 - \sqrt{15}=0.12701665\ldots$
and a new phenomenon takes place.
The two stable order-$2$ orbits (they were simultaneously
stable in the overlap region) cross each other, and the
pattern of orbits around crossing is pretty sophisticated,
see Fig.\ref{a1_8s}.
Most interesting is preservation of two small overlap sections,
where two different stable order-$2$ orbits continue to
coexist, but exhibit non-trivial monodromy under a travel
in the complex-$c$ plane
around the cusps at zeroes of discriminant $D_2$.

\Fig{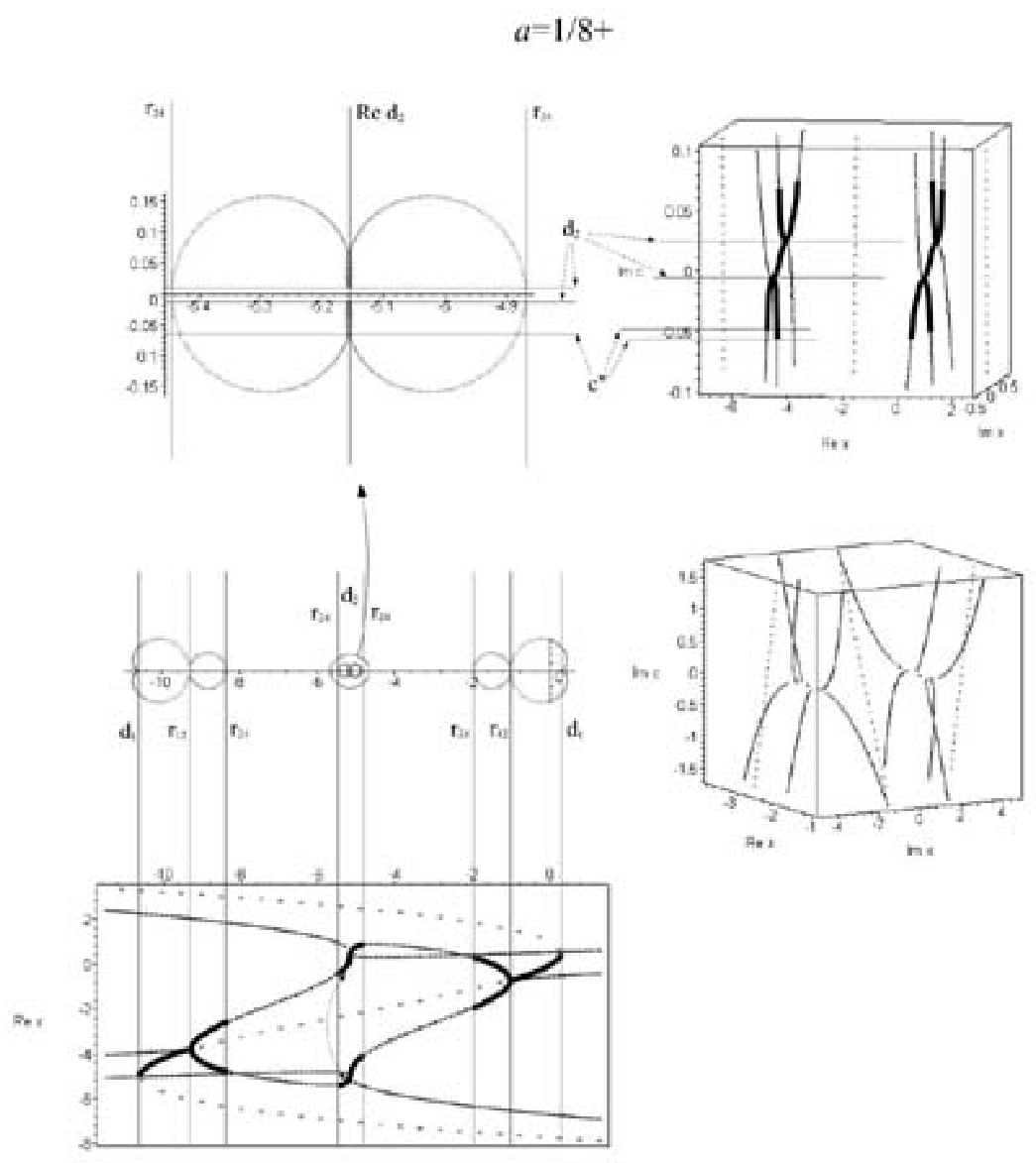}
{500,491}
{\footnotesize
Domains of the orders $p=1,2$ of the
Mandelbrot set at $a=1/8+1/224 = 0.12946(428571)$.
The two root domains $(2)_+$ and $(2)_-$ just {\it collided}
at $a=4 - \sqrt{15}=0.12701665\ldots$
(after a period of overlap, originated at
$a=0.1163898556\ldots$, see Fig.\ref{a1_9s}),
and their clusters are already united into a single new
cluster, so that the union of $(2) = (2)_+ \cup (2)_-$,
$(2)_+\cap (2)_- \neq \empty$,
is now the new root domain.
Arrows in the picture at the low left corner show the
action of evolution $x \rightarrow f(x)$ on the points
of the order-$2$ orbit.
Enlarged picture in the left upper corner shows in
more detail the new root cluster $(2)$.
It has two narrow regions of self-overlap along the
${\rm Re}(c) = {\rm Re}(d_2)$ vetical axis, where two
different order $(2)$ orbits are simultaneously stable.
In all other points of the $(2)$ cluster only one order $2$
orbit is stable, but non-trivial monodromy occurs when we
go around the cusps at
$d_2^\pm = {\rm Re}(d_2)\pm{\rm Im}(d_2)$ (points $D$):
if we pick up the single stable order-$2$ orbit,
say, at $c={\rm Re}(d_2)$ (point $O$)
and carry it into the upper overlap region from the right
(counter-clock-wise), we obtain one of the two orbits,
stable in that region, but if we carry the same orbit
into the same region from the left (clock-wise), we obtain
{\it another} stable orbit. If we continue to carry our
orbits in the same direction and leave the overlap domain,
the orbits loose stability. In other words, the two stable
orbits in the overlap domain are permuted when carried around
the cusp, and each orbit continues to be stable outside the
overlap domain only if it is carried away in one out of two
possible directions: either to the right or to the left.
Another end of the overlap domain (point $E$), which is on the
boundary of the $(2)$ cluster is not a cusp and not a
singularity. As any point on the boundary of an elementary
domain it has in its infinitesimal vicinity an infinite number
of zeroes of various resultants $R_{2,2m}$, where the
stable order-$2$ orbit exchanges stability with some order $2m$
orbit, see \cite{DM}. However, the set of relevant $\{m\}$ changes
irregularly as we move along the boundary, and accordingly,
for the point $E$ this set depends irregularly on $a$.
}

\subsubsection{Colliding clusters}

After collision of two root $(2)_\pm$ domains and formation
of a unified cluster with the root $(2)$, the two other
clusters, growing from the $(1)_\pm$ roots, continue to
move towards each other and soon collide with the $(2)$
cluster, sandwiched in between them.
Now this is indeed a {\it collision}, not just {\it overlap},
and, in variance with collision of the $(2)_\pm$ domains
it now originates at the highest leaves (at $p=\infty$)
rather than at the root domains.
Full description of this process is impossible with the
knowledge about the $p=1,2$ orbits only,
thus our illustrations will be necessarily incomplete.
Still, a lot is seen even with these limited tools.

Immediately after collision of the $(2)_\pm$ domains
in Fig.\ref{a1_8s}
they begin growing and soon become comparable in size
with the $(2,1)_\pm$ domains, belonging to approaching
clusters.
Even earlier the overlapping region inside $(2)$
shrinks down and disappears.
Finally, when the zeroes $r_{24}$ of the resultant $R_{24}$,
marking the closest points of the $(2)$ and $(2,1)_\pm$
domains, coincide, collision wave, going down from the
upper leaves of the clusters, reach the $p=2$ level:
cluster collision gets seen at the level of our
consideration. This happens at $a=0.1497297977\ldots$,
see Fig.\ref{a3_20}.\footnote{
Like in footnote \ref{estcluoverlap} one can try to
{\it estimate} the moment of {\it clusters} collision.
However, this time the clusters shape deviates considerably
from that in Fig.\ref{Mand2}, therefore such estimate is
less reliable.
}
Figs.\ref{a1_7s} and \ref{a1_5} show Mandelbrot sets soon
after that and a little later, when continuing approach
of $(1)_\pm$ domains (which are now {\it two roots} of
a single cluster!) starts pushing the unified $(2,1)$ domain
outside of the region between them.
This push-away process leads to the next bifurcation
at $a= \frac{5-\sqrt{21}}{2} = 0.20871215\ldots$
where the two zeroes $r_{12}$ of another resultant $R_{12}$
coincide, marking collision of the $(1)_\pm$ domains:
collision wave reached the $p=1$ level.
At this moment the $(2)$ cluster is ripped into two
disconnected pieces, see Fig.\ref{a1_5s}.
Clearly, just the same push-away and ripping processes
took place with all the higher-order domains
$(2^k,2^{k-1},\ldots,2,1)$ in between the moment of clusters
collision till it reached the $p=2$ level at $a=0.2087\ldots$

\Fig{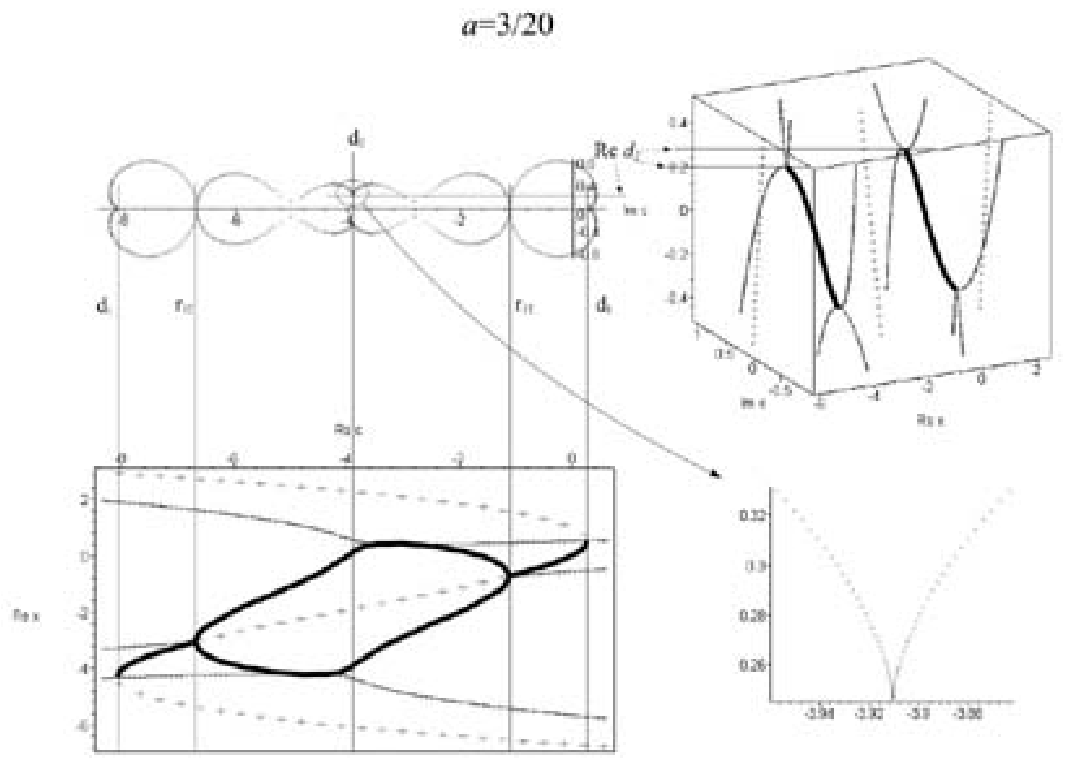}
{500,348}
{\footnotesize
Domains of the orders $p=1,2$ of the Mandelbrot set
${\cal M}_{2,3}(a)$ at $a=3/20=0.15$.
The $(2)$ domain just merged with the $(2,1)_\pm$
at $a=0.1497297977\ldots$ to form a single descendant
$(2,1)$ domain.
This unified domain has a characteristic
four-sausage structure,
remembering about its recent formation from four
distinct elementary domains $(2)_\pm$ and $(2,1)_\pm$.
}

\Fig{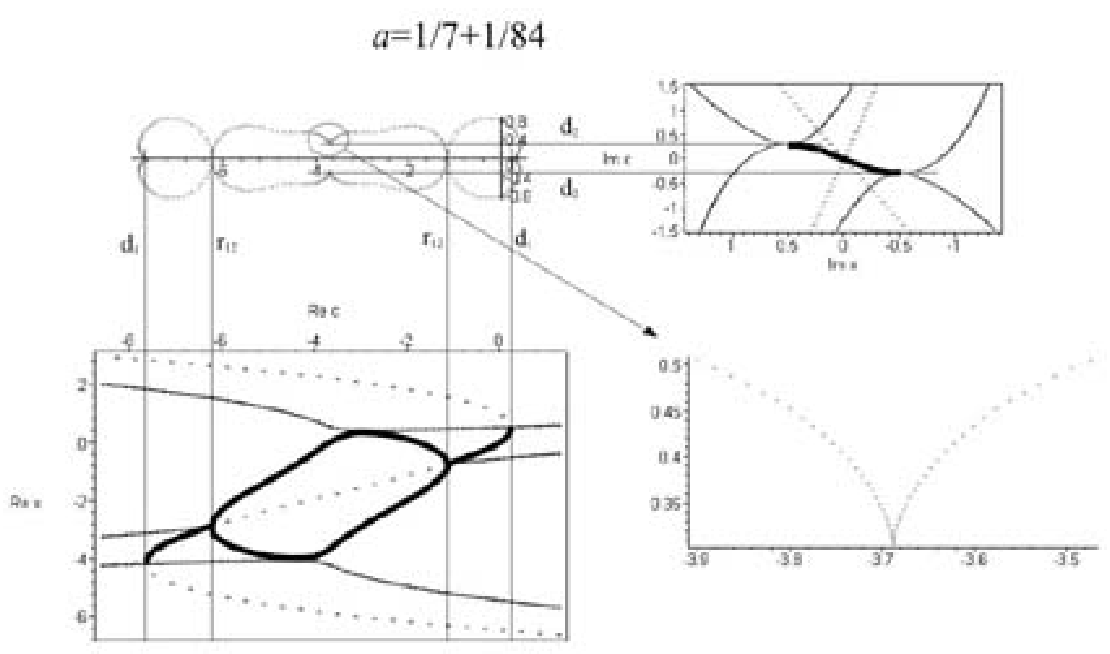}
{500,298}
{\footnotesize
Domains of the orders $p=1,2$ of the Mandelbrot set
${\cal M}_{2,3}(a)$ at $a=1/7+1/84 = 0.15(476190)$.
The shape of the unified $(2,1)$ domain evolved
from the four-sausage to a two-sausage one: the memory
about the difference between the $(2)$ and $(2,1)$
domains is almost erased, but distinction between
$+$ and $-$ is still preserved.
}

\Fig{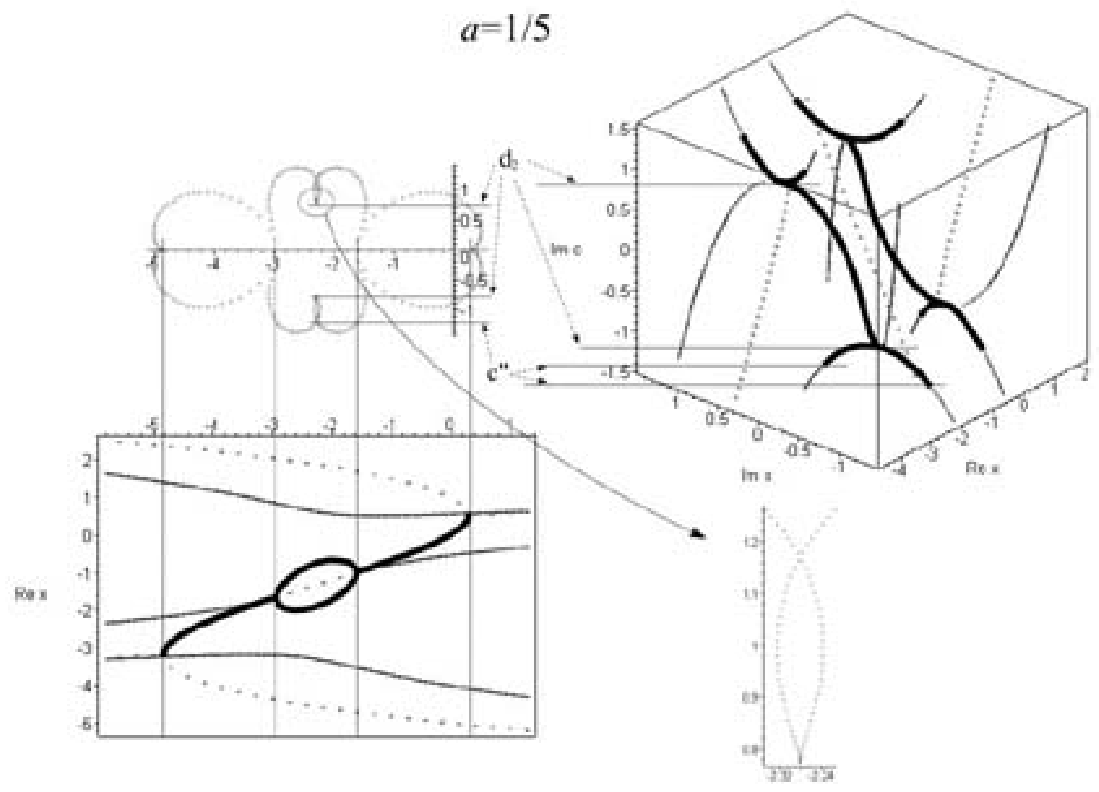}
{500,355}
{\footnotesize
Domains of the orders $p=1,2$ of the
Mandelbrot set ${\cal M}_{2,3}(a)$ at $a=1/5=0.2$.
The $(2,1)$ domain is being pressed away by approaching
$(1)_\pm$ domains, which are now the two roots of a
single cluster.
Note re-appearance of the narrow overlap regions
inside the $(2,1)$ domain and non-trivial monodromy
of the order $2$ orbits when they are carried along
a circle in the Mandelbrot set, surrounding one
of the cusps $d_2^\pm$.
}

\subsubsection{Mandelbrot sets with the topology of ${\cal M}_3$
(in the vicinity of $a=1$)}

The further evolution of ${\cal M}_{2,3}(a)$ with increasing
$a$ consists mostly of continuous deformation of the shape
of emerged unified root $(1)$ domain with two cusps:
from a bone-like region in Fig.\ref{a1_5s} it grows into
nearly oval one (deviating from oval only near the
cusps) in Figs.\ref{a1_4}, \ref{a1_3} and finally at $a=1$,
when the cusps extend their region of influence, acquires the
standard form of Fig.\ref{a1_1},
familiar from Fig.\ref{0Man3}.

\Fig{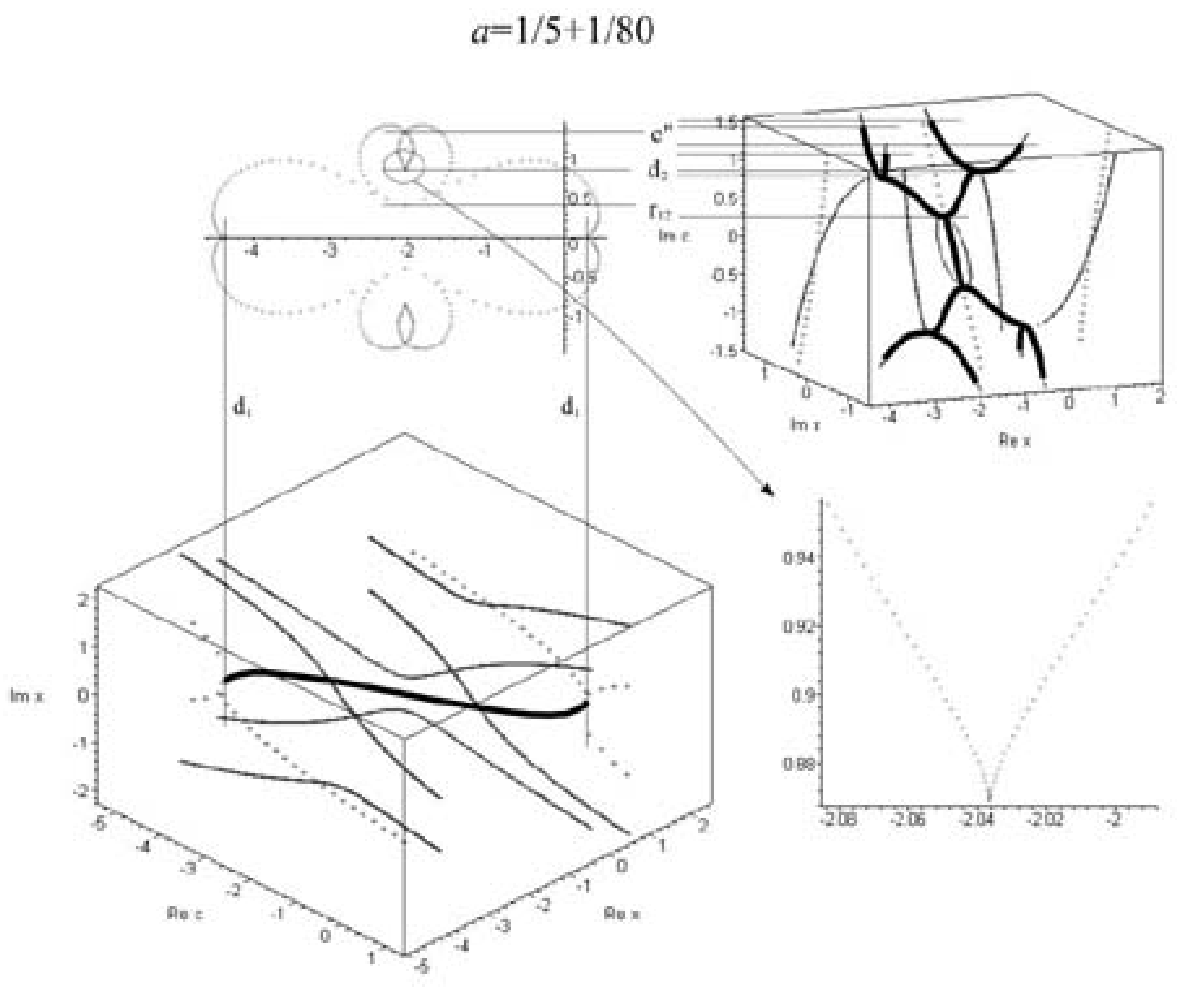}
{500,381}
{\footnotesize
Domains of the orders $p=1,2$ of the
Mandelbrot set ${\cal M}_{2,3}(a)$ at $a=1/5+1/80 = 0.2125$.
The two root domains $(1)_\pm$ just collided at
$a= \frac{5-\sqrt{21}}{2} = 0.20871215\ldots$ and formed a
single root domain $(1)$. It has a typical bone-like shape,
with two pieces of the former descendant $(2,1)$ domain
attached at the merging point.
Overlap regions inside these $(2,1)_\pm$ fragments are
now pretty large and cusps are not well seen, still
the picture in the low right corner
shows that they are still there.
$3_Rd$ view in the low left corner is taken at ${\rm Im}(c)=0$,
where only order $p=1$ orbit can be stable inside the root
$(1)$ domain.
}

\Fig{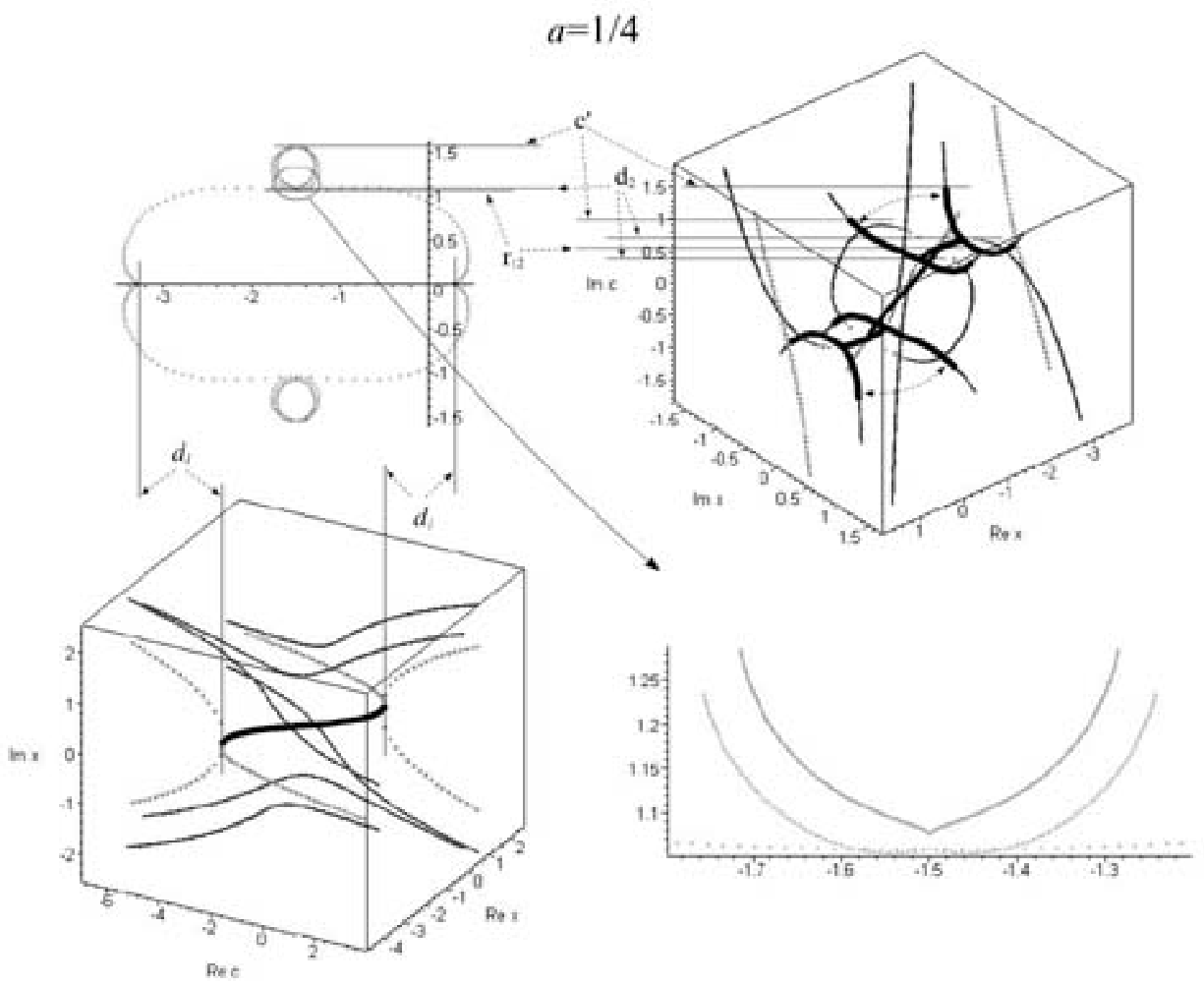}
{500,399}
{\footnotesize
Domains of the orders $p=1,2$ of the
Mandelbrot set ${\cal M}_{2,3}(a)$ at $a=1/4$.
In this picture we are near the point
$a = \frac{9-\sqrt{65}}{4}=0.23443556\ldots$
where inside-out reshuffling of the $(2,1)$
domains takes place (two zeroes $r_{24}$ of $R_{24}$
coincide at this transition point).
Overlap regions, seen in Fig.\ref{a1_5s}, grew up
to the full size of the $(2,1)$ domains, and
boundaries of the regions changes roles with boundaries
of the domains. Later the former boundary domains will
get closer and form new small overlap regions in
Fig.\ref{a1_3} that will later disappear at $a=0.42\ldots$}

\Fig{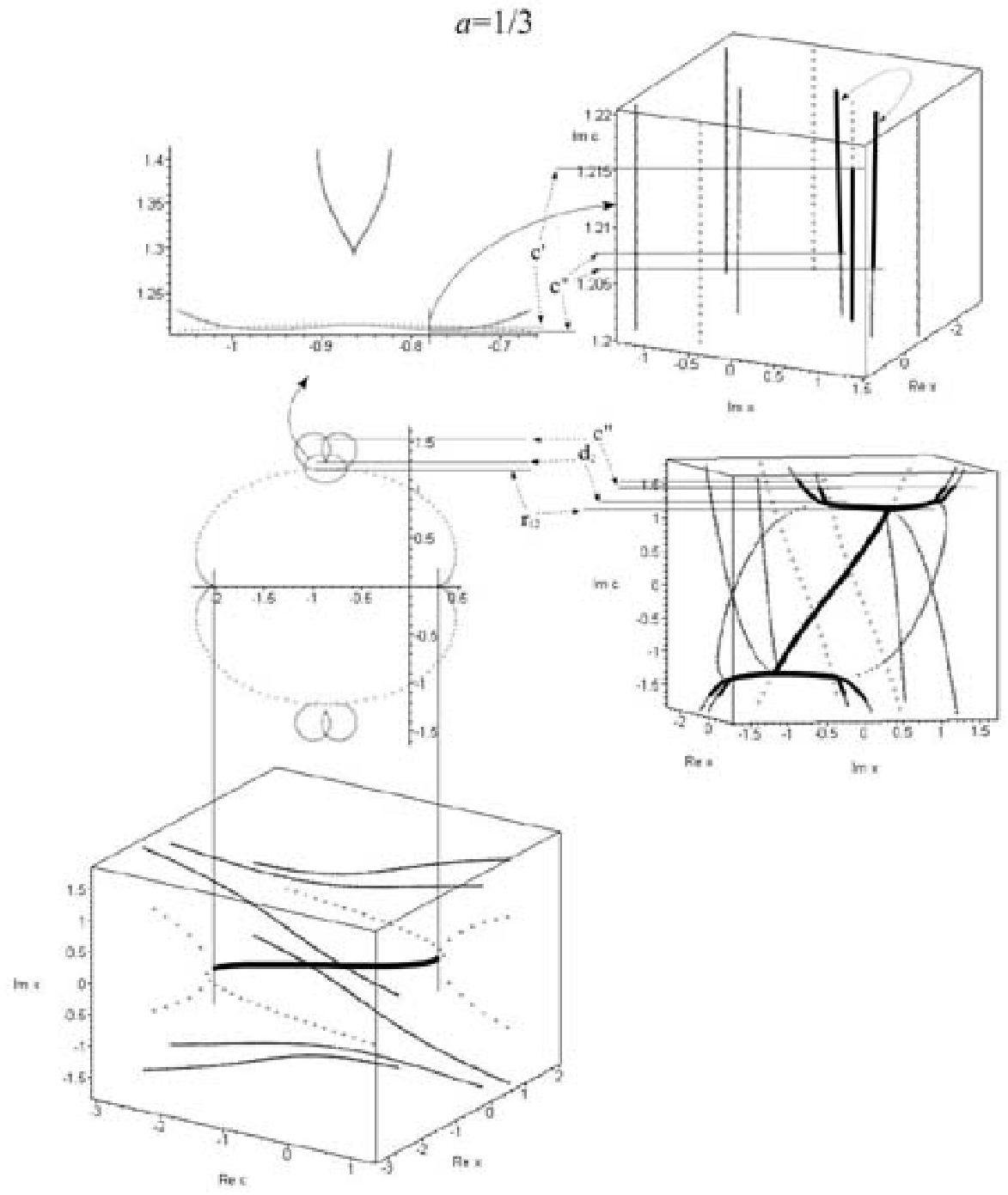}
{500,567}
{\footnotesize
Domains of the orders $p=1,2$ of the
Mandelbrot set ${\cal M}_{2,3}(a)$ at $a=1/3$.
The root $(1)$ domain acquired an almost ellipsoidal
form (outside the cusp regions).
The interesting phenomenon, seen already in Fig.\ref{a1_4},
 is appearance of additional
overlap regions, marked by arrows in the upper left corner:
between the root $(1)$ and descendant
$(2,1)_\pm$ domains, where order-$1$ and order-$2$
orbits are simultaneously stable.
This time there is no interesting monodromy: when we
leave the overlap region in the direction of $(1)$ the
order-$1$ orbit remains stable, when we go deep into
$(2,1)_\pm$, the stable one is an order-$2$ orbit.
}

\Fig{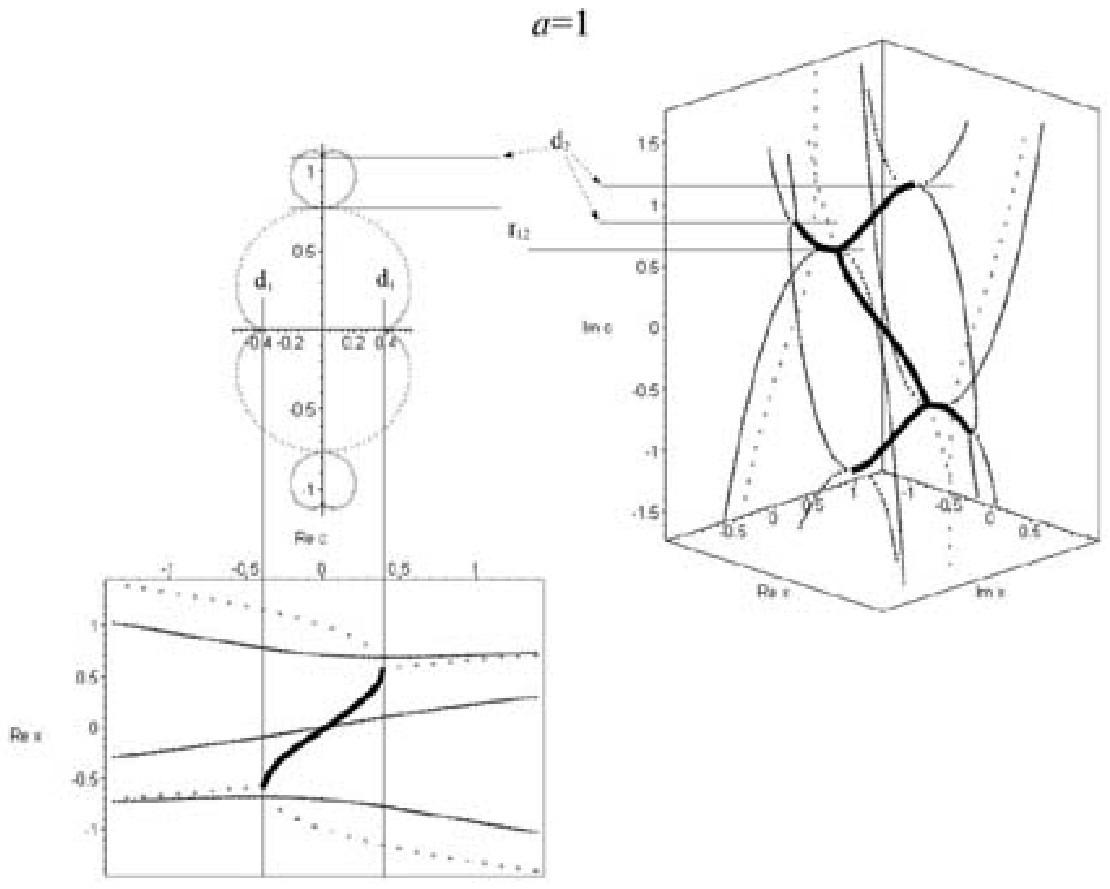}
{500,438}
{\footnotesize
Domains of the orders $p=1,2$ of the
Mandelbrot set ${\cal M}_{2,3}(a)$ at $a=1$.
This is the central part of the standard Mandelbrot
set ${\cal M}_3$, shown in Fig.\ref{0Man3}.
All overlap regions disappeared, the $(1)$ domain
is ideal cubic cardioid, other elementary domains
acquire nearly cardioid shapes.
}


Still, while nothing equally drastic happens after the
cluster collision, evolution is not quite event-less.
In above pictures one can see that the overlap regions
in the $(2,1)$ domain(s) appear and disappear, signaling
about the motion of {\it orbits} in Julia sheaf with the
change of $a$.
In particular, an interesting inside-out reshuffling is shown in
Fig.\ref{a1_4}.
Moreover, in Fig.\ref{a1_3} one can see
that overlap occurs even between the $(1)$ and $(2,1)$ domains.
We emphasize once again, that no bifurcations (phase transitions,
orbit crossing or reshuffling) are associated with the overlaps,
still they affect the shape and even the very presentation
of the Mandelbrot set (when overlaps exist, it is not a clever
idea to draw it all in black, like we did in
Figs.\ref{Mand2}-\ref{symmebre}) and in fact this is a signal
that the phase portrait gets richer: a non-trivial pattern
of attractors and repulsers  occurs, nothing to say that
the vicinities of unstable orbits are not necessarily
attracted to infinity, as implicitly assumed in some algorithms,
mentioned in the first paragraphs  of s.\ref{interpol}.

Events, encountered in the evolution of Mandelbrot set
${\cal M}_{2,3}(a)$ from $a=0$ to $a=1$, i.e. in interpolation
between Figs.\ref{Mand2} and \ref{0Man3}, are collected
in the following table:

\vspace{-0.5cm}
\centerline{
{\footnotesize
\begin{tabular}{|c|c|c|}
\hline
&&\\
$a$ & typical feature & picture \\
&&\\
\hline\hline
&&\\
$a<0$&$\ldots$&\\
&&\\
\hline\hline
&&\\
$a=0$&the standard Mandelbrot Set ${\cal M}_2$ &
Fig.\ref{Mand2}\\
&&\\
\hline
&&\\
$0<a<0.1164\ldots$ &two root domains of type $(1)$;
& Fig.\ref{a1_10}\\
&two descendant domains $(2,1)$, attached to them;&\\
&two isolated root domains $(2)$; & \\
&two $(4,2,1)$ domains and two $(4,2)$ domains,&\\
&attached to $(2,1)$ and $(2)$ at real values of $c$ &\\
&&\\
\hline
&&\\
$a=0.1164\ldots$&{\it projections} of two domains $(2)$ meet,&\\
&i.e. two zeroes of $R_{24}$ coincide,& \\
&responsible for stability of two
{\it different} order-$2$ orbits&\\
&&\\
\hline
&&\\
$0.1164\ldots < a < 0.1270\ldots$
&{\it projections} of two domains $(2)$ overlap;&\\
&two {\it stable} order-$2$ orbits coexist& Fig.\ref{a1_9s}\\
&in the region of overlap&\\
&&\\
\hline
&&\\
$a=4-\sqrt{15}=0.1270\ldots$&cusps of overlapping domains $(2)$
merge,&\\
&i.e. two zeroes of $D_2$ coincide&\\
&&\\
\hline
&&\\
$0.127\ldots< a < 0.130\ldots$&overlap of the two domains $(2)$ &
Fig.\ref{a1_8s}\\
&splits into two isolated components&\\
&&\\
\hline
&&\\
$a = 0.130\ldots$&overlap region shrinks down&\\
&&\\
\hline
&&\\
$0.130\ldots< a < 0.1497\ldots
$&only one unified root domain $(2)$  exists&\\
&&\\
\hline
&&\\
$a=0.1497\ldots$&$(2,1)$ domains merge with the $(2)$ domain,
& Fig.\ref{a3_20}\\
&i.e. two pairs of zeroes of $R_{24}$ coincide,&\\
&each pair responsible for stability&\\
&of {\it the same} order-$2$ orbit&\\
&&\\
\hline
&&\\
$0.1497\ldots < a < 6.6800\ldots$
&only one $(2,1)$ domain exists;& Fig.\ref{a1_7s}\\
&no $(2)$ domains &\\
&&\\
\hline
&&\\
$a=0.1875\ldots$ &coexisting stable order-$2$ orbits re-emerge&
Fig.\ref{a1_5}\\
&&\\
\hline
&&\\
$a = \frac{5-\sqrt{21}}{2} = 0.2087\ldots$
&two $(1)$ domains meet and&Fig.\ref{a1_5s}\\
&the $(2,1)$ domain splits into two,&\\
&two zeroes of $R_{12}$ coincide&\\
&&\\
\hline
&&\\
$0.2087\ldots < a < 0.42\ldots $
&overlap regions, where two stable order-$2$ orbits
& Figs.\ref{a1_4} \& \ref{a1_3}\\
&or order-$2$ and order-$1$ orbits can coexist;&\\
$0.2087\ldots < a < 4.7916\ldots $ &one $(1)$ domain;&\\
&two attached $(2,1)$ domains;&\\
&no $(2)$ domains&\\
&&\\
\hline
&&\\
a = 0.42\ldots &overlap region shrinks down&\\
&&\\
\hline
&&\\
$a=1$&the standard Mandelbrot set ${\cal M}_3$ &
Figs.\ref{a1_1} \& \ref{0Man3}\\
&&\\
\hline\hline
&&\\
$a>1$&$\ldots$&\\
&&\\
\hline
\end{tabular}
}}

\bigskip

In s.\ref{doverlap} we briefly discussed what happens
beyond the realm of this table: for negative values of $a$.
The evolution of ${\cal M}_{2,3}(a)$ can be also continued
to the region where $a>1$ (where $b=a-1<0$).
This evolution appears to be reverse of what we already
considered: the Mandelbrot set of Fig.\ref{0Man3} at $a=1$
passes through the same stages of Figs.\ref{a1_3},
\ref{a1_4}, \ref{a1_5s}
(at $a=3,\ 4, \ 4.706\ldots$ respectively) and so on.
In particular, at $a= \frac{5+\sqrt{21}}{2} = 4.7912878\ldots$
the single root $(1)$ domain is split into two, while
two descendant $(2,1)$ domains merge into one.
For illustration we show in Fig.\ref{a5_1} the counterpart
of Fig.\ref{a1_5}.
The full picture will be shown in s.\ref{UMS} below.


\Fig{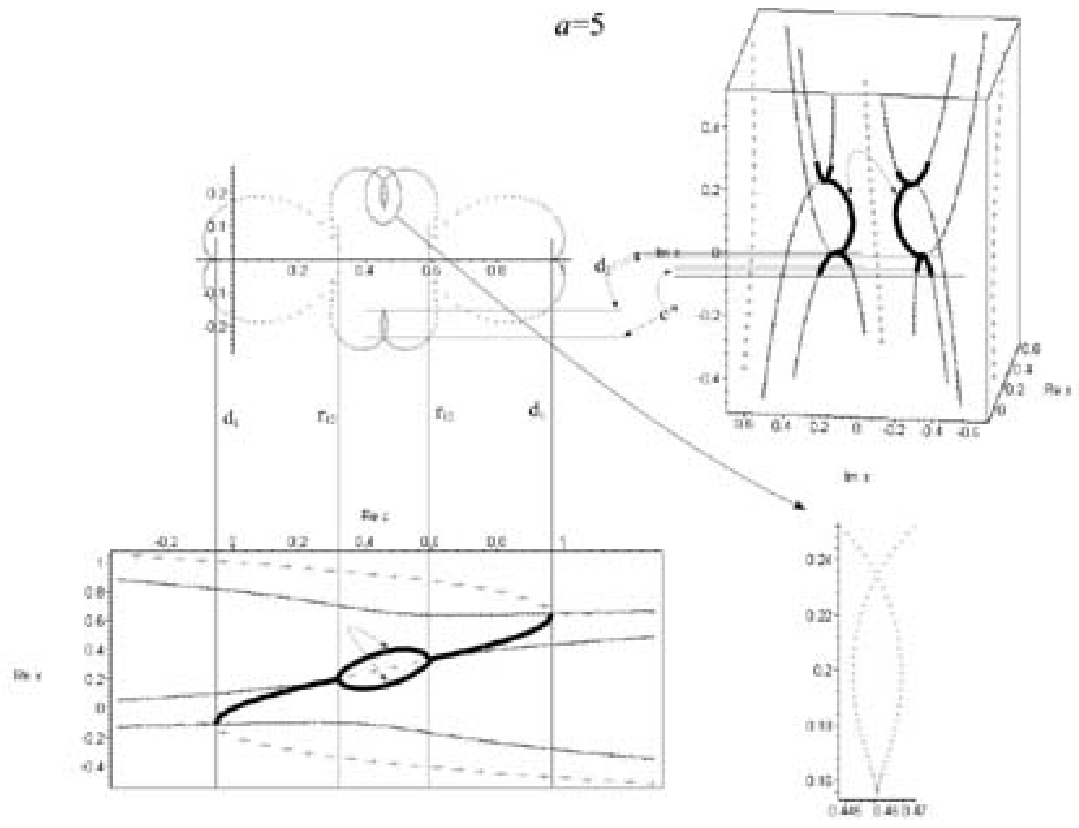}
{500,393}
{\footnotesize
Domains of the orders $p=1,2$ of the
Mandelbrot space ${\cal M}_{2,3}(a)$ at $a=5$.
This picture is direct analogue of Fig.\ref{a1_5},
and serves as an illustration of the symmetry property
(\ref{symmemaf}):
that the same sequence of bifurcations
happens to the Mandelbrot set ${\cal M}_{2,3}(a)$ on the way
from $a=1$ to $a=-1$ along the
real-$a$ line through $a=\infty$ as on direct way, presented in
Figs.\ref{a1_10}--\ref{a1_1}.
}

\subsection{First steps towards UMS \label{UMS}}

Let us now return to interpretation of Mandelbrot sets as
sections of a single Universal Mandelbrot Set (UMS).
It implies that all what we observe about particular
collection of Mandelbrot sets, like our one-parameter
family ${\cal M}_{2,3}(a)$, can be re-interpreted as result
of particular view on one and the same solid structure:
variation of patterns is result of the changing view,
the structure is always the same.\footnote{We can not
avoid stressing analogy with the well-known
{\it projection} approach
to {\it integrable} systems, see \cite{UFN3}}.

For example, the entire collection of the $p=1$ domains
in Figs.\ref{a1_10}-\ref{a5_1} which consist of a single
component for $a$ in between $\frac{5\mp \sqrt{21}}{2}$
and split into two components outside this segment,
see Fig.\ref{p1domsdia}.A,
can be alternatively described as the image of a single
cardioid-like domain in the slices, evolving with the
change of the section in Fig.\ref{p1domsdia}.B.
Moreover, in this approach one can even start from
an ordinary circle, not from a cardioid,
see Fig.\ref{cardcircdia}.
In fact these pictures are nothing but approximate drawings,
attempting to capture the properties of exact formula
\be
c = -\frac{b}{3a}\left(1+\frac{2b^2}{9a}\right) \pm
\left(1+\frac{2b^2}{9a} -\frac{e^{i\theta}}{3}\right)
\frac{\sqrt{b^3 +
3ae^{i\theta}\phantom{5^{5^5}}\hspace{-0.45cm}}}{3a},
\label{p1doms}
\ee
just now we interpret it is an evolution of
(degenerate) {\it elliptic} mappings
\be
(c-c_0)^2 = k(u-u_1)(u-u_2)(u-u_3),\ \ \
{\rm with}\ a-{\rm dependent\ parameters}\ c_0, k\ {\rm and}\ u_i;
\ \ {\rm actually}\ u_2=u_3
\label{ellip}
\ee
of a complex-$u$ plane with a unit circle on it
into a complex-$c$ plane, where the image of the unit circle
looks differently: like our $p=1$ domains, evolving and
even bifurcating (splitting and merging) under the change
of the mapping.

\Fig{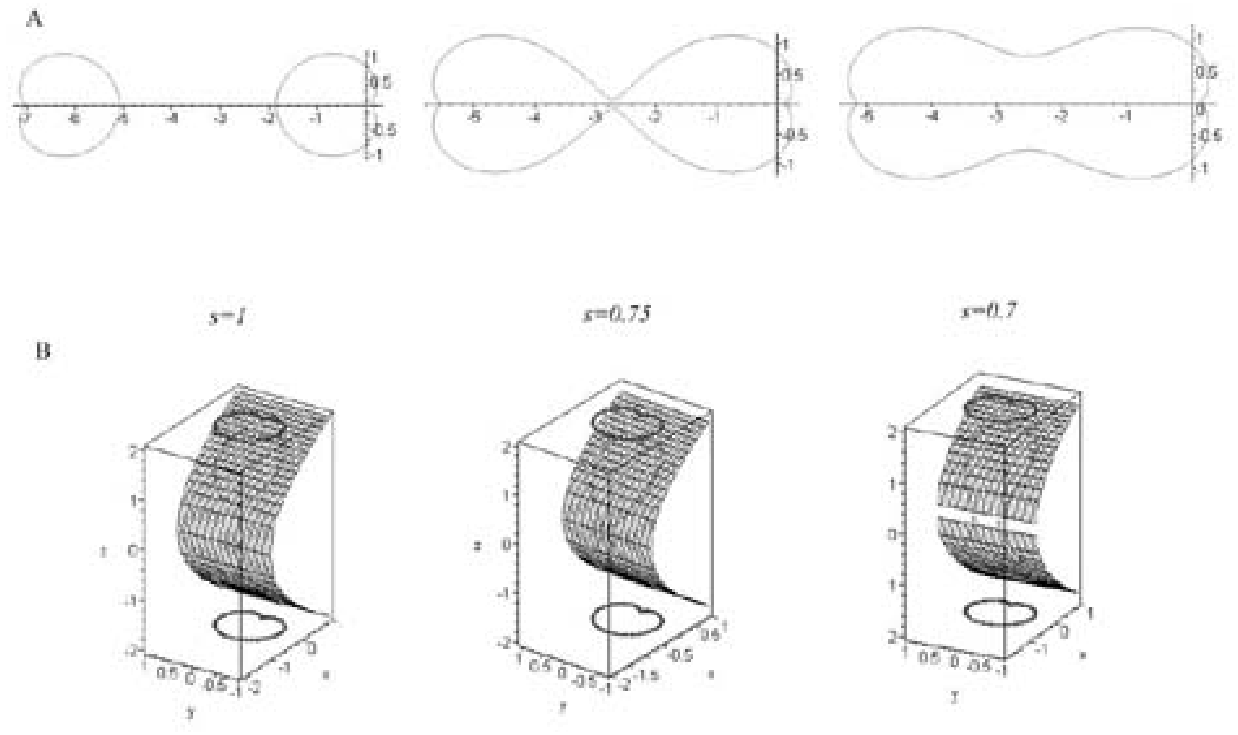}
{500,315}
{\footnotesize
{\bf A.} The shapes of $p=1$ elementary domains for different
values of $a$ in the family $ax^3+(1-a)x^2+c$,
as given by eq.(\ref{p1doms}).\ \
{\bf B.} The same (qualitatively) pictures arise in the
sections of a single cardioid-like cylinder by different
parabolic-like sections ($x-(z^2/2-s)=0$ for various $s$) with a complex-$c$ plane. \ \
{\bf C.} The cylinder can be even made circlic??? with the
help of the circle-cardioid relation shown in
Fig.\ref{cardcircdia}.
The corresponding sections are now generic conics (quadrics???),
not obligatory paraboloid.
The true sections behind eq.(\ref{p1doms}) are {\it cubic},
see (\ref{ellip}),
not a quadric, which are a little less convenient to draw.
}

\Fig{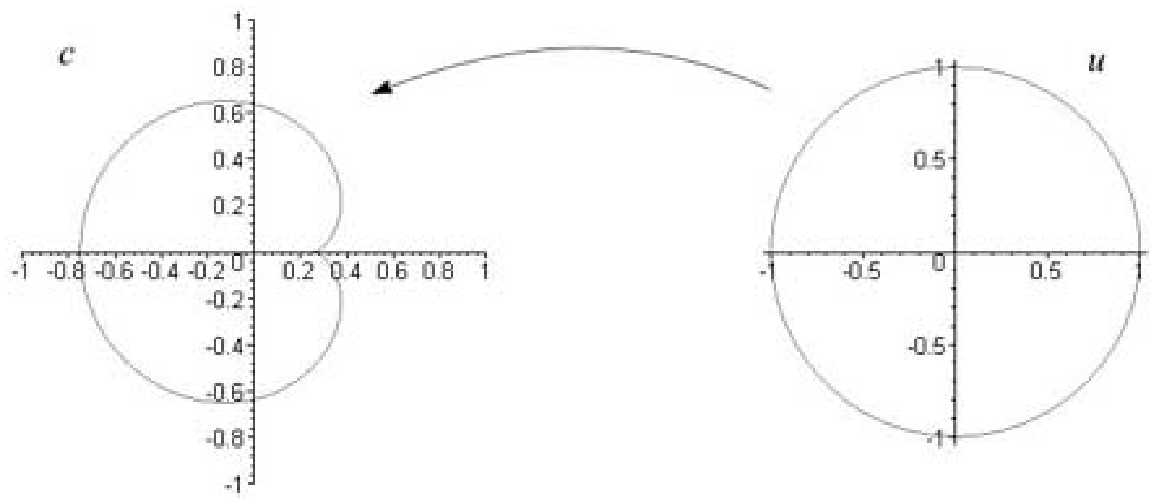}
{500,188}
{\footnotesize
Cardioid cylinder of the previous Figure \ref{p1domsdia}
can be substituted by an ordinary rotation-symmetric cylinder
at expense of a more sophisticated slicing.
Cardioid as a square of a circle.
Analytically this correspondence is represented by
$c = \frac{u}{2}\left(1-\frac{u}{2}\right)$ or
$1-4c=(u-1)^2$.
}

At this early stage of investigation of UMS it is unclear,
what is its best and simplest possible representation.
In particular, nothing as simple as Fig.\ref{p1domsdia} is
immediately available when the order-$2$ orbits are taken into
account.
Therefore, instead of playing with different realizations,
we show in Figs.\ref{tubeD3}, \ref{tubeD3L1longrange1}
the most straightforward views of the
Universal Mandelbrot Set, directly in the $a-c$ coordinates.
Unfortunately, only $3_R$-dimensional section of the full
$2_C$-dimensional pattern can be drawn in a picture
and it can be rotated (what is very informative!) only on
computer screen, see \cite{maplesamples}
for details about this option.

\Fig{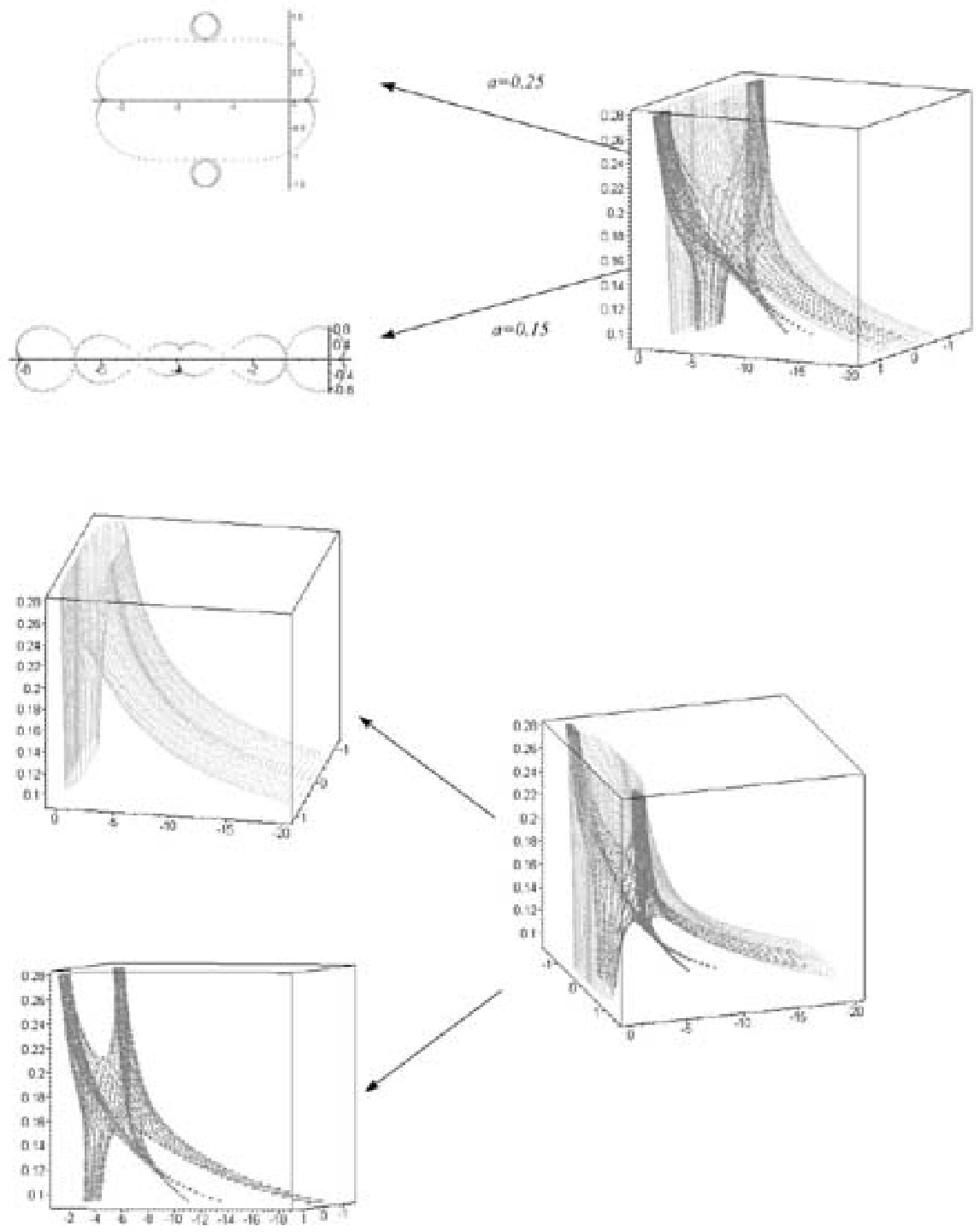}
{500,604}
{\footnotesize
Collection of order-$1$ and -$2$ domains of
Mandelbrot sets ${\cal M}_{2,3}(a)$ for the family
$f(x) = ax^3+(1-a)x^2+c$ with various values of $a$
in the region $0.10\leq a\leq 0.28$,
where most bifurcations are taking place,
represented as slices of a single $3_Rd$ entity,
which can be also considered as a $3_Rd$ section of UMS.
Horizontal is the plane of complex $c$, vertical is the
line of real $a$.
This picture provides a concise summary of all the properties,
described throughout s.\ref{23interp}.
The lower part of the figure represents separately the $p=1$ and
$p=2$ domains: all pictures are also shown from different angles,
what can help to appreciate the beauty of the structure.
}

\Fig{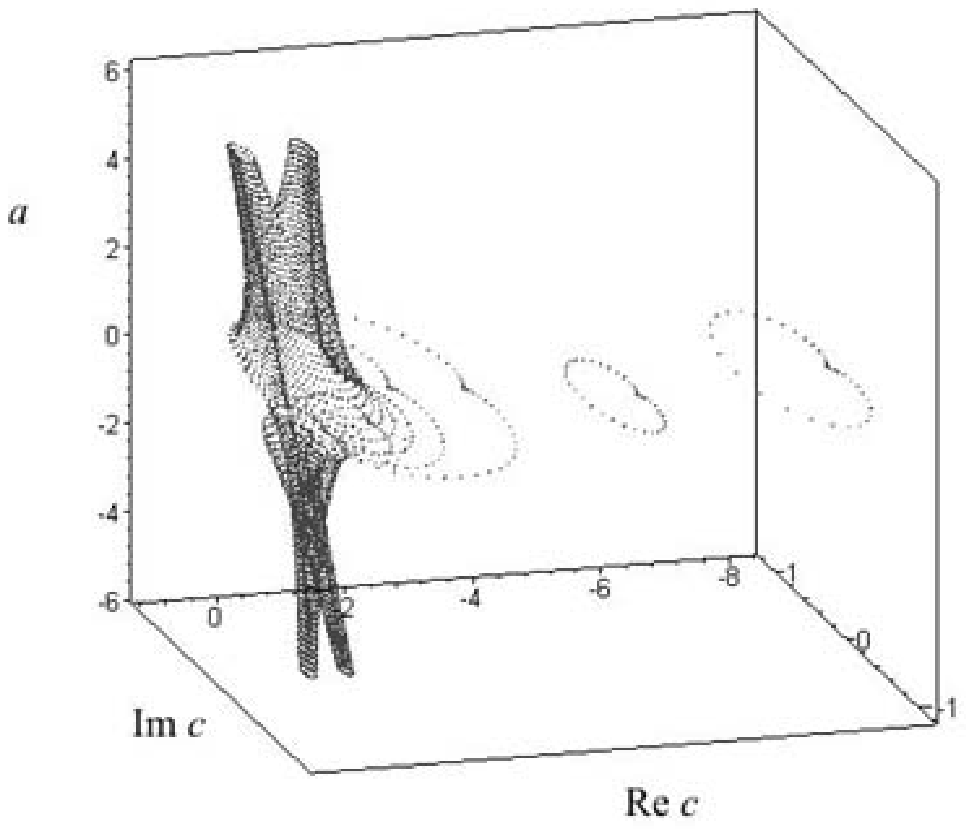}
{450,395}
{\footnotesize
The same collection of Mandelbrot sets ${\cal M}_{2,3}(a)$
for the family $f(x) = ax^3+(1-a)x^2+c$
with all real values of $a$
(including those represented in Figs.\ref{a1_10}-\ref{a5_1}).
Only central domains of order $1$ are shown.
}

In Fig.\ref{tubesD4} we give some more examples of $2_C$-sections
of the UMS. In particular, we demonstrate that topologies of this
section can be very different and they can be investigated by
already available tools. As a small illustration, the chain of
pictures in Fig.\ref{tubesD4},B shows how a loop in particular
section of the UMS can be (on Fig.\ref{tubesD4},A) contracted. Of
course, we are very far from calculation of homologies of the UMS,
but the way is already open.

\Fig{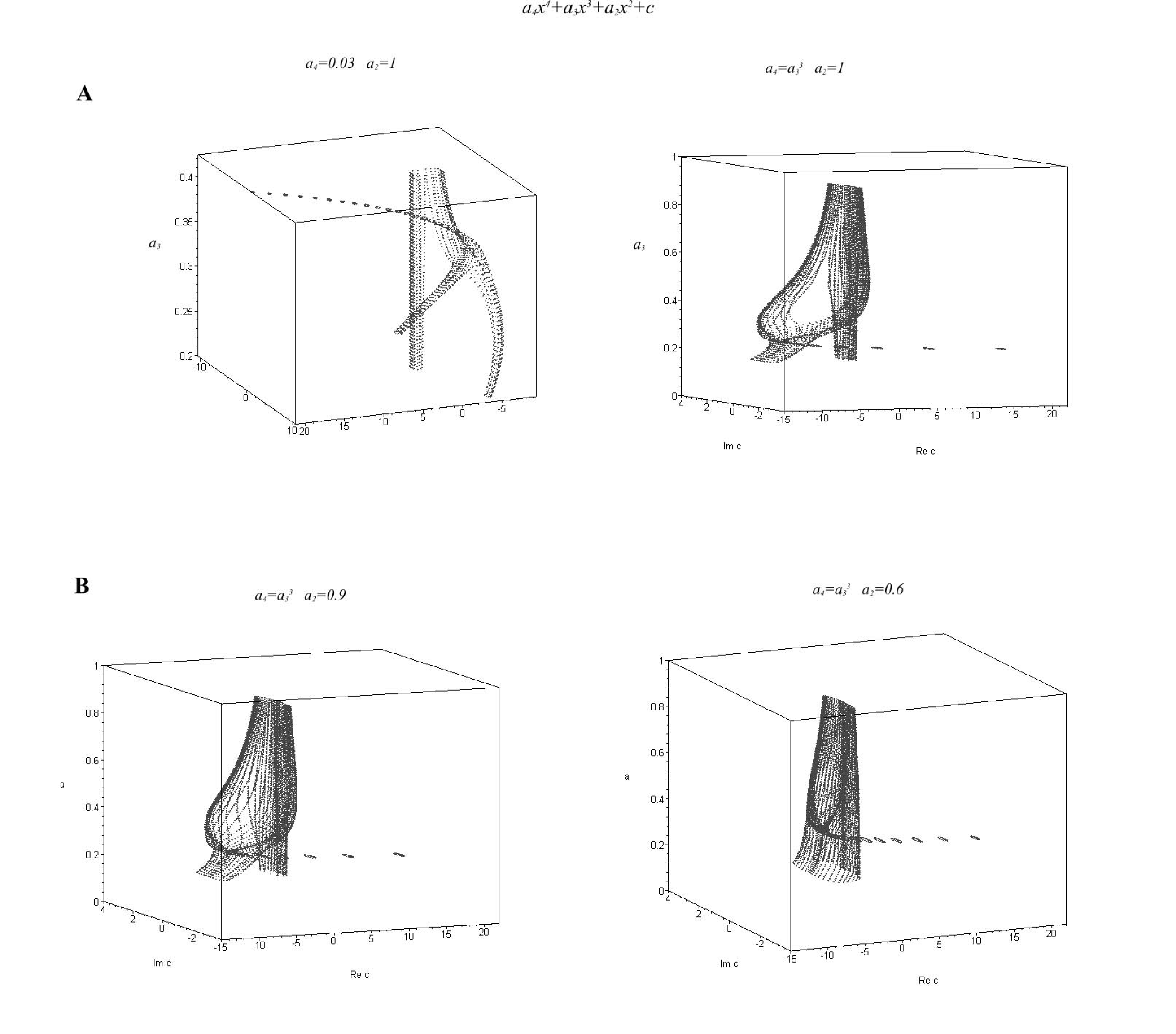}
{500,440}
{\footnotesize
Examples of other $3_Rd$ sections of the Universal Mandelbrot Set.
Only central domains of order $1$ are shown.
}

\section{Small-size approximation (SSA) \label{ssasec}}

\setcounter{equation}{0}

We now switch from transparent exact results
to subtle approximate methods.

\subsection{On status of the SSA}

The shape of elementary domains can be also considered in the
"small-size" approximation (see s.4.9.3 of \cite{DM}
and its less accurate predecessor in \cite{LL}
and many other text-books).
In SSA we expand all functions of $x$ and $c$ in powers
of their deviations from the critical (or, simply, mean)
values and
leave only the first three (constant, linear and the next)
terms in these expansions.
In what follows $x_{cr}=0$, but $c_{cr}$ will have different
values. "Next" normally means quadratic, but for homogeneous
polynomials $P_d(x) = x^d$ (giving rise to $Z_{d-1}$-symmetric
Mandelbrot sets) it will actually mean $x^d$.

SSA would be very natural, if typical deviations of $x$ and $c$
from the mean values were small. However, while it can seem
reasonable for the study of elementary domains --
except for the first few, they are indeed pretty small in
the $c$-plane (most are not even seen in
Figs.\ref{Mand2}--\ref{0Man4}),-- actually this assumption is wrong:
as we already know from Figs.\ref{a1_10}--\ref{a5_1},
the $x$-variables in solutions to eqs.(\ref{shapeeq}) are not small.
At today's level of knowledge justification of SSA comes only
{\it a posteriori}, say, from comparison with experimental
data in s.\ref{accu}).
SSA seems adequate for {\it phenomenological} description of
experimentally observed \cite{pdb,Mand}
self-similarity (fractal or scaling) property of the Mandelbrot
sets, but no clear {\it theoretical} reason for this adequateness
is known.
The algebro-geometrical approach of \cite{DM} only adds to the
mystery: the (experimentally) obvious scaling properties
of the universal discriminantal variety call for clear conceptual
explanations.

In any case, today SSA is the only available approach to
evaluation of Feigenbaum indices and other characteristics of
elementary domains when their order $p \rightarrow \infty$.
Surprisingly or not, it does a very good job in this field:
see s.\ref{Feig} below.

\subsection{SSA for the Mandelbrot Set}

The {\bf first step} of the SSA in application to the Mandelbrot
Set (i.e. to the family $f(x;c) = x^2+c$) is to expand
\be
F_p(x) = f_p(c) - x + x^2\gamma_p(c)  + O(x^{4})
\label{approx}
\ee
It is used here that that $f'(x=0)=0$, otherwise expansion in powers
of $x$ should be substituted by that in powers of $x-x_{cr}$,
where $f'(x_{cr})=0$.

The {\bf second step} is to substitute (\ref{approx}) into
(\ref{shapeeq}). This gives:
\be
2x\gamma_p(c)  = e^{i\phi}
\label{xvsphi}
\ee
and
\be
2f_p(c)\gamma_p(c) = 2 x\gamma_p(c) \Big(1-x\gamma_p(c)\Big)
\ \stackrel{(\ref{xvsphi})}{=}\
e^{i\phi}\left(1 - \frac{1}{2}\,e^{i\phi}\right)
\label{shape}
\ee
Eq.(\ref{shape}) provides the answer: the l.h.s. depends on the
shape of the function $f(x;c)$, i.e. on $c$, and (\ref{shape})
describes a curve $c(\phi)$.

Actually this curve can have many disconnected components,
and the {\bf third step} is to consider a particular component,
surrounding a particular root $c_p$ of the Mandelbrot function
$f_p(c) = f(x_{cr},c)$:
\be
f_p(c_p) = 0
\ee
Expand (\ref{shape}) around $c_p$:
$$
c = c_p+\sigma,
$$ \vspace{-0.25cm}
\be
f_p(c) = f_p(c_p+\sigma) = \dot f_p\,\sigma\left(1 +
\frac{\ddot f_p}{2\dot f_p}\, \sigma + O(\sigma^2)\right)
\label{expans}
\ee
\vspace{-0.25cm}
$$
\gamma_p(c) = \gamma_p(c_p+\sigma) =
\gamma_p\left(1 + \frac{\dot\gamma_p}{\gamma_p}\,\sigma +
O(\sigma^2)\right)
$$
From now on we denote through $f_p$ and $\gamma_p$
the values of the corresponding functions at $c=c_p$:
$\gamma_p \equiv \gamma_p(c_p)$ etc.
Substituting (\ref{expans}) into (\ref{shape}), we get:
\be
\frac{\sigma}{r_p}\left(1 -\frac{\xi_p}{2}
\frac{\sigma}{r_p}\right) \approx
r_p e^{i\phi} \left(1 - \frac{1}{2}\,e^{i\phi}\right)
\label{shape1}
\ee
with
\be
r_p = \frac{1}{2\dot f_p\gamma_p}
\label{diam}
\ee
and
\be
\xi_p = -\frac{1}{\dot f_p\gamma_p}\left(
\frac{\ddot f_p}{2\dot f_p} + \frac{\dot\gamma_p}{\gamma_p}
\right)
\label{xip}
\ee

Eq.(\ref{shape1}) is our final SSA answer for the shape of
the elementary domain of the Mandelbrot Set, surrounding
a point $c_p$. We see that the complex-valued $r_p$ defines the
{\it size} and {\it orientation} of the domain, while its
{\it shape} is fully controlled by the value of $\xi_p$:
for $|\xi_p|\ll 1$ we get a cardioid, while for $|\xi_p-1|\ll 1$
it turns into a circle.

\bigskip

Thus our problem is reduced to:

-- the check of accuracy of the small-size approximation:\ \
we do this in s.\ref{accu} by comparing the values of $r_p$,
predicted by (\ref{diam}) with their actual values for the
family $f(x;c) = x^2+c$, measured with the help of the
{\it Fractal Explorer} \cite{FE} or defined from the roots of
the relevant resultants;

-- evaluation of parameter $\xi_p$ with the help of (\ref{xip}):
\  in s.\ref{shapes2} we show that indeed
in the small-size approximation $\xi_p=1$ for elementary domains,
which are {\it not} roots of any clusters (i.e. are descendants
of some lower-level domains);

-- demonstration that higher-order cardioids emerge in the special
case of maps with $Z_{d-1}$ symmetry: in this case the symmetry
requires that $\gamma_p=0$ and (\ref{shape1}) gets substituted
by a more sophisticated expression (\ref{shaped}), investigated
in s.\ref{dfam} (emerging shapes are somewhat less ideal than for
$d=2$, deviations can reach tens of percents).

\subsection{Comments}

Note that at the third step we kept terms up to $\sigma^2$
in expansions of the $c$-dependent functions, like we did
at the first step with the $x$ functions.
As usual for this type of method to work successfully
it is important to correlate all approximations:
attempt to make one part of calculation more accurate than
another {\it decreases} the total accuracy.
Note also that keeping linear terms only, without quadratic
corrections, would make eq.(\ref{xvsphi}) senseless, and
according to the just formulated mnemonical rule one is
forced to keep $\sigma^2$ terms as well.
And indeed, neglecting them would provide a disaster in
description of descendant domains: we already learned
in s.\ref{minuscusp} that their
characteristic difference from the root domains is that
$\dot F_p(x,c) = \dot f_p(c) + x^2 \dot\gamma(c) + O(x^4)$
should vanish somewhere at the boundary, while in neglect of the
$\sigma^2$ terms this would be very difficult to achieve
while keeping $\dot f_p = \dot f_p(c_p) \neq 0$ in the center
$c_p$ of the domain.
In fact, the difference between descendant and root domains
is exactly in $\sigma^2$-terms: their relative magnitude
is measured by parameter $\xi_p$, and it is negligibly small
for root domains and close to unity for descendant ones.

\section{On accuracy of the small-size approximation for
the family $f = x^2 + c$ \label{accu}}

\setcounter{equation}{0}


Numerical characteristics of the lowest (in divisor forest)
elementary domains of the Mandelbrot set ${\cal M}_2$ are
represented in the following table (positions of these domains
are marked by arrows in Fig.\ref{6th}).

\Figeps{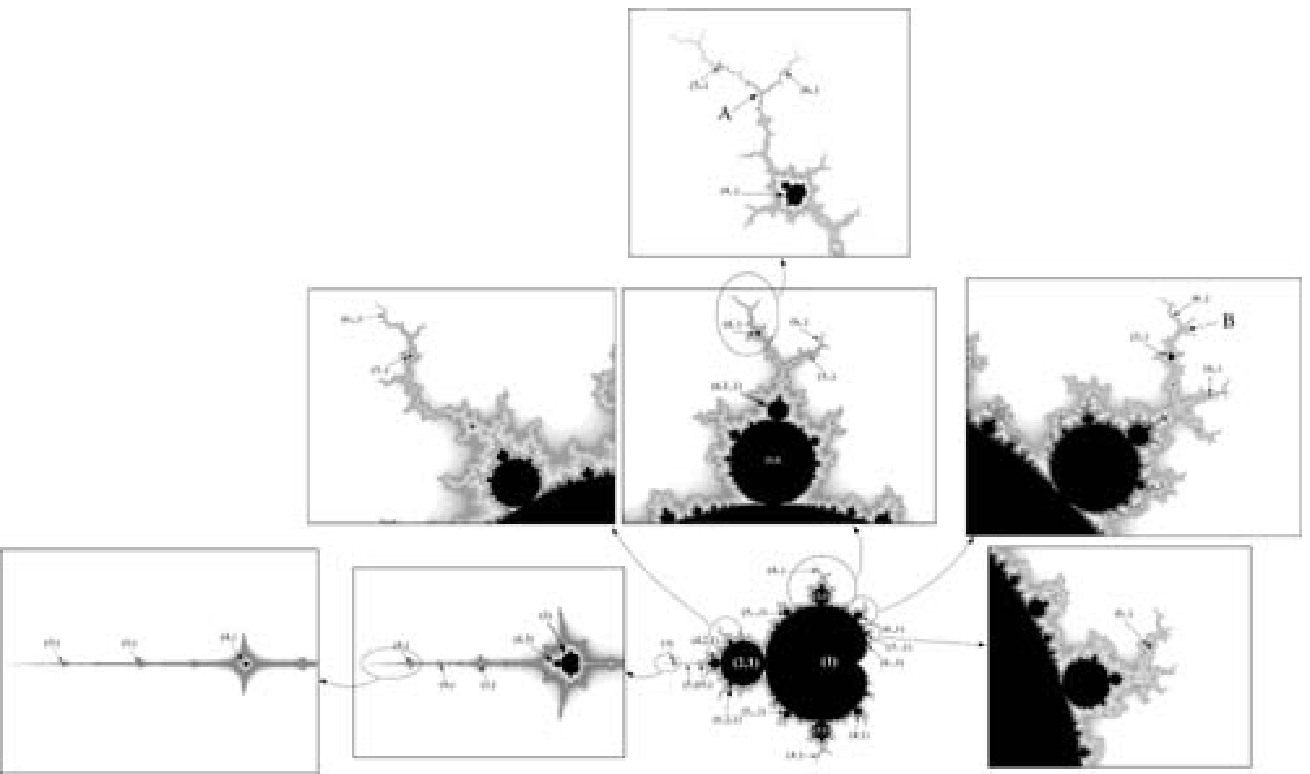}
{500,294}
{\footnotesize Mandelbrot Set from
Fig.\ref{Mand2} with arrows, pointing at particular elementary
domains, represented in the table in s.\ref{accu}. {\it Trails}
are also seen in pictures with increased resolutions. Trails are
densely populated by clusters. The theory of trails remains an
open subject. An interesting task is to study how a trail between
some two clusters is formed, when we travel across Universal
Mandelbrot Space as in s.\ref{23interp}, and these two clusters
emerge from a splitting of a single cluster. Another interesting
problem is to find the locations of the {\it crossing points},
where many trails meet together, examples of such crossings are
points ${\bf A}$ and ${\bf B}$.}

\bigskip

{\footnotesize
$$
\hspace{-0.6cm}
\begin{array}{|c|c|c|c|c|c|c|c|c|}
\hline
&&&&&&&&\\
p & 1 & 2 & 3 & 3 & 4 & 4 & 4 & 4 \\
&&&&&&&&\\
\hline
&&&&&&&&\\
c_p
&0&-1&-1.754877667&-0.1225611669&-1.940799807
&-0.1565201668&0.2822713908&-1.310702641\\
&&&&\pm 0.7448617670i&
&\pm 1.032247109i&\pm 0.5300606176i&\\
&&&&&&&&\\
\hline
&&&&&&&&\\
{\rm distance}
&{\bf 0}&{\bf 1}&{\bf 0}&{\bf 1}&{\bf 0}&{\bf 0}&{\bf 1}&{\bf 2}\\
{\rm from\ the\ root}
&(1)&(2,1)&(3)&(3_\pm,1)&(4_1)&(4_{2\pm})&(4_{\pm},1)&(4,2,1)\\
&&&&&&&&\\
\hline\hline
&&&&&&&&\\
\dot f_p
&1&-1&-5.649435914&-1.67528205&-25.53361247
&-9.826826127&-2.273407347&1.734079638\\
&&&&\mp 1.1245590i&
&\mp 1.391722418i&\mp 2.878229429i&\\
&&&&&&&&\\
\hline
&&&&&&&&\\
\ddot f_p
&0&2&17.89661552&-5.9483077&247.9985718
&-13.82424296&-30.39333448&-15.56341649\\
&&&&\pm 6.7473541i&
&\pm 86.41684550i&\mp 9.578562508i&\\
&&&&&&&&\\
\hline
&&&&&&&&\\
\gamma_p
&1&-2&-9.29887185&-1.35056409&-39.49472178
&-10.55453437&-0.142465954&4.88872230\\
&&&&\mp 2.249118i&
&\mp 5.448066568i&\mp 3.098932717i&\\
\hline
&&&&&&&&\\
\dot \gamma_p
&0&2&22.91612617&-7.458063&-39.49472178
&-42.68642404&-14.72856928&-21.82510289\\
&&&&\pm 3.767907i&
&\pm 80.90620560i&\mp 16.79943562i&\\
\hline\hline
&&&&&&&&\\
\xi_p
&{ 0}&{ 1}&{ 0.07706201109}
&{ 1.0212516}&{ 0.01367007063}
&{ 0.05842559742}&{ 1.011598455}&{ 1.055967271}\\
{\rm from}\ (\ref{xip})
&&&&{\pm 0.04763015i}&
&{\pm 0.08449808566i}&{\pm 0.07045065295i}&\\
&{\bf \ll 1}&{\bf =1} &{\bf \ll 1}&
{\bf \approx 1}&{\bf \ll 1}&{\bf |\xi_p|\ll 1}
&{\bf \approx 1}&{\bf \approx 1}\\
&&&&&&&&\\
\hline
&&&&&&&&\\
{\bf 2r_p} = (\dot f_p \gamma_p)^{-1}
&{\bf 1}&{\bf 0.5}&{\bf 0.0190355}
&{\bf -0.009518}&{\bf 0.00099163}
&{\bf 0.0069178}&{\bf -0.066394}
&{\bf 0.1179602} \\
{\rm from}\ (\ref{diam})
&&&&{\bf \mp 0.18867i}&
&{\bf \mp 0.0049095i}&{\bf \mp 0.057585i}&\\
&&&&&&&&\\
\hline\hline
&&&&&&&&\\
c_{p+}
&0.25&-0.75&-1.75&-0.125&-1.940550789
&-0.1547246055&0.25&-1.25\\
&&&&\pm 0.6495190528i&
&\pm 1.031047228i&\pm 0.5i&\\
&&&& (=\pm 3\sqrt{3}i/8)&&&&\\
\hline
&&&&&&&&\\
c_{p-}
&-0.75&-1.25&-1.768529153&-0.1157354238
&-1.941537753&-0.1613575037&0.3161758500&-1.368098940\\
&&&&\pm 0.8379990280i&&\pm 1.036031085i&\pm 0.5574717760i&\\
\hline
&&&&&&&&\\
c_{p+}-c_{p-}
&{\bf 1}&{\bf 0.5}&{\bf 0.018529153}&{\bf -0.0092645762}
&{\bf 0.000986964}&{\bf 0.0066328982}
&{\bf -0.0661758500}&{\bf 0.118098940}\\
&&&&{\bf \mp 0.1884799750i}&&{\bf \mp 0.004983857i}
&{\bf \mp 0.057471776i}&\\
&&&&&&&&\\
\hline
&&&&&&&&\\
\kappa &{\bf 4}&{\bf 2}&{\bf 4}&{\bf 2}
&{\bf 4}&{\bf 4}&{\bf 2}&{\bf 2}\\
(c_{p+}-c_{p})\kappa
&{\bf 1}&{\bf 0.5}&{\bf 0.019510668}&{\bf -0.0048776662}
&{\bf 0.000996072}&{\bf 0.0071822452}&{\bf -0.0645427816}
&{\bf 0.121405282}\\
&&&&{\bf \mp 0.1906854280i} &&{\bf \mp 0.004799524i}
&{\bf \mp 0.0601212352i}&\\
&&&&&&&&\\
\hline
\end{array}
\hspace{+0.6cm}
$$
}

\noindent
The entries in the last two rows 
should be both compared with $2r_p$, which is
calculated within SSA in the middle part of the table.
Since the shapes of root and descendant domains
are different, parameters $\kappa$ in the last row
are also different: $\kappa = 4$ for cardioid-shape
root domains and $\kappa=2$ for circle-shape descendant
domains.

{\footnotesize
$$
\begin{array}{|c|c|c|c|c|c|c|}
\hline
&&&&&&\\
p
&5&5&5&5&5&5\\
&&&&&&\\
\hline
&&&&&&\\
c_p
&-1.625413725&-1.860782522&-1.985424253
&0.3592592248&-0.04421235770&-0.1980420994\\
&&&&\pm 0.6425137371i&\pm 0.9865809763i&\pm 1.100269537i\\
&&&&&&\\
\hline
&&&&&&\\
{\rm distance}
&{\bf 0}&{\bf 0}&{\bf 0}&{\bf 0}&{\bf 0}&{\bf 0}\\
{\rm from\ the\ root}
&(5_1)&(5_2)&(5_3)&(5_{4\pm})&(5_{5\pm})&(5_{6\pm})\\
&&&&[\ldots,8,4_{\pm},1]&[\ldots,12,6,3_\pm,1]
&[\ldots,12,6,3_\pm,1]\\
&&&&&&\\
\hline\hline
&&&&&&\\
\dot f_p
&-12.346786&27.952811&-106.51134
&-15.264582&-1.4606030&-34.451623\\
&&&&\mp 4.049590i&\mp 15.65788i&\pm 8.245915i\\
&&&&&&\\
\hline
&&&&&&\\
\ddot f_p
&32.445339&-211.46971&3811.7679
&-270.4039&-190.8108&421.6135\\
&&&&\pm 144.8293i&\mp 151.4911i&\pm 736.0759i\\
&&&&&&\\
\hline
&&&&&&\\
\gamma_p
&-19.95443&45.84473&-161.34688
&-9.4954135&6.050675&-41.054355\\
&&&&\mp 9.112127i&\mp 17.26642i&\mp 3.02090i\\
&&&&&&\\
\hline
&&&&&&\\
\dot \gamma_p
&17.8489&-272.6037&5625.2834
&-237.5248&-108.0265&207.2558\\
&&&&\mp 0.045178i&\mp 230.0270i&\pm 911.8688i\\
&&&&&&\\
\hline\hline
&&&&&&\\
\xi_p
&{0.008963643}&{0.007591840}&{0.00306997}
&{0.02825982}&{0.03863736}&{0.00311728}\\
{\rm from}\ (\ref{xip})
&&&&{\pm 0.1305419i}&{\mp 0.06450516i}
&{\pm 0.02358242i}\\
&{\bf \ll 1}&{\bf \ll 1}&{\bf \ll 1}
&{\bf |\xi_p|\ll 1}&{\bf |\xi_p|\ll 1}&{\bf |\xi_p|\ll 1}\\
&&&&&&\\
\hline
&&&&&&\\
{\bf 2r_p} = (\dot f_p \gamma_p)^{-1}
&{\bf 0.004058884}&{\bf 0.0007803423}&{\bf 0.0000581894}
&{\bf 0.00250125}&{\bf -0.0033726}&{\bf 0.00067682}\\
{\rm from}\ (\ref{diam})
&&&&{\bf\mp 0.00411026i}&{\bf\pm 0.00083981i}
&{\bf\pm 0.00011025i}\\
&&&&&&\\
\hline\hline
&&&&&&\\
c_{p+}
&-1.624396989&-1.860586973&-1.985409691&0.3599331332
&-0.04506136598&-0.1978729467\\
&&&&\pm 0.6415066668i&\pm 0.9868115622i&\pm 1.100298438i\\
&&&&&&\\
\hline
&&&&&&\\
\kappa &{\bf 4}&{\bf 4}&{\bf 4}&{\bf 4}
&{\bf 4}&{\bf 4}\\
(c_{p+}-c_{p})\kappa
&{\bf 0.004066944}&{\bf 0.000782196}&{\bf 0.000058248}
&{\bf 0.0026956336}&{\bf -0.00339603312}&{\bf 0.0006766108}\\
&&&&{\bf\mp 0.0040282812i}&{\bf\pm 0.0009223436i}
&{\bf\pm 0.000115604i}\\
&&&&&&\\
\hline
\end{array}
$$
}
{\footnotesize
$$
\hspace{-1.0cm}
\begin{array}{|c|c|c|c|c|c|c|c|}
\hline
&&&&&&&\\
p
&5&5&5&6&6&6&6\\
&&&&&&&\\
\hline
&&&&&&&\\
c_p
&-1.256367930&0.3795135880&-0.5043401754
&-1.476014643&-1.907280091&-1.966773216&-1.996376138\\
&\pm 0.3803209635i&\pm 0.3349323056i&\pm 0.5627657615i
&&&&\\
&&&&&&&\\
\hline
&&&&&&&\\
{\rm distance}
&{\bf 0}&{\bf 1}&{\bf 1}
&{\bf 0}&{\bf 0}&{\bf 0}&{\bf 0}\\
{\rm from\ the\ root}
&(5_{7\pm})&(5_{1\pm},1)&(5_{2\pm},1)
&(6_1)&(6_2)&(6_3)&(6_4)\\
&[\ldots,12,6_\pm,2,1]&&&&&&\\
\hline\hline
&&&&&&&\\
\dot f_p
&-9.720195&-2.700383&-3.449207
&9.557119&-73.91417&135.0997&-431.94389\\
&\mp 11.81330i&\mp 5.404227i&\mp 1.112266i
&&&&\\
&&&&&&&\\
\hline
&&&&&&&\\
\ddot f_p
&-59.44225&-48.15074&9.823111&-106.133658
&491.1284&-2562.839&60341.247\\
&\pm 171.0187i&\mp 78.38748i&\pm 30.87527i
&&&&\\
&&&&&&&\\
\hline
&&&&&&&\\
\gamma_p
&-5.71067&0.809902&-2.871463&20.23824
&-115.0273&207.856&-649.8852\\
&\mp 21.95456i&\mp 3.399590i&\mp 2.098768i
&&&&\\
&&&&&&&\\
\hline
&&&&&&&\\
\dot \gamma_p
&-165.8458&2.830436&0.04567&-150.6058
&576.3060&-3609.262&90085.134\\
&\pm 183.4574i&\mp 47.49326i&\pm 29.22689i
&&&&\\
&&&&&&&\\
\hline\hline
&&&&&&&\\
\xi_p
&{ 0.0176798}&{ 1.00083}&{ 0.984283}
&{ 0.067182}&0.00098004&0.00095613&0.0007426\\
{\rm from}\ (\ref{xip})
&{\mp 0.0451181i}&{\pm 0.0866142i}
&{\mp 0.000539978i}&&&&\\
&{\bf |\xi_p|\ll 1}&{\bf \approx 1}&{\bf \approx 1}&
{\bf\ll 1}&{\bf\ll 1}&{\bf\ll 1}&{\bf\ll 1}\\
&&&&&&&\\
\hline
&&&&&&&\\
{\bf 2r_p} = (\dot f_p \gamma_p)^{-1}
&{\bf -0.001692541}&{\bf -0.04612246}&{\bf 0.04556086}
&{\bf 0.005170116}&{\bf 0.000117617}&{\bf 0.000035611}
&{\bf 0.0000035623}\\
{\rm from}\ (\ref{diam})
&{\bf \mp 0.00233202i}&{\bf \mp 0.0107757i}&
{\bf \mp 0.0627926i}&&&&\\
&&&&&&&\\
\hline\hline
&&&&&&&\\
c_{p+}
&-1.256801994&0.3567627458&-0.4817627458&-1.47469537780
&-1.90725067795&-1.96676431090&-1.99637524690\\
&\pm 0.3797412022i&\pm 0.3285819450i&\pm 0.5316567552i
&&&&\\
&&&&&&&\\
\hline
&&&&&&&\\
\kappa &{\bf 4}&{\bf 2}&{\bf 2}&{\bf 4}
&{\bf 4}&{\bf 4}&{\bf 4}\\
(c_{p+}-c_{p})\kappa
&{\bf -0.0017363}&{\bf -0.04550168}&{\bf 0.04515486}
&{\bf  0.00527706}&{\bf 0.000117652}
&{\bf 0.000035620}&{\bf 0.000003564}\\
&{\bf \mp 0.00231905i}&{\bf \mp 0.01270072i}
&{\bf \mp 0.06221801i}&&&&\\
&&&&&&&\\
\hline
\end{array}
\hspace{+1.0cm}
$$
}

{\footnotesize
$$
\begin{array}{|c|c|c|c|c|c|c|}
\hline
&&&&&&\\
p
&6&6&6&6&6&6\\
&&&&&&\\
\hline
&&&&&&\\
c_p
&0.4433256334&0.3965345700&0.3598927390
&-0.01557038602&-0.1635982616&-0.2175267470\\
&\pm 0.3729624167i&\pm 0.6041818105i&\pm 0.6847620202i
&\pm 1.020497366i&\pm 1.097780643i&\pm 1.114454266i\\
&&&&&&\\
\hline
&&&&&&\\
{\rm distance}
&{\bf 0}&{\bf 0}&{\bf 0}&{\bf 0}&{\bf 0}&{\bf 0}\\
{\rm from\ the\ root}
&(6_{5\pm})&(6_{6\pm})&(6_{7\pm})&(6_{8\pm})
&(6_{9\pm})&(6_{10\pm})\\
&[\ldots,10,5_{1\pm},1]&[\ldots,8,4_\pm,1]&[\ldots,8,4_\pm,1]
&[\ldots,6,3_\pm,1]&[\ldots,6,3_\pm,1]&[\ldots,6,3_\pm,1]\\
&&&&&&\\
\hline\hline
&&&&&&\\
\dot f_p
&-22.131316&-7.730347&-45.44965&-47.08247&-13.70794&-94.05752\\
&\mp 8.549589i&\mp 29.76415i&\pm 11.11460i&\mp 42.3870i
&\mp 59.0423i&\pm 65.91697i\\
&&&&&&\\
\hline
&&&&&&\\
\ddot f_p
&-868.64319&-427.073&-431.8717&-3003.383&-2600.199&7694.999\\
&\mp 90.27354i&\mp 1083.050i&\pm 2162.074i&\pm 437.063i
&\mp 913.171i&\pm 2405.039i\\
&&&&&&\\
\hline
&&&&&&\\
\gamma_p
&-8.156137&6.87116&-37.7109&-28.2862&7.68865&-125.4789\\
&\mp 11.41327i&\mp 23.4912i&\mp 9.24215i&\mp 64.5198i
&\mp 69.5663i&\pm 40.9618i\\
&&&&&&\\
\hline
&&&&&&\\
\dot \gamma_p
&-389.998&151.331&-1097.368&-3129.19&-2370.323&7560.924\\
&\mp 307.274i&\mp 866.959i&\pm 1391.979i&\mp 931.571i
&\mp 1950.970i&\pm 5049.389i\\
&&&&&&\\
\hline\hline
&&&&&&\\
\xi_p
&0.00405&0.07215&-0.01588&0.014630&0.010020&-0.001788\\
{\rm from}\ (\ref{xip})
&\pm 0.16159i&\mp 0.01058i&\pm 0.034629i&\pm 0.005837i
&\mp 0.01208i&\pm 0.006621i\\
&{\bf |\xi_p| \ll 1}&{\bf |\xi_p| \ll 1}&{\bf |\xi_p| \ll 1}
&{\bf |\xi_p| \ll 1}&{\bf |\xi_p| \ll 1}&{\bf |\xi_p| \ll 1}\\
&&&&&&\\
\hline
&&&&&&\\
{\bf 2r_p} = (\dot f_p \gamma_p)^{-1}
&{\bf 0.0007487}&{\bf -0.0013280}&{\bf 0.00055046}
&{\bf -0.00007044}&{\bf -0.00023408}&{\bf 0.000039602}\\
{\rm from}\ (\ref{diam})
&{\bf \mp 0.0029099i}&{\bf \pm 0.00004046i}
&{\bf \mp 0.000000276i}&{\bf \mp 0.0002127i}
&{\bf\mp 0.00002776i}&{\bf\pm 0.00005275i}\\
&&&&&&\\
\hline\hline
&&&&&&\\
c_{p+}
&0.44355069141&0.39619446067&0.3600296164
&-0.01558797607&-0.163657003678&-0.217516881\\
&\pm 0.372246821i&\pm 0.60419336859i&\pm 0.684763498i
&\pm 1.0204438975i&\pm 1.0977739136i&\pm 1.114467467i\\
&&&&&&\\
\hline
&&&&&&\\
\kappa &{\bf 4}&{\bf 2}&{\bf 2}&{\bf 4}
&{\bf 4}&{\bf 4}\\
(c_{p+}-c_{p})\kappa
&{\bf 0.000900232}&{\bf -0.001360437}&{\bf 0.00054751}
&{\bf -0.000070360}&{\bf -0.000234968}&{\bf 0.000039463}\\
&{\bf\mp 0.00286238i}&{\bf\pm 0.00004623i}
&{\bf \pm 0.00000591i}&{\bf\mp 0.00021387i}
&{\bf\mp 0.00002692i}&{\bf\pm 0.00005280i}\\
&&&&&&\\
\hline
\end{array}
$$
}

{\footnotesize
$$
\begin{array}{|c|c|c|c|c|c|c|}
\hline
&&&&&&\\
p
&6&6&6&6&6&6\\
&&&&&&\\
\hline
&&&&&&\\
c_p
&-0.5968916446&-1.284084926&0.3890068406
&-1.772892903&-0.1134186559&-1.138000667\\
&\pm 0.6629807446i&\pm 0.4272688960i&\pm 0.2158506509i
&&\pm 0.8605694725i
&\pm 0.2403324013i \\
&&&&&&\\
\hline
&&&&&&\\
{\rm distance}
&{\bf 0}&{\bf 0}&{\bf 1}&{\bf 1}&{\bf 2}&{\bf 2}\\
{\rm from\ the\ root}
&(6_{11\pm})&(6_{12\pm})&(6_\pm,1)&(6,3)
&(6,3_{\pm},1)&(6_\pm,2,1)\\
&[\ldots,10,5_{2\pm},1]&[\ldots, 12,6_\pm,2,1]&&&&\\
&&&&&&\\
\hline\hline
&&&&&&\\
\dot f_p
&-21.17430&-49.4808&-2.84507&6.02446709
&2.710416&3.0373845\\
&\mp 1.239334i&\mp 43.51796i&
\mp 8.81348i&&\pm 2.01496i&\pm 0.674370i\\
&&&&&&\\
\hline
&&&&&&\\
\ddot f_p
&258.9018&-401.263&-13.2205&-699.2133940
&53.81286&-9.154561\\
&\pm 390.3375i&\pm 2355.001i&\mp 228.2727i
&&\mp 56.37906i&\mp 50.431515i\\
&&&&&&\\
\hline
&&&&&&\\
\gamma_p
&-19.86933&-40.9497&1.51341&19.09634167&3.35657&3.882374 \\
&\mp 7.65439i&\mp 85.4133i&
\mp 3.51012i&&\pm 5.72766i&\pm 4.659455i\\
&&&&&&\\
\hline
&&&&&&\\
\dot \gamma_p
&122.1821&-1796.0107&47.3746&-1106.4705637
&71.1246&28.408033\\
&\pm 435.7183i&\pm 2809.8473i&
\mp 74.1184i&&\mp 30.8637i&\mp 64.35131i\\
&&&&&&\\
\hline\hline
&&&&&&\\
\xi_p
&0.062667&0.005776&0.99288&1.00806&1.03542&1.0497\\
{\rm from}\ (\ref{xip})
&\pm 0.034437i&\mp 0.006297i&
\pm 0.09882i&&\pm 0.013004i&\pm 0.0438i\\
&{\bf |\xi_p| \ll 1}&{\bf |\xi_p| \ll 1}
&{\bf \approx 1}&{\bf \approx 1}&{\bf \approx 1}&{\bf \approx 1}\\
&&&&&&\\
\hline
&&&&&&\\
{\bf 2r_p} = (\dot f_p \gamma_p)^{-1}
&{\bf 0.0020161}&{\bf -0.000043399}
&{\bf -0.0281207}
&{\bf 0.008692230}&{\bf -0.0048602}&{\bf 0.0242924}\\
{\rm from}\ (\ref{diam})
&{\bf \mp 0.0009153i}&{\bf \mp 0.00015422i}&
{\bf \pm 0.0026745i}&&{\bf\mp 0.044335i}&{\bf \mp 0.047098i}\\
&&&&&&\\
\hline\hline
&&&&&&\\
c_{p+}
&-0.5963742242&-1.284095877
&0.375&-1.768529153&-0.1157354238&-1.125\\
&\pm 0.6627532528i&\pm 0.4272302898i&\pm 0.2165063509i
&&\pm 0.837999027i&\pm 0.2165063509i\\
&&&&&&\\
\hline
&&&&&&\\
\kappa &{\bf 4}&{\bf 4}&{\bf 2}&{\bf 2}&{\bf 2}&{\bf 2}\\
(c_{p+}-c_{p})\kappa
&{\bf 0.00206968}&{\bf -0.00004380}&{\bf -0.028013681}&{\bf 0.008727500}
&{\bf -0.0046335357}&{\bf 0.0260013340}\\
&{\bf\mp 0.000909967i}&{\bf\mp 0.00015442i}
&{\bf\pm 0.001311400i}&&{\bf\mp 0.045140890254i}
&{\bf\mp 0.0476521007i}\\
&&&&&&\\
\hline
\end{array}
$$
}

\bigskip

A few comments to the table are now in order:

We present the values of parameters with high accuracy,
which strongly exceeds our needs in the present paper,
but this data can be used in the future investigations.
One should not be surprised by the high accuracy of experimental
data: since it is computer experiment over {\it Platonian} entity,
accuracy is unlimited. Moreover the numbers in the last column
can be also reproduced as resultants zeroes \cite{DM}:
though it is a difficult calculation (beyond capabilities of MAPLE
on an ordinary laptop already for $p > 6$), its accuracy is in
principle unlimited.

$c_+$ is closer to the root than $c_-$:
$c_+$ is the point of merging with the parent domain
or the cusp position if the domain is itself a root,
while $c_-$ is the "opposite" point, i.e. the
merging point with the next descendant of the order $2$.

Starting from $p=5$ there are many root domains of the type
$(5)$, $(6)$ etc, and -- starting from $p=3$ -- many descendants
$(3,1)$, $(4,1)$,\ldots, $(6,2,1)$ etc:
$\alpha$-parameters begin to emerge. The two domains $(3_\pm,1)$
differ by complex conjugation only, but in other cases
systematization in $\alpha$-sector is less straightforward,
their sizes and orientations depend essentially on $\alpha$.
Still, because of the symmetry of the Mandelbrot Set
under complex conjugation, the domains with centers at
non-real $c$ come in pairs.
Such complex conjugate domains are
always labeled by indices $\pm$.

Positions of the domains in divisor forest are shown
in the second line of the third row.
For root domains the original direction of the {\it trail},
connecting it to the central cluster,
is also shown in square brackets in the third row.
Of course, all root domains with centers at real values
of $c$ belong to the trail, originating at
$[\ldots, 2^n, \ldots, 16,8,4,2,1]$, and it is not
mentioned in the table.

\bigskip

From this table we observe:

-- the good accuracy of the relation (see the last three columns)
$$
2r_p\ \approx\ c_{p+}-c_{p-}
$$
between the theoretically-predicted
(in the small-size approximation) complex-valued size
$r_p$ of an elementary domain and the difference between
experimentally found extreme points $c_{p+}$ and $c_{p-}$;

-- the correlation between the value of $\xi_p$ and
the distance of elementary domain from the root of the
corresponding cluster (the corresponding columns are boldfaced):
$\xi_p$ is tiny for the roots (distance $= 0$)
and close to unity for all descendants (distance $\geq 1$).

\section{Why $\xi_p\approx 1$ for descendants:
{\it la raison d'etre} for circles \label{shapes2}}

\setcounter{equation}{0}

\subsection{Approach to description of descendants}

We can study descendants of a given elementary domain within the
same small-size approximation (SSA),
simply iterating approximate expression
$$
f^{\circ p}(x) \approx f_p(c) + \gamma_p(c) x^2
$$
to
\be
f^{\circ (2p)}(x) \approx f_{2p}(c) + \gamma_{2p}(c)x^2 \approx
f_p(c) + \gamma_p(c)\Big(f_p(c) + \gamma_p(c)x^2\Big)^2 \approx
f_p(c)\Big(1+f_p(c)\gamma_p(c)\Big) + 2f_p(c)\gamma_p^2(c)x^2
\label{f2pc}
\ee
and so on.
Thus in this framework
\be
f_{2p}(c) \approx f_p(c)\Big(1+f_p(c)\gamma_p(c)\Big),\nn \\
\gamma_{2p}(c) \approx 2f_p(c)\gamma^2_p(c)
\label{2it}
\ee
$c_{2p}$ is a non-trivial root of this new $f_{2p}(c)$,
\be
f_p(c_{2p}) \gamma_p(c_{2p}) = -1
\label{gampc2p}
\ee
This procedure -- if at all justifiable -- can be valid only for
$c_{2p}$, associated with a {\it descendant} domain of $c_p$
(but not a {\it root} domain of some new cluster),
since it relies on SSA
and assumes that $c_{2p}$ is very close to $c_p$.
The shift $\sigma_{2p}\equiv c_{2p}-c_p$ can actually be found
in SSA by solving (\ref{gampc2p}) iteratively:
\be
\frac{\sigma_{2p}}{2r_p}
\left(1 - \xi_p\frac{\sigma_{2p}}{2r_p}\right) = -1
\label{c2p}
\ee

Now we are going to demonstrate that $\xi_{2p}$, evaluated for
{\it such} $c_{2p}$ within SSA, is indeed equal to unity
(this is no more than a consistency check, because validity of
the SSA itself will not be theoretically justified).
Afterwards this calculation is extended to
descendant $c_{mp}$ for all $m$.
Further, eq.(\ref{c2p}) and its generalizations for $\sigma_{mp}$
are used in s.\ref{Feig} to evaluate SSA approximations of various
Feigenbaum indices.
Finally, in s.\ref{dfam}, we briefly consider the case of
specific $Z_{d-1}$-symmetric $f(x;c) = x^d+c$ families.

\subsection{Evaluation of $\xi_{2p}$ for
a descendant \label{calcul}}

This is a rather straightforward calculation.
From (\ref{f2pc}) we obtain -- in the small-size approximation,
after substitution of $c=c_{2p}$ and (\ref{gampc2p}), and
after expanding functions of $c_{2p}$ in powers of
$\sigma_{2p}=c_{2p}-c_p$ from (\ref{c2p}) --
the set of recurrent expressions:
\be
\dot f_{2p}(c) = \dot f_p(c)\Big(1 + 2f_p(c)\gamma_p(c)\Big) +
f^2_p(c)\dot\gamma_p(c)
\ \ \ \stackrel{\ c=c_{2p}}{\Longrightarrow}\ \ \
\dot f_{2p} \equiv \dot f_{2p}(c_{2p}) = \nn \\
= -\dot f_p(c_{2p}) -
f_p(c_{2p})\frac{\dot\gamma_p}{\gamma_p}(c_{2p}) \approx
-\dot f_p\left(1+\sigma_{2p}\left[\frac{\ddot f_p}{\dot f_p} +
\frac{\dot\gamma_p}{\gamma_p}\right]\right)
\label{appr1}
\ee
\be
\gamma_{2p}(c) = 2f_p(c)\gamma^2_p(c)
\ \ \ \stackrel{(\ref{gampc2p})}{\Longrightarrow}\ \ \
\gamma_{2p} \equiv \gamma_{2p}(c_{2p})
= -2\gamma_p(c_{2p}) = -2\gamma_p\left(1+\sigma_{2p}
\frac{\dot\gamma_p}{\gamma_p}\right)
\label{appr2}
\ee
\be
\ddot f_{2p}(c) = \ddot f_p(c)\Big(1 + 2f_p(c)\gamma_p(c)\Big) +
2\dot f_p^2(c)\gamma_p(c) + 4f_p(c)\dot f_p(c)\dot\gamma_p(c)
+ f^2_p(c)\ddot\gamma_p(c) \ \ \ \Longrightarrow\ \ \
\ddot f_{2p} = 2\dot f^2_p\gamma_p
\ee
\be
\dot\gamma_{2p}(c) = 2\dot f_p(c)\gamma^2_p(c) +
4f_p(c)\gamma_p(c)\dot\gamma_p(c) \ \ \ \Longrightarrow\ \ \
\dot\gamma_{2p} = 2\dot f_p(c)\gamma^2_p(c)
\label{appr4}
\ee
Substituting these expressions into (\ref{diam}) and (\ref{xip}),
we obtain:
\be
r_{2p} = \frac{1}{2\dot f_{2p}\gamma_{2p}} \approx
\frac{1}{4\dot f_{p}\gamma_{p}
\left\{1+\sigma_{2p}\left(\frac{\ddot f_p}{\dot f_p} +
2\frac{\dot\gamma_p}{\gamma_p}\right)\right\}}
= \frac{r_p}{2\left(1 - \xi_p\frac{\sigma_{2p}}{r_p}\right)}
\label{r2p}
\ee
and
\be
\xi_{2p} = -\frac{1}{\dot f_{2p}\gamma_{2p}}
\left(\frac{\ddot f_{2p}}{2\dot f_{2p}} +
\frac{\dot\gamma_{2p}}{\gamma_{2p}}\right) \approx
-\frac{1}{\dot f_{2p}\gamma_{2p}}\left(
\frac{2\dot f^2_p\gamma_p}{-2\dot f_p} +
\frac{2\dot f_p(c)\gamma^2_p(c)}{-2\gamma_p}\right) =
2\frac{\dot f_{p}\gamma_{p}}{\dot f_{2p}\gamma_{2p}}
\approx 1
\label{xi2p}
\ee
as required.

\subsection{The rules of SSA}

Note, that within SSA we consider $r_p\ddot f_p/\dot f_p$ and
$r_p\dot\gamma_p/\gamma_p$ as {\it small parameters} and
ignore their quadratic powers as well as higher derivatives.
This is needed for self-consistency of the SSA,
even despite individual corrections need not be small
(especially for domains which are {\it not} the {\it first}
descendants, i.e. when $\xi_p\approx 1$ is {\it not small})
-- however, if included, they should come {\it together}
with {\it other} corrections to the SSA, which were also ignored.
Actually, as we saw in s.\ref{accu} the summary effect of
{\it all} corrections is small, but the theoretical reason
for this conspiracy in the case of higher descendants remains
to be identified.

\bigskip

It deserves formulating the rules of SSA explicitly:

\bigskip

$\bullet$
Expand in powers of $x$ and leave the first two non-trivial terms
(constant and $x^2$ in the case of $x^2+c$ family) -- for generic
value of $c$.

$\bullet$
Expand in powers of $\sigma = c-c_{crit}$ and leave only the first
corrections $\sim \ddot f /\dot f$ and $\sim \dot\gamma/\gamma$.

$\bullet$
If two different but two close $c_{crit}$ appear in the problem
(say, centers of two adjacent elementary domains),
expand in powers of their difference, leaving only the first two
powers of the difference.

$\bullet$
Combining all these expansions, keep only the first two corrections
in expressions for the final quantities, in practice this means
keeping all powers of $\dot f$ and $\gamma$ and ignore everything
beyond the first powers of $\ddot f /\dot f$ and $\gamma/\gamma$.

\subsection{Position and radius of arbitrary descendant domain}

Generic descendant domain has a parent of order $p$
and has itself a multiple order $mp$. It is attached to the parent
at a zero of the resultant $R_x\Big(F_{mp}(x)/F_p(x),F_p(x)\Big)$.
Parent can be itself a descendant and a chain of ancestors lead
to a root domain of the cluster, however, only the first term in
this chain -- the mother domain, of which the domain of interest
is an {\it immediate} descendant, -- is relevant in the SSA-based
calculations.

It is easy to check that generalization of the SSA relation
(\ref{2it}) to arbitrary $m$ is
\be
f_{mp}(c) \approx
\frac{1}{\gamma_p(c)}f_m\Big(f_p\gamma_p(c)\Big)
\ee
\vspace{-0.4cm}
\be
\gamma_{mp}(c) \approx \gamma_p(c)\
\gamma_m\Big(f_p\gamma_p(c)\Big)
\label{mit}
\ee
Now descendant root $c_{mp,p}$ of $f_{mp}$
is defined by the choice of immediate descendant
$c_{m,1}$ for $f_m$: from
\be
f_m(c_{m,1}) = 0
\label{parcon}
\ee
we have
\be
f_p\gamma_p(c_{mp,p}) = c_{m,1}
\label{cmpp}
\ee
For comparison with s.\ref{calcul} one should keep in mind that
for $f=x^2+c$ there is a single order-two critical point
$c_{2,1}=c_2=-1$.

Note, that not all the zeroes $c_m$ of (\ref{parcon}) describe
{\it immediate} descendants $(m,1)$ of the central domain $(1)$:
some provide the new root domains $(m)$ or higher descendants
$(m,m_1,\ldots)$ with nontrivial divisors $m_1\neq 1$ of $m$.
These extra zeroes (especially associated with domains from
the different clusters) should not be used in the following
calculations, because they correspond to remote domains and
SSA has no reason to work for them.

Repeating for generic $m$ the calculations,
performed s.\ref{calcul} for particular case of $m=2$, we obtain:
\be
\dot f_{mp}(c) \approx
\frac{1}{\gamma_p(c)}\dot f_m\Big(f_p\gamma_p(c)\Big)
\Big(\dot f_p\gamma_p(c) + f_p\dot\gamma_p(c)\Big) -
\frac{\dot\gamma_p(c)}{\gamma^2_p(c)}f_m\Big(f_p\gamma_p(c)\Big)
\ee
In combination with (\ref{mit}) this implies that
$$
\frac{1}{2r_{mp,p}} = \dot f_{mp}\gamma_{mp}(c_{mp,p}) \approx
\dot f_m\gamma_m(c_{m,1})
\Big(\dot f_p\gamma_p + f_p\dot\gamma_p\Big)(c_{mp,p}) \approx
\dot f_m\gamma_m(c_{m,1})\Big(\dot f_p\gamma_p + \sigma_{mp,p}
\left(\ddot f_p\gamma_p + 2\dot f_p\dot\gamma_p\right)\Big)(c_p)
$$
In the first transformation we omitted one term with
$f_m(c_{m,1})=0$, and in the second transformation we defined
functions at $c_{mp,p}$ through their values at $c_p$,
keeping only the first non-trivial term of Taylor expansion
in powers of $\sigma_{mp,p} \equiv c_{mp,p}-c_p$.
This shift is defined in a similar manner  from (\ref{cmpp}):
$$
\sigma_{mp,p} \dot f_p\gamma_p(c_p) + \frac{1}{2}\sigma_{mp,p}^2
\left(\ddot f_p\gamma_p + \dot f_p\dot\gamma_p\right)
(c_p) \approx c_{m,1}
$$
(we remind that $f_p(c_p)=0$).
Substituting for remaining parameters
$\dot f_m\gamma_m = (2r_m)^{-1}$, $\dot f_p\gamma_p = (2r_p)^{-1}$
and $\ddot f_p\gamma_p + 2\dot f_p\dot\gamma_p =
-\xi_p (2r_p^2)^{-1}$,
we obtain for the counterparts of (\ref{c2p}) and (\ref{r2p}):
\be
\frac{\sigma_{mp,p}}{2r_p}\left(1 - \xi_p\frac{\sigma_{mp,p}}{2r_p}
\right) \approx c_m
\label{cmp}
\ee
and
\be
r_{mp} \approx 2r_mr_p\left(1 - \xi_p\frac{\sigma_{mp,p}}{r_p}\right)^{-1}
\label{rmp}
\ee

\subsection{Evaluation of generic $\xi_{mp,p}$}

For evaluation of $\xi_{mp,p}$ we need also
\be
\ddot f_{mp}(c) \approx
\frac{1}{\gamma_p(c)}\ddot f_m\Big(f_p\gamma_p(c)\Big)
\left(\dot(f_p\gamma_p)\right)^2 +
\frac{1}{\gamma_p(c)}f_m\Big(f_p\gamma_p(c)\Big)
\ddot(f_p\gamma_p) - \nn \\ -
2\frac{\dot\gamma_p(c)}{\gamma_p^2(c)}
\dot f_m\Big(f_p\gamma_p(c)\Big)
\dot(f_p\gamma_p) +
2\frac{\dot\gamma_p^2(c)}{\gamma_p^3(c)}
\dot f_m\Big(f_p\gamma_p(c)\Big) + O(\ddot\gamma)
\ee
and
\be
\dot\gamma_{mp}(c) \approx
\gamma_p(c)\dot\gamma_m\Big(f_p\gamma_p(c)\Big)
\Big(\dot f_p\gamma_p(c) + f_p\dot\gamma_p(c)\Big) +
\dot\gamma_p(c)\gamma_m\Big(f_p\gamma_p(c)\Big)
\ee
At $c=c_{mp,p}$ the terms with $f_m(c_{m,1})=0$ do not
contribute, and we obtain
$$
\left(\frac{1}{2}\ddot f_{mp}\gamma_{mp} +
\dot f_{mp}\dot\gamma_{mp}\right)(c_{mp,p}) \approx
\left(\frac{1}{2}\ddot f_m\gamma_m + \dot f_m\dot\gamma_m
\right)(c_{m,1})
\Big(\dot f_p\gamma_p + f_p\dot\gamma_p\Big)^2(c_{mp,p})
\ + $$ \vspace{-0.25cm}
\be
+\ \dot f_m\gamma_m(c_{m,1})
 \left(\frac{1}{2}\ddot f_p\gamma_p + \dot f_p\dot\gamma_p
\right)(c_{mp,p})
\ee
In order to get $\xi_{mp,p}$ we divide by the square of
\be
\dot f_{mp}\gamma_{mp} \approx \dot f_m\gamma_m(c_{m,1})
\Big(\dot f_p\gamma_p + f_p\dot\gamma_p\Big)(c_{mp,p})
\ee
and change sign, so that (\ref{xi2p}) generalizes to
and
\be
\xi_{mp} \approx -\frac{1}{\dot f_{mp}\gamma_{mp}}
\left(\frac{\ddot f_{mp}}{2\dot f_{mp}} +
\frac{\dot\gamma_{mp}}{\dot\gamma_{mp}}\right)(c_{mp})
\ \approx\ -\frac{1}{\dot f_{m}\gamma_{m}}
\left(\frac{\ddot f_{m}}{2\dot f_{m}} +
\frac{\dot\gamma_{m}}{\dot\gamma_{m}}\right)(c_{m}) -
\nn \\ -
\frac{1}{\dot f_m\gamma_m}
\left(\frac{\ddot f_{p}}{2\dot f_{p}} +
\frac{\dot\gamma_{p}}{\dot\gamma_{p}}\right)
\frac{1}{(\dot f_p\gamma_p +f_p\dot\gamma_p)^2(c_{mp})}
\approx \xi_m +
\frac{2r_m\xi_p}
{\left(1-\xi_p\frac{\sigma_{mp,p}}{r_p}\right)^2}
\ \stackrel{m\neq 1}{\approx}\ \xi_m
\label{ximp}
\ee
Keeping the second term at the r.h.s. is beyond the
accuracy of the SSA and it should be neglected
(we ignored it in (\ref{xi2p}), but kept in (\ref{ximp})
to preserve  formal consistency with the case
$m=1$, when $r_1=\frac{1}{2}$, $\xi_1=0$ and, of course,
$\sigma_{p,p}\equiv 0$).

From (\ref{ximp}) it is clear that if $\xi_{m,1}\approx 1$ for
direct descendant $(m,1)$ of order $m$ of the central root
domain ($c_1=0$), then in the SSA
$\xi_{mp,p}\approx 1$ for all other descendants,
at all levels in all clusters.
In s.\ref{calcul} we exploited the fact that for $m=2$
the r.h.s. is extremely simple: for $f_2(c) = c(c+1)$,
$\gamma_2(c) = 2c$ and $c_2=-1$ it is obviously unity.
In s.\ref{accu} we saw that $\xi_{m,1}$ is indeed close
to unity for $m\leq 6$, and it is natural to believe that
this remains true for all $m$, however no theoretical
explanation of this fact is yet available.
Still, if accepted, it implies that $\xi_{mp,p}\approx 1$
for all $p>1$.

\section{Feigenbaum indices \label{Feig}}

\setcounter{equation}{0}

\subsection{The case of period-doubling, $m=2$}

It is now time to solve quadratic equation (\ref{r2p}):
the distance between the centers of a parent domain
$(p,\ldots)$ and its immediate descendant $(2p,p,\ldots)$ is
\be
\sigma_{2p} \approx \left\{\begin{array}{ccc}
-2r_p & {\rm if}\  \xi_p \approx 0 & {\rm i.e.\ for\
the\ } first\ {\rm descendant} \\ \\
(1-\sqrt{5})r_p & {\rm if}\  \xi_p \approx 1 & {\rm i.e.\ for\
a\ } higher\ {\rm descendant}\end{array}\right.
\ee
Substituting this into (\ref{r2p}), we obtain:
the radius of descendant domain $(2p,p,\ldots)$ is
\be
r_{2p} = \left\{\begin{array}{ccc} \frac{1}{2}r_p &
{\rm if}\  \xi_p \approx 0 & {\rm i.e.\ for\
the\ } first\ {\rm descendant} \\ \\
\frac{r_p}{2\sqrt{5}} & {\rm if}\  \xi_p \approx 1 &
{\rm i.e.\ for\ a\ } higher\ {\rm descendant}\end{array}\right.
\ee
Thus we get for the Feigenbaum doubling parameter
$\delta_2 = \lim_{p\rightarrow\infty} (r_p/r_{2p})$
\be
\delta_2 \approx 2\sqrt{5} = 4.4721\ldots
\ee
(exact value is known to be $\delta_2 = 4.6692\ldots$).
Note that $\delta_2$ approximately acquires this value already
for the second descendant of the root, far before the
$p \rightarrow \infty$ limit.

Consistency requires that
\be
c_{2p} + r_{2p} = c_p - r_p
\ee
i.e.
\be
-\sigma_{2p} = r_{2p} + r_p \approx
r_p\left(1+\frac{1}{2\sqrt5}\right)
\ee
This is indeed almost true:
$1+\frac{1}{2\sqrt5} \approx \sqrt{5}-1$\ \
($1.2236\ldots \approx 1.2361\ldots$).

The west-limit point $c_{\infty}^{(2)}$
 of the central cluster
(superscript is $(2)$ because the point is obtained
by a sequence of doublings of the order of the orbits)
can be represented as
\be
c_{\infty}^{(2)} \approx -\frac{3}{4} - 2r_2 - 2r_4 - \ldots
= -\frac{3}{4} - 2\sum_{k=1}^\infty r_{2^km} =
- \frac{3}{4} - \frac{1/2}{1-\frac{1}{2\sqrt{5}}} =
-1.3940\ldots
\ee
(we remind that for the first descendant
$r_2=\frac{1}{2}r_1 = \frac{1}{4}$) or, alternatively, as
\be
c_{\infty}^{(2)} = c_2 + \sum_{k=1}^\infty \sigma_{2^{k+1}} \approx
c_2 - \frac{r_2(\sqrt{5}-1)}{1-\frac{1}{2\sqrt{5}}} =
-1 -\frac{1}{2}\frac{5-\sqrt{5}}{2\sqrt{5}-1} = -1,3980\ldots
\ee
The difference between these two values characterize accuracy of
the SSA, and within such error they coincide with
exact value $c_{\infty}^{(1)} = -1.4012\ldots$.

\subsection{The general case (arbitrary $m$)}

Solving (\ref{cmp}), we obtain:
\be
\sigma_{mp,p} = \left\{\begin{array}{ccc}
2c_mr_p & {\rm if}\  \xi_p \approx 0 & {\rm i.e.\ for\
the\ } first\ {\rm descendant} \\ \\
(1-\sqrt{1-4c_m})r_p  &
{\rm if}\  \xi_p \approx 1 &
{\rm i.e.\ for\ a\ } higher\ {\rm descendant}\end{array}\right.
\label{sigmampp}
\ee
\be
r_{mp} = \left\{\begin{array}{ccc} 2r_mr_p  &
{\rm if}\  \xi_p \approx 0 & {\rm i.e.\ for\
the\ } first\ {\rm descendant} \\ \\
\frac{2r_mr_p}{\sqrt{1-4c_m}} & {\rm if}\  \xi_p \approx 1 &
{\rm i.e.\ for\ a\ } higher\ {\rm descendant}\end{array}\right.
\ee
Thus we obtain for the Feigenbaum parameter
$\delta_m = \lim_{p \rightarrow \infty} (r_p/r_{mp})$
\be
\delta_m \approx \frac{\sqrt{1-4c_m}}{2r_m}
\ee
and for the complex-valued ratio $\varepsilon_m = \sigma_{mp,p}/r_p$
we get
\be
\varepsilon_m = 1-\sqrt{1-4c_m}
\ee

When $(p,\ldots)$ is itself a descendant domain and has circle
rather than cardioid shape, the consistency condition
\be
|\sigma_{mp}| = |r_p| + |r_{mp}|,
\label{d=r+r}
\ee
expressing the distance between centers of two touching circles
through their radiuses,
implies that
\be
|\varepsilon_m| = 1 + |\delta_m|^{-1}
\label{modcons}
\ee
or
\be
|1-\sqrt{1-4c_m}| = 1 + \frac{2|r_m|}{|\sqrt{1-4c_m}|}
\ee
A more detailed consistency condition includes not only distances,
like (\ref{d=r+r}), but also exact position (the phase $\phi_m$)
of the touching point between the circles $(mp,p,\ldots)$ and
$(p,\ldots)$:
\be
c_{mp} + r_{mp} = e^{i\phi_m} r_p + c_p
\ee
This means that
\be
\varepsilon_m =  e^{i\phi_m} -\frac{1}{\delta_m}
\label{complcons}
\ee

The end-point of an infinite sequence of descendant domains
$(p)$, $(mp,p)$, $(m^2p,mp,p)$, $\ldots$
is given by
\be
c_{\infty}^{(p|m)} = c_p + \sum_{k=1}^\infty \sigma_{m^kp} =
c_{mp} + \varepsilon_m\sum_{k=1}^\infty r_{m^kp} =
c_{mp} + \frac{\varepsilon_m r_{mp}}{1-\delta_m^{-1}} =
c_p + 2r_p\left(c_m + \frac{r_m\varepsilon_m}{1-\delta_m^{-1}}
\right)
\ee
In particular, for the central cluster with $(p) = (1)$
\be
c_{\infty}^{(1|m)} =
c_m + \frac{r_m\varepsilon_m}{1-\delta_m^{-1}}
\label{endpointmm}
\ee

Within SSA the only input in all these formulas for a given $m$
consists of two complex numbers: $c_m = c_{m,1}$ and $r_m = r_{m,1}$,
characterizing the properties of the next-to-root domain
$(m,1)$ in the central cluster.
These $c_m$ and $r_m$ are entries of the table in s.\ref{accu}.
Taking $c_m$ and $r_m$ from that table, we now make a new one,
comparing predictions of eqs.(\ref{sigmampp})-(\ref{endpointmm})
with experimental data.
Numbers in square brackets in the last column are positions
of the limiting points, {\it measured} with the help of
{\it Fractal Explorer}.

\bigskip

{\footnotesize
$$
\hspace{-1.4cm}
\begin{array}{|c|c|c|c||c|c|c||c|c||c|}
\hline
&&&&&&&&&\\
{\rm domain}&c_m&2r_m&e^{i\phi_m}
&\sqrt{1-4c_m\phantom{5^5}\hspace{-0.4cm}}
&\varepsilon_m&\delta_m&
(\ref{modcons})&(\ref{complcons})& c_{\infty}^{(1|m)}\\
(m,1)&&&&&(\sigma_{mp}/r_p)&(r_p/r_{mp})&&&
{\rm from}\ (\ref{endpointmm}) \\
&&&&&&&&&\\
\hline
&&&&&&&&&\\
(2,1)&-1&0.5&-1&\sqrt{5}&1-\sqrt{5}&2\sqrt{5}&
\sqrt{5}-1&1-\sqrt{5}&-1 -\frac{1}{2}\frac{5-\sqrt{5}}{2\sqrt{5}-1}\\
&&&&&&&
\approx 1+\frac{1}{2\sqrt{5}}&\approx -1-\frac{1}{2\sqrt{5}}
&= -1.3980\\
&&&&&&&1.2361\approx 1.2236&&[-1.401]\\
&&&&&&&&&\\
\hline
&&&&&&&&&\\
(3,1)&-0.123&-0.01&-0.5&1.55&-0.55&4.61&1.107
&-0.553+0.959i&-0.020\\
&+0.745i&-0.19i=&+0.87i&-0.96i&+0.96i&+8.42i=&\approx 1.104
&\approx -0.550+0.957i&+0.785i\\
&&0.19\cdot e^{1.03\frac{i\pi}{2}}&\phi_3=\frac{2\pi}{3}&&
&9.59\cdot e^{1.02\frac{i\pi}{3}} &&&[-0.0234+0.7836i]\\
&&&&&&&&&\\
\hline
&&&&&&&&&\\
(4,1)&+0.282&-0.066&&0.999&0.001&-0.2848
&1.061&0.001+1.06i&0.3115\\
&+0.530i&-0.06i=&i&-1.061i&+1.061i&+16.34i=
&\approx 1.061&\approx 0.001+1.06i&+0.4932i\\
&&0.089\cdot e^{-1.02\frac{3i\pi}{4}}&\phi_4=\frac{2\pi}{4}&&
&16.34\cdot e^{1.01 \frac{i\pi}{2}}&&&[0.3098+0.4947i]\\
&&&&&&&&&\\
\hline
&&&&&&&&&\\
(5_{1},1)&0.380&-0.046&0.309&0.677&0.323&-9.06
&1.041&0.323+0.989i&0.377\\
&+0.335i&-0.011i=&+0.951i&-0.989i&+0.989i&+23.7i=
&\approx 1.039&\approx 0.323+0.988i&+0.311i\\
&&0.047\cdot e^{-0.93\pi i}&\phi_{5_1}=\frac{2\pi}{5}&&
&25.35\cdot e^{3.08\frac{i\pi}{5}}&&&[0.3770+0.3117i]\\
&&&&&&&&&\\
\hline
&&&&&&&&&\\
(5_{2},1)&-0.504&0.046&-0.809&1.841&-0.841&20.25
&1.040&-0.841+0.612i&-0.503\\
&+0.563i&-0.063i=&+0.588i&-0.612i&+0.612i&+14.44i=
&\approx 1.040&\approx -0.842+0.611i&+0.605i\\
&&0.078\cdot e^{-0.997\frac{3\pi i}{10}}
&\phi_{5_2}=\frac{4\pi}{5}&&
&24.87\cdot e^{0.99\frac{i\pi}{5}}&&&[-0.5031+0.6048i]\\
&&&&&&&&&\\
\hline
&&&&&&&&&\\
(6,1)&0.389&-0.028&0.5&0.487&0.513&-20.57
&1.028&0.513+0.891i&0.380\\
&+0.217i&+0.003i=&+0.866i&-0.891i&+0.891i&+29.61i =
&\approx 1.028&\approx 0.516+0.889i&+0.206i\\
&&0.028\cdot e^{-0.97\pi i}&\phi_6\frac{2\pi}{6}
&&&36.05\cdot e^{2.08\frac{i\pi}{3}}
&&&[0.3810+0.2047i]\\
&&&&&&&&&\\
\hline
\end{array}
\hspace{1.4cm}
$$
}

\noindent
Of course, one can consider limiting points of other
sequences, not obligatory of the type $[\ldots,m,m,m]$.
One of the open questions is if there is any
difference between periodic (after some step) and
aperiodic, i.e. "rational" and "irrational" sequences.
Another important class consists of sequences
$[\ldots,2,2,\ldots,2,m_r,\ldots,m_1]$, ending by
$2$'s only -- they describe {\it normals} to the
cluster's boundary and serve as origins of {\it trails},
connecting the cluster with its neighbors.

\section{Cardioids and resultant zeroes}

As explained in \cite{DM}, a boundary of domain
$(p,\ldots)$ is densely populated with a countable
set of its merging
points with descendant domains $(mp,p,\ldots)$,
located at zeroes of the resultants $R(G_{mp},G_p)$
with all integer $m$.
Since within SSA the boundaries are well approximated
by cardioids and circles, and merging points are
characterized by the angles $\phi_m = \frac{2\pi}{m}$,
one can expect that simple approximations exist for
locations of the resultant zeroes in terms of
$c_p$, $r_p$ and  $e^{ik\phi_m}$ with $k = 1,\ldots,m-1$.
This is indeed the case, at least for the Mandelbrot Set,
i.e. the family $\{f(x) = x^2+c\}$.

For example, the zeroes of $R(G_m,G_1)$
-- they can be found among the values of $c_{p+}$
in tables in s.\ref{accu} -- are given by
\be
c^{(1)}_{m_k}=\frac{e^{{2\pi i k}/{m}}}{2}
\left(1-\frac{e^{{2\pi i k}/{m}}}{2}\right)
\ee
The first values of this quantity are:

{\footnotesize
$$
\hspace{-0.9cm}
\begin{array}{|c|c|c|c|c|c|c|c|c|c|}
\hline
&&&&&&&&&\\
m_k&(2)&(3)&(4)&(5_1)&(5_2)&(6)&(7_1)&(7_2)&(7_3)\\
&&&&&&&&&\\
\hline
&&&&&&&&&\\
c^{(1)}_{m_k}&-0.75&-0.1249999999&0.25
&0.3567627456&-0.4817627458
&0.375&0.3673751344&0.1139817500&-0.6063568845\\
&&+0.6495190530i&+0.5i&0.3285819454i&0.5316567550i
&+0.2165063510i&+0.1471837632i&+0.5959348910i&+0.4123997402i\\
&&&&&&&&&\\
\hline
\end{array}
\hspace{0.9cm}
$$
}

\bigskip

\noindent
When $k$ are not shown, it is equal to unity, $k=1$.
Similarly, the zeroes of $R(G_{2m},G_2)$,
belonging to the boundary of descendant domain $(2,1)$,
are given by:
\be
c^{(2,1)}_{m_k} \approx c_{2} + r_2\,e^{{2\pi i k}/{m}}
= -1 + \frac{e^{{2\pi i k}/{m}}}{4}
\ee

{\footnotesize
$$
\hspace{-1.4cm}
\begin{array}{|c|c|c|c|c|c|c|c|c|c|}
\hline
&&&&&&&&&\\
m_k&(2)&(3)&(4)&(5_1)&(5_2)&(6)&(7_1)&(7_2)&(7_3)\\
&&&&&&&&&\\
\hline
&&&&&&&&&\\
c^{(2,1)}_{m_k}&-1.25&-1.125&-1&-0.9227457516&-1.202254249
&-0.875&-0.8441275496&-1.055630233&-1.225242217\\
&&+0.2165063510i&+0.25i&+0.2377641291i&+0.1469463130i
&+0.2165063510i&+0.1954578706i&+0.2437319780i&+0.1084709348i\\
&&&&&&&&&\\
\hline
\end{array}
\hspace{1.4cm}
$$
}

\bigskip

\noindent
and zeroes of $R(G_{3m},G_3)$, belonging to the boundary of
descendant domain $(3_\pm,1))$ -- by
\be
c^{(3_\pm\!,1)}_{m_k} \approx c_{3} + r_3\,e^{\pm{2\pi i k}/{m}}
= -0.1226\pm 0.7449i
- (0.0047\pm 0.0943i){e^{\pm {2\pi i k}/{m}}}
\ee

{\footnotesize
$$
\begin{array}{|c|c|c|c|c|c|c|c|c|c|}
\hline
&&&&&&&&&\\
m_k&(2)&(3)&(4)&(5,1)&(5,2)&(6)&(7,1)&(7,2)&(7,3)\\
&&&&&&&&&\\
\hline
&&&&&&&&&\\
c^{(3_\pm\!, 1)}_{m_k}
&-0.118&-0.039&-0.028&-0.034&-0.063
&-0.043&-0.052&-0.030&-0.077\\
&\pm 0.839i&\pm 0.788i&\pm 0.740i&\pm 0.711i&\pm 0.818i
&\pm 0.694i&\pm 0.682i&\pm 0.761i&\pm 0.828i\\
&&&&&&&&&\\
\hline
\end{array}
$$
}

\bigskip

\noindent
while those belonging to the boundary of
the root domain $(3)$ are
\be
c^{(3)}_{m_k} \approx c_{3}
+ r_3\,e^{{2\pi i k}/{m}}
\left(1-\frac{e^{{2\pi i k}/{m}}}{2}\right)
= -1.7549 + 0.0095{e^{{2\pi i k}/{m}}}
\left(1-\frac{e^{{2\pi i k}/{m}}}{2}\right)
\ee

{\footnotesize
$$
\begin{array}{|c|c|c|c|c|c|c|c|c|c|}
\hline
&&&&&&&&&\\
m_k&(2)&(3)&(4)&(5_1)&(5_2)&(6)&(7_1)&(7_2)&(7_3)\\
&&&&&&&&&\\
\hline
&&&&&&&&&\\
c^{(3)}_{m_k}
&-1.769&-1.757&-1.750&-1.748&-1.764&-1.748
&-1.748&-1.753&-1.767\\
&&+0.012i&+0.009i&+0.006i&+0.010i&+0.004i
&+0.003i&+0.011i&+0.008i\\
&&&&&&&&&\\
\hline
\end{array}
$$
}

\bigskip

\noindent
Since domains $(3,1)$ and $(3)$ are circle and cardioid
only approximately, accuracy in the last two tables is
relatively low and we do not keep as many digits as in the
first two tables.
Still, the numbers in the tables reproduce actual
positions of resultant zeroes at percent-level accuracy,
standard for the SSA in the case of the Mandelbrot Set.
Thus, not only the shapes of elementary domains
are nicely represented by cardioids and circles,
but all the merging points of stable orbits at the
boundaries (zeroes of the corresponding resultants,
\cite{DM}) can be easily found by the SSA methods.

\section{The case of $Z_{d-1}$-symmetric maps $f(x;c)=x^d+c$
\label{dfam}}

\setcounter{equation}{0}

\subsection{SSA in the case of $Z_{d-1}$-symmetry}

In this case all the iterated maps are expanded in powers
of $x^d$ and in SSA we truncate them as follows:
$f^{\circ p}(x;c) = f_p(c) + x^d\gamma_p(c) + O(x^{2d})$.
Then the boundary of elementary domain, surrounding
a root $c_p$ of $f_p(c)$, is defined by
\be
\left\{\begin{array}{c}
f_p(c) = x\Big(1 - x^{d-1}\gamma_p(c)\Big) \\
d\cdot x^{d-1}\gamma_p(c) =  e^{i(d-1)\phi}
\end{array}\right.
\ee
or, as generalization of (\ref{shape}),
\be
f_p(c) \Big(d\gamma_p(c)\Big)^{\frac{1}{d-1}} =
e^{i\phi}\left(1 - \frac{e^{i(d-1)\phi}}{d}\right)
\label{shaped}
\ee
Now we need to expand the l.h.s. in powers of $\sigma = c-c_p$
and leave the first $d$ terms of the expansion.

\subsection{Example of the $p=2$ domains for $d=3$ and $d=4$}

We consider here
the first descendants of the central elementary domain $c=0$
in the case of $d=3$ and $d=4$:
relations like (\ref{ximp}) should be used to extend the
result to all other descendants.
Also we restrict example to $p=2$ only.

From $f^{\circ 2}(x,c) = (x^3+c)^3+c$ we read:
\be
f_2(c) = c(c^2 + 1),\nn \\
\gamma_2(c) = 3c^2
\ee
and the critical values $c_2 = \pm i$.
Eq.(\ref{shaped}) now states:
\be
3c^2(c^2+1) = u\Big(1-\frac{1}{3}u^2\Big)
\label{shaped32}
\ee
with $u = e^{i\phi}$, we need to substitute $c = \pm i + \sigma$
and check that (\ref{shaped32}) is approximately -- modulo terms
$\sim O(u^4)$ -- solved by
\be
\sigma = r_2u\left(1- \left[\frac{1}{2}-a\right]u - bu^2\right)
\label{cuca}
\ee
with negligibly small $a$ and $b$.
Substitution of this ansatz into (\ref{shaped32}) gives
$a = \frac{1}{12} \ll 1$ and
$b = \frac{1}{24} \ll 1$.
Also from the same calculation $r_2 = \frac{i}{6}$,
and this is in good accordance with reality: the first descendant
domain $(2,1)_+$ in Fig.\ref{0Man3} is bounded by the points
$c_+ = 1,09i$ and $c_-=0.77i$, so that
$c_+ - c_- = 0.32i \approx 2r_2 = 0.33i$.

Similarly, for $d=4$ we have
\be
2^{4/3}c^2(c^3+1) = u\left(1 - \frac{1}{4}u^3\right)
\ee
and for $c = \omega_3 + \sigma$, $\omega_3 = e^{i\pi /3}$
we obtain
\be
\sigma = \frac{\omega}{6\cdot 2^{1/3}}(\omega u)
\Big(1-a(\omega u) + \frac{b}{3}(\omega u)^2 - c(\omega u)^3\Big)
\label{d4card}\label{quca}
\ee
with $a = 2^{-4/3} = 0.40$, $b = 11\cdot 4^{-1/3}/9 = 0.77$
and $c = 4/81 \ll 1$.
The biggest "diameter" of this elementary domain is
$2\frac{1}{6\cdot 2^{1/3}}\left(1+\frac{b}{3}\right) \approx 0.33$,
in good agreement with $c_- = -1.10$, $c_+ = -0.78$ for the
$(2,1)$ domain in Fig.\ref{0Man4}. The distance between the two
cusps of this deformed cardioid is approximately $0.4$ of the
biggest diameter, what is also in agreement with Fig.\ref{0Man4}
(ordinate of the cusp, which is shown by arrow in the picture,
is $0.063$, and $2\cdot 0.063/0.33 \approx 0.4$).
Since $b < 1$ in (\ref{d4card}), the cusps have finite angles,
what is {\it not} confirmed by Fig.\ref{Mand2}: the true value
of $b$ is close to unity -- the difference $1-b\approx 0.23$ is
inaccuracy of SSA in this example.

\section{Conclusion \label{conc}}

\setcounter{equation}{0}

In this paper we {\it calculated} the shapes of {\it elementary
domains} of the Mandelbrot set \cite{Mand}, following the
general algebro-geometric approach of \cite{DM}.
We explained the qualitative features of these shapes,
found the origin and number of cusps, explicitly showed how they
change when one Mandelbrot set is deformed into another inside
the unifying Universal Mandelbrot set.
We showed that the nearly ideal cardioid and circle shapes of
these domains in ${\cal M}_2$ (Fig.\ref{Mand2})
are nicely described in the
small-size approximation, based on truncating the relevant
polynomials to the first orders in deviations $x-x_{cr}$ and
$c-c_{cr}$ from their critical values.
It is not a big surprise, but some conspiracy is needed --
and was indeed found in the behavior of parameter $\xi_p$,
which is not always small, as one could naively expect --
to explain the coexistence of {\it different} structures:
cardioids of different orders.

We did not give a {\it theoretical} justification of the
{\it small-size approximation} -- next-order corrections were
not estimated -- instead its percents-order accuracy was
demonstrated by comparison of its predictions with the
properties of the actual Mandelbrot set (measured with the
help of the {\it Fractal Explorer} \cite{FE}).
Accuracy is actually much higher than one could expect from
the over-simplified calculations in \cite{LL}, for example the
small-size-approximation of the ordinary Feigenbaum constant
$\delta_2^{(SSA)} = 2\sqrt{5} = 4.4714\ldots$ is much closer
to experimental value $\delta_2 = 4.6692\ldots$\ than
$\delta_2^{(LL)} = 4+\sqrt{3} = 5.7321\ldots$\ of ref.\cite{LL}.
The systematic approach allows to find {\it all}
Feigenbaum indices in the same way, moreover other
characteristics, including continuous, like {\it shapes}
of elementary domains, not only their {\it sizes},
are straightforwardly {\it calculated}.

We demonstrated that
characteristics of elementary domains in ${\cal M}_2$
are nicely encoded by two parameters
like $r_p$ and $\xi_p$, which  by recursive formulas like
\be
r_{mp} \approx r_p\cdot\delta^{-1}(c_{m,1})
\ee
and
\be
\xi_p \approx \left\{
\begin{array}{cc}
0 & {\rm for\ the\ root\ domain}\\
1 & {\rm for\ other\ domains}
\end{array} \right.
\ee
are expressed through the size $r_{root}$  of the root domain
in the given cluster and through the critical values $c_{m,1}$
-- positions of centers of immediate descendants of the central
root domain.
However, these remaining parameters need to be evaluated
from sophisticated algebraic equations.
As explained in \cite{DM}, the equations emerge from universal
structures in particular {\it section} of
the Universal Mandelbrot set (UMS).
Naturally, some characteristics of such arbitrary section look
arbitrary -- at least from its {\it internal} perspective.
Hopefully, a better understanding of $r_{root}$ and $c_m$
distributions can be found at the level of UMS, but this remains
beyond the scope of the present paper.

\bigskip

It remains to emphasize that investigation of Mandelbrot sets
is not just an interesting problem by itself, it is crucial
for understanding of the future physics, which is going to
deal with essentially multi-phase systems, far from equilibrium
and from the trivial end-points of renormalization groups.
One of the main lessons of Mandelbrot theory \cite{DM} is that
phase transitions are not just rare isolated events, concentrated
on smooth hypersurfaces in the space of coupling constants.
Examples of {\it such} phase transitions are given by particular
merging points between two elementary domains (say, between
$(2,1)$ and $(1)$) -- these isolated points in particular
${\cal M}_d$ in Fig.\ref{hompols} form a nice
complex-codimension-one hypersurface in UMS (partly represented
in Fig.\ref{tubeD3}).
However, the true picture -- Figs.\ref{Mand2}-\ref{0Man4} --
is very different: the entire variety of various phase transitions
(mergings of {\it all} elementary domains of {\it all} orders)
is not just a collection of particular transition lines.
Instead they form a profound new structure, moreover they
tend to condense and fill entire boundaries of elementary domains,
i.e. dimension of the phase transitions variety increases as
compared to the naive one
(and actually its {\it real}, not {\it complex} codimension
in the space of complex couplings, is one!).
Within particular slices like particular Mandelbrot sets,
different phases now get fully disconnected, and
analytical continuation between them, if at all possible,
essentially depends on the properties of the new fundamental entity:
the UMS, which scientists even did not begin to study!
It is the UMS that is behind the sophisticated phase structure
\cite{AMM} of stringy  $\tau$-functions -- effective actions of
various multi-phase systems, classical or quantum.
It is the UMS that one encounters in various problems,
from baby-universe
creation in modern cosmological models to optimization of cooling
processes in various solid-state technologies.
Still, despite its central role in the mysteries of uncertainty,
there is no mystery in the UMS itself: it is one of the most
important and structurized mathematical objects -- the universal
discrminantal variety, a would-be classical topic of algebraic
geometry, which, however, did not attract much attention so far.
We believe that time has come for its investigation and this
paper is just a modest example of how one can approach the
fundamental problems of this kind: very simple methods are quite
effective and produce answers, which are not easy to foresee, and
numbers, which are not easy to guess. This looks like a real and
wonderful science to do.

\section{Appendix. Some elementary MAPLE programs for UMS studies
\label{MAProgs}}

We did our best to illustrate quantitative considerations of
Universal Mandelbrot Set and its particular sections with modest
illustrations.
However, the number of illustrations in a printed text is
necessarily restricted and can be non-sufficient for full
visualization of the object.
In order to cure this problem we collect in this appendix
a set of sample MAPLE programs, which were used to generate
some illustrations in the text.
One can easily play with these simple programs, change parameters,
accuracy of calculation and output formats in order to extract
more information, numerical and visual.
Programs are super-primitive, transparent and easy to modify,
they work fast and smoothly on ordinary PC's.
One can straightforwardly copy them into MAPLE file
(with $.mws$ extension) and use.
When substituting desired parameters instead of the question marks,
one should better do it in rational rather than decimal form,
say $a=1/10$ rather than $a=0.1$.

\subsection{Cardioids \label{MAPcard}}

MAPLE program for cardioid studies consists of just four lines:
\begin{verbatim}
> r:=1:
> a:=?: b:=-(1+2*a)/3;
> f:=r*(exp(I*t)+a*exp(I*2*t)+b*exp(I*3*t));
> plot([Re(f),Im(f),t=-Pi..Pi],scaling=CONSTRAINED);
\end{verbatim}
(cubic case is presented, generalization is obvious).
It remains to substitute various $a$ and $b$
instead of the question marks (say, $a = 0.4 + 0.2*I$)
and enjoy the pictures.
For looking at more details, especially at the critical
values $t=0$ and $t=\pi$, where cusps can occur
(or $t = \frac{2\pi k}{d-1}$ in general case) one can
enhance resolution:
\begin{verbatim}
> plot([Re(c),Im(c),t=Pi-0.01..Pi+0.01],scaling=constrained);
> plot([t,Re(c)/Im(c),t=Pi-0.01..Pi+0.01]);
> plot([t,Im(c)/Re(c),t=-0.01..+0.01]);
\end{verbatim}

Examples of output of this program are shown in
Figs.\ref{cardi} and \ref{cardicusp}.

\subsection{UMS through discriminants and resultants}

\begin{verbatim}
> F1:=f(x)-x:
> F2:=f(f(x))-x:
> F3:=f(f(f(x)))-x:
> F4:=f(f(f(f(x))))-x:
> F5:=f(f(f(f(f(x)))))-x:
> F6:=f(f(f(f(f(f(x))))))-x:
> G1:=F1:
> G2:=simplify(F2/G1):
> G3:=simplify(F3/G1):
> G4:=simplify(F4/(G2*G1)));
> G6:=simplify(F6/(G1*G2*G3));
> ...
> D2:=discrim(G2,x);
> D3:=discrim(G3,x);
> ...
> R24:=resultant(G2,G4,x);
> R36:=resultant(G3,G6,x);
> ...
\end{verbatim}

\subsection{Domains $(1)$, $(2)$ and $(2,1)$ of
${\cal M}_{ax^3+(1-a)x^2+c}$}

Parameter $M$ in the program defines the number $T$ of points
in the picture. The bigger $M$ the more detailed will be the
plot, but computer time will also increase.
To make sure that the program is working we added the line
$print("k=",k)$, one can safely omit it.

\bigskip

\begin{verbatim}
> with(plots):
>
> unassign('a','b','u','z','t','c'):
>
> a:=?:
> b:=1-a:
>
> D1:=factor(discrim(a*x^3+b*x^2+c-x,x));
> R21:=factor(resultant(a^3*x^6+2*a^2*x^5*b+a^2*x^4+2*a^2*x^3*c+a*b^2*x^4+2*a*x^3*b+a*x^2+2*x^2*c*a*b+x*c*a+a*c^2+b^2*x^2+x*b+c*b+1,c-x+a*x^3+b*x^2,x)):
> D2:=factor(simplify(discrim(a^3*x^6+2*a^2*x^5*b+a^2*x^4+2*a^2*x^3*c+a*b^2*x^4+2*a*x^3*b+a*x^2+2*x^2*c*a*b+x*c*a+a*c^2+b^2*x^2+x*b+c*b+1,x)/R21));
>
> ## various choices of s and MID
> #s:=evalf(solve(D1,c)):
> s:=evalf(solve(D2,c));
> #s:=evalf(solve(R21,c));
> MID:=s[1];
> #MID:=(s[1]+s[2])/2.;
>
> P:= x -> a*x^3+b*x^2:
> u:=exp(I*t):
>
> zp:=(x,t)->(-b+root[2](b^2+3*a*exp(I*t)/(3*a*x^2+2*b*x)))/(3*a):
> zm:=(x,t)->(-b-root[2](b^2+3*a*exp(I*t)/(3*a*x^2+2*b*x)))/(3*a):
> sp:=solve(P(zp(x,t))+zp(x,t)-P(x)-x,x):
> sm:=solve(P(zm(x,t))+zm(x,t)-P(x)-x,x):
>
> Tp:=0: Tm:=0:
> M:=200:
>      for k to M do
>
> t:=evalf(2*Pi*k/M):
> N:=ArrayNumElems(Array([sp]));
> for i to N do
>
> wp:=allvalues(sp[i]): wm:=allvalues(sm[i]):
> n:=ArrayNumElems(Array([wp])): # nm:=ArrayNumElems(Array([wm])): print(n,nm);
>   for j to n do
>
> Tp:=Tp+1: Tm:=Tm+1:
> if n >1 then
> Xp:=evalf(wp[j]):   Xm:=evalf(wm[j]):
> else
> Xp:=evalf(wp):      Xm:=evalf(wm):
> end if:
> Pp[Tp]:=evalf(zp(Xp,t)-(a*Xp^3+b*Xp^2)):
> Pm[Tm]:=evalf(zm(Xm,t)-(a*Xm^3+b*Xm^2)):
>
>   xp1:=Xp:                       xm1:=Xm:
>   zp1:=a*xp1^3+b*xp1^2+Pp[Tp]:   zm1:=a*xm1^3+b*xm1^2+Pm[Tm]:
>   chp:=a*zp1^3+b*zp1^2+Pp[Tp]-xp1:
>   chm:=a*zm1^3+b*zm1^2+Pm[Tm]-xm1:
>   ap:= evalf(Re(chp)^2+Im(chp)^2):  am:=evalf(Re(chm)^2+Im(chm)^2):
>
> # MAGNIFY (Enhanced resolution for vicinity of a chosen value of 'c')
> ## CENTER POSITION
> zz:=s[1];
> ### version of defining zz
> #zz:=MID+I*0.:
> ## RADIUS
> rr:=0.3;
> if rr>0 then
>   if (ap>10^(-5)) or abs(Pp[Tp]-zz)>rr then
>     Tp:=Tp-1:
>   else
>   fi:
>   if (am>10^(-5)) or abs(Pm[Tm]-zz)>rr then
>     Tm:=Tm-1:
>   else
>   fi:
> else
>   if (ap>10^(-5)) then  Tp:=Tp-1: fi:
>   if (am>10^(-5)) then  Tm:=Tm-1: fi:
> fi:
>
>   od:
> od:
>      od:
>
> pp:=pointplot({seq([Re(Pp[n]),Im(Pp[n])],n=1..Tp)},
           scaling=CONSTRAINED,color=red,symbol=circle,symbolsize=5):
> pm:=pointplot({seq([Re(Pm[n]),Im(Pm[n])],n=1..Tm)},
           scaling=CONSTRAINED,color=red,symbol=circle,symbolsize=5):
>
> display({pp},{pm});a;
\end{verbatim}


\subsection{3D tubes}

\begin{verbatim}
> unassign('a','b','u','z','t','c'):
>
> a:=b->b^3:
> c:=b->1.:
> # |f'| VALUE
> MD:=1:
> sp:=solve(4*a(b)*x^3+3*b*x^2+2*c(b)*x-MD*exp(I*t),x):
>
> Tp:=0: Tm:=0:
> M:=00:
> M1:=15:M2:=60:
> zmi:=.2:zma:=.8:
>      for k1 to M1 do
>        print("k=",k1);
>      for k2 to M2 do
>
> t:=evalf(2*Pi*k1/M1):
> b:=zmi+(zma-zmi)*k2/M2:
>
> N:=ArrayNumElems(Array([sp]));
> for i to N do
>
> wp:=allvalues(sp[i]):
> n:=ArrayNumElems(Array([wp])):
>   for j to n do
> Tp:=Tp+1:
> if n >1 then
> Xp:=evalf(wp[j]):
> else
> Xp:=evalf(wp):
> end if:
> u:=evalf(Xp-(a(b)*Xp^4+b*Xp^3+c(b)*Xp^2)):
> Pp[Tp]:=array([Re(u),Im(u),b]):
>   xp1:=Xp:
>   zp1:=a(b)*xp1^4+b*xp1^3+c(b)*xp1^2+Pp[Tp][1]+I*Pp[Tp][2]:
>   chp:=a(b)*xp1^4+b*xp1^3+c(b)*xp1^2+Pp[Tp][1]+I*Pp[Tp][2]-xp1:
>   ap:= evalf(Re(chp)^2+Im(chp)^2):
>   if (ap>10^(-5)) then
> Tp:=Tp-1:
> fi:
>   od:
> od:
>      od:
>      od:



> # PREPARE ARRAY FOR 3D PLOT
> L:=1:
> N:=Tp+Tm;
> B:=array(1..N):
> k:=0:j:=0:
>
> for i to Tp do
> k:=k+1:
> B[k]:=Pp[i];
> od:
>
> for i to Tm do
> k:=k+1:
> B[k]:=Pm[i];
> od:
>
> j:=j+1:
> print(k,j,B[k]);



> # PLOT
> with(linalg):
> with(plots):
> with(plottools):
> setoptions3d(color=BLUE,symbol=CROSS,symbolsize=3);
> p:=pointplot3d(B,axes=BOXED):
> display(p);

\end{verbatim}

\subsection{Fragments of Julia sheaf ${\cal J}_{ax^3+(1-a)x^2+c}$:
orbits of orders $1$ and $2$ vs $c$ and $a$}


\begin{verbatim}

> unassign('x','a','b','c'):
> f:=x->a*x^3+b*x^2+c;
> simplify(diff(f(f(x)),x));
> fp1:=x->diff(f(x),x):
> fp:=x->diff(f(f(x)),x):
> G1:=f(x)-x;
> F2:=f(f(x))-x:
> G2:=simplify(F2/G1);


> ################################################
> a:=1/10;
> b:=1.-a;
>
> D1:=factor(discrim(a*x^3+b*x^2+c-x,x)):
> R21:=factor(resultant(a^3*x^6+2*a^2*x^5*b+a^2*x^4+2*a^2*x^3*c+
  a*b^2*x^4+2*a*x^3*b+a*x^2+2*x^2*c*a*b+x*c*a+a*c^2+b^2*x^2+x*b+c*b+1,c-x+a*x^3+b*x^2,x)):
> D2:=factor(discrim(a^3*x^6+2*a^2*x^5*b+a^2*x^4+2*a^2*x^3*c+
  a*b^2*x^4+2*a*x^3*b+a*x^2+2*x^2*c*a*b+x*c*a+a*c^2+b^2*x^2+x*b+c*b+1,x))/R21:
>
> # GET MIDDLE POINT
> s:=evalf(solve(D1,c)):
> ## versions of defining 's'
> #s:=evalf(solve(D2,c));
> #s:=evalf(solve(R21,c));
> MID:=(s[1]+s[2])/2.;
>
> # CHOOSE C VALUE
> c:=-6.24+I*0.:
> ## version of defining 'c'
> #c:=MID+I*0.;
>
> s1:=solve(G1,x);
> s2:=solve(G2,x);
>
> N1:=ArrayNumElems(Array([s1]));
> N:=ArrayNumElems(Array([s2]));
>
> # GET PAIRS
> k:=0:
> for i to N do
> for j from i+1 to N do
>   if abs(f(s2[i])-s2[j])<0.0001 then
> k:=k+1:
> P[k][1]:=i:
> P[k][2]:=j:
>   fi:
> od:
> od:
>
> print(P);


> # SHOW ROOT POSITION
> with(plots):
> p0:=pointplot({[Re(s1[1]),Im(s1[1])],[Re(s1[2]),Im(s1[2])],[Re(s1[3]),Im(s1[3])]},
  color=BLACK,symbol=CROSS,symbolsize=15):
> p11:=pointplot({[Re(s2[P[1][1]]),Im(s2[P[1][1]])]},color=red):
> p12:=pointplot({[Re(s2[P[1][2]]),Im(s2[P[1][2]])]},color=red):
> p21:=pointplot({[Re(s2[P[2][1]]),Im(s2[P[2][1]])]},color=green):
> p22:=pointplot({[Re(s2[P[2][2]]),Im(s2[P[2][2]])]},color=green):
> p31:=pointplot({[Re(s2[P[3][1]]),Im(s2[P[3][1]])]},color=blue):
> p32:=pointplot({[Re(s2[P[3][2]]),Im(s2[P[3][2]])]},color=blue):
> display({p0,p11,p12,p21,p22,p31,p32});
\end{verbatim}

\subsubsection{Stability of orbits}

\begin{verbatim}

> # GET STABILITY INFO
> print("ORDER 1");
> F1:=fp1(x):
> for i to N/2 do
> x:=s1[i];
> print(abs(F1),x);
> od:
> unassign('x');
>
> print("ORDER 2");
> F:=fp(x):
> ## versions of defining 'F'
> #F:=2*b*x;
> #F:=3*a*x^2+2*b*x;
>
> for i to N/2 do
> x:=s2[P[i][1]];
> print(abs(F),x,f(x));
> od:
> unassign('x');

\end{verbatim}

\subsubsection{Attraction pattern}

\begin{verbatim}
> unassign('a','b','c'):
> f:=x->a*x^3+b*x^2+c;
> F2:=factor(f(f(x))-x);
>
> a:=1/3: b:=1-a:
> R21:=factor(resultant(a^3*x^6+2*a^2*x^5*b+a^2*x^4+2*a^2*x^3*c+
  a*b^2*x^4+2*a*x^3*b+a*x^2+2*x^2*c*a*b+x*c*a+a*c^2+b^2*x^2+x*b+c*b+1,c-x+a*x^3+b*x^2,x));
> D2:=factor(discrim(a^3*x^6+2*a^2*x^5*b+a^2*x^4+2*a^2*x^3*c+
  a*b^2*x^4+2*a*x^3*b+a*x^2+2*x^2*c*a*b+x*c*a+a*c^2+b^2*x^2+x*b+c*b+1,x));
>
> r12:=evalf(solve(R21,c));
> d2:=evalf(solve(D2,c));


> ND:=4:
> k:=0:
> for t to ND+1 do
> c:=d2[3]+0.01*exp(I*2*Pi/ND*(t-1)+I*Pi/2);
> C[t]:=c;
> sx:=evalf(solve(a^3*x^6+2*b*a^2*x^5+a^2*x^4+2*a^2*x^3*c+a*b^2*x^4+
  2*b*a*x^3+a*x^2+2*b*a*x^2*c+a*x*c+a*c^2+b^2*x^2+b*x+c*b+1,x));
>   for j to 6 do
> R[(t-1)*6+j]:=sx[j];
> k:=k+1:
>   od:
> print("t=",t);
> od:
> print(k);

> # COMPUTE PATHS WITH RANDOM START
> rf:=rand(-100..100):
> L:=1:
> c:=evalf(C[L+1]);
> k:=0:
> BL:=1:EL:=20:
> LM:=2.8:
> for i from BL to EL do
>
> cx:=rf()/30+I*rf()/30:
> CX:=cx:
> N:=1:
> j1:=0:
> ep:=0.0003:
>   for j to N do
> cx:=evalf(CX+0*exp(I*2*Pi*j/N)*ep):
> abs(cx-CX);
> for i1 to 50 do
> cx:=f(cx);
>   if abs(Re(cx))<LM and abs(Im(cx))<LM then
>   k:=k+1;
>   S[k]:=cx;
>   fi:
> od:
>   od:
>
> od:
> print(k);


> # PLOT DATA
> with(plots):
> pr:=pointplot({seq([Re(R[n]),Im(R[n])],n=L*6+1..L*6+6)},
  scaling=CONSTRAINED,color=blue,symbol=cross,symbolsize=25):
> ps1:=pointplot({seq([Re(S[n]),Im(S[n])],n=1..k)},
  scaling=CONSTRAINED,color=red,symbol=circle,symbolsize=5):
> display({pr,ps1});

\end{verbatim}

\section{Acknowledgements}

This work is partly supported by Russian Nuclear Ministry, by RFBR
grants 07-01-00644 and 07-02-00645, by NWO 047.011.2004.026,
ANR-05-BLAN-0029-01 and E.I.N.S.T.E.IN 06-01-92059 projects and by
the Russian President's grant for support of the scientific
schools LSS-8004.2006.2.

\end{document}